# Local Pitch Angles of Galactic Structures in Nearby Galaxies with JWST: I. Spiral Arms and Interarm Substructures in Disk-Dominated Systems

Stella Faria-Costa[1⋆], Irapuan Rodrigues[1,2] and Ivânio Puerari[3]
[1] Instituto de Pesquisa e Desenvolvimento, Universidade do Vale do Paraíba, Av. Shishima Hifumi - 2911, 12244-000, São José dos Campos, São Paulo, Brazil
[2] Laboratório Nacional de Astrofísica, Rua Estados Unidos 154, 37504-364, Itajubá, Minas Gerais, Brazil
[3] Instituto Nacional de Astrofísica, Óptica y Electrónica, Calle Luis Enrique Erro 1, 72840, Santa María Tonantzintla, Puebla, México

## ABSTRACT
We present a spatially resolved study of spiral arms and feather substructures in seven nearby galaxies observed with JWST. Using eight images per galaxy, from four NIRCam and four MIRI filters, we combine a two-dimensional Fourier transform (2D-FT) with a circular-window correlation method to measure both global and local pitch angles. The 2D-FT identifies the dominant azimuthal mode and the corresponding $p_{\max}$, allowing us to reconstruct the spiral pattern and define arm and interarm masks. We then apply cosine filters with different pitch angles and window sizes in the $(\ln(R), \theta)$ plane to map the pitch-angle distribution across multiple spatial scales. NIRCam images provide the most robust measurements of the main spiral pattern, whereas MIRI data reveal dusty interarm features, particularly feathers, with greater contrast at the smallest window size ($D_w = 0.25$). Across the sample, the main arms typically have $|\alpha| \simeq 15°$–$30°$, while feathers show systematically larger values, often above 50°. These results demonstrate that pitch angle is scale dependent and that different structural components within the same galaxy can occupy distinct angular regimes. NIRCam and MIRI therefore offer complementary views of spiral structure, with the former tracing the large-scale stellar backbone and the latter exposing finer dust substructures. Together with the companion study of barred galaxies, this work provides a common framework for using local pitch angles to investigate distinct dynamical environments in nearby galaxies.

## 1 INTRODUCTION

Spiral galaxies exhibit a rich diversity of structural patterns, varying in the number of arms, their degree of winding, the smoothness of the stellar and gaseous distribution, surface brightness, and the presence or absence of bars (Carroll & Ostlie 2007). These morphological characteristics are not merely visual traits: spiral arms play a crucial role in regulating the dynamics and long-term evolution of galactic disks, channeling gas flows, shaping stellar orbits, and influencing star formation.

Beyond their main arms, many spiral galaxies also display secondary filamentary structures, including feathers and spurs, which extend from the primary arms into the interarm regions. These features are important tracers of the physical processes that shape galactic disks. Theoretical models and simulations suggest that such substructures may arise from gravitational and magneto-Jeans instabilities, shearing motions, or other dynamical processes (Balbus 1988; Kim & Ostriker 2002; Wada & Koda 2004; Dobbs et al. 2006; Dobbs & Bonnell 2006; Shetty & Ostriker 2006). Their identification and characterization can therefore provide valuable constraints on the interplay between galactic dynamics, interstellar medium structure, and star formation.

The physical framework underlying spiral structure was first formalized by C.C. Lin and Frank Shu in the 1960s, through the density-wave theory, which proposes that spiral arms arise from long-lived, quasi-stationary density waves propagating through the disk (Lin & Shu 1964). As interstellar gas and dust encounter these density waves, they are compressed, triggering star formation and enhancing the visibility of spiral arms and their substructures. In this context, analysing both the large-scale geometry of spiral arms and smaller-scale arm and interarm substructures provides a way to investigate the interplay between galactic dynamics and star formation.

Logarithmic spirals are commonly used as an approximation to the geometry of spiral arms in disk galaxies (Groot 1925, 1926; Reynolds 1925; Danver 1942; Rots 1975; Boeshaar & Hodge 1977). In an ideal logarithmic spiral the pitch angle is constant with radius, whereas real galaxies can depart from this idealized geometry and exhibit radial and local variations. The pitch angle is defined as the angle between the tangent to a circle centered on the galactic center and the tangent to a logarithmic spiral at a given radius. In this framework, small absolute values of the pitch angle are associated with tightly wound spirals, as in Sa galaxies, whereas large absolute values correspond to more open spirals, as in Sc galaxies (Davis et al. 2012).

With respect to the appearance and number of arms, spiral galaxies can be classified as grand design (well-defined and symmetric arms), flocculent (poorly defined, discontinuous, and irregular arms), or multi-armed (intermediate characteristics) (Elmegreen & Elmegreen 1987). Spiral arms are not smooth, continuous features. Typically, an arm is composed of many substructures commonly referred to as feathers, spurs, and branches, which give the arm its patchy and divaricate appearance (La Vigne et al. 2006). The term spurs refers to stellar interarm features that extend outward from the main arms of the galaxy, but differ from branches, which occur when a spiral arm splits into two or more arms. Spurs extend from the outer side of the arm into the interarm region and are seen in photographic plates as stellar features, generally exhibiting pitch angles larger than those measured in the main spiral arm structure (Weaver 1970). Similarly, feathers are defined as gaseous or dusty features projecting from the main stellar arms, observed by Lynds (1970) as thin dust lanes with large pitch angles crossing the luminous arms.

⋆ E-mail: stellafariacosta@gmail.com (SFC)

Historically, these features and substructures have been frequently observed in many spiral galaxies and have been the subject of study for several decades, with the aim of understanding, explaining, and describing them through observational measurements (Aalto et al. 1999; Corder et al. 2008; Schinnerer et al. 2017) or theoretical models (Balbus 1988; Kim & Ostriker 2002; Wada & Koda 2004; Dobbs et al. 2006; Dobbs & Bonnell 2006; Shetty & Ostriker 2006).

According to Weaver (1970), the Sun is not located in one of the main spiral arms, but rather in a spur or offshoot originating near the Sagittarius arm, indicating the presence of various spurs and interarm features, similar to those observed in other galaxies.

In Elmegreen (1980), the spurs of the spiral arms were analyzed in B- and I-band photographs of the NGC 628, NGC 2403, NGC 3031, NGC 4321, NGC 5194, NGC 5236, and NGC 5457 galaxies. The study revealed that these spurs appear to be long-lived features, with pitch angles around 60°. Russell & Roberts (1992) carried out a study using a modified numerical method originally developed to analyze the morphology of global and local structures in prototype galaxies. This method was applied to the NGC 1232 and NGC 5457 galaxies. The authors combined partitioning techniques with a least-squares geometric regression algorithm to isolate local segments of arms, spurs, feathers, and secondary features, and to measure their pitch angle and length. The results revealed spurs with pitch angles between 30° and 50°, certainly larger than those of the main spiral arms, highlighting that prominent spurs tend to exhibit even larger pitch angles.

La Vigne et al. (2006) conducted a survey of feathers in the arms of 223 spiral galaxies using images from the *Hubble Space Telescope* (HST). Galaxies of types Sb and Sc showed a significant presence of feathers, whereas Scd and Sd galaxies did not exhibit clearly defined feathers. Consistent with earlier work, neighboring feathers tend to have similar shapes and pitch angles. In addition, the study provides a detailed distinction between the terms feathers, spurs, and branches, which can be useful for clarifying the nomenclature of spiral galaxy substructures.

Near-infrared (NIR) and mid-infrared (MIR) observations provide complementary views of star formation and the interstellar medium (ISM) in nearby galaxies. In particular, PHANGS–JWST observations have demonstrated that mid-infrared emission traces both gas column density and dust heating, including emission associated with star-forming regions (Leroy et al. 2023). Since the earliest stages of star formation can be heavily obscured by dust, infrared observations are particularly valuable for identifying young stellar populations that are difficult to detect at optical wavelengths

The unprecedented spatial resolution of JWST enables the identification of spiral substructures at scales that were previously unresolved in the infrared. We exploit the complementary capabilities of NIRCam and MIRI: while NIRCam is more sensitive to stellar light tracing the underlying gravitational potential, MIRI reveals dust and warm ISM structures closely associated with ongoing or recent star formation. This dual-instrument approach allows us to simultaneously characterize the geometry of the main spiral arms and the finer-scale features such as feathers, providing a more complete picture of the disk structure.

The *Physics at High Angular resolution in Nearby GalaxieS* (PHANGS) survey conducts high-resolution observations of nearby galaxies through the collaboration of several facilities, including the *Atacama Large Millimeter/submillimeter Array* (ALMA) (Leroy et al. 2021), the *Very Large Telescope* (VLT) (Emsellem et al. 2022), the HST (Lee et al. 2022), and the JWST (Lee et al. 2023). The PHANGS–JWST Cycle 1 Treasury survey targets 19 nearby spiral galaxies ($d < 20$ Mpc) in eight JWST filters, covering both the NIR and MIR, with wavelengths ranging from 2 $\mu$m to 21 $\mu$m. These observations reveal detailed dust and gas structures. Using the 3.3 $\mu$m PAH emission feature in NGC 7496, Rodríguez et al. (2023) identified candidate dust-embedded star clusters, many of which were absent from the optically selected PHANGS–HST cluster catalogue.

Out of the 19 galaxies observed by the PHANGS–JWST project, we selected 13 for the structural analysis presented across this two-paper study. The selected galaxies were organized into two groups based on the presence of bars and the dominant morphological features within the observed field. The first group comprises SA and SAB galaxies, in which large-scale spiral arms are the primary structures and can be directly analyzed. The second group consists of strongly barred (SB) galaxies, whose central regions are dominated by bar-driven features such as dust lanes, where the classical definition of spiral arms is no longer directly applicable.

In this work, we focus on the first group, where the pitch angle can be interpreted in the context of spiral density waves and associated substructures such as feathers. The second group, explored in a companion paper (Faria-Costa et al. 2026), extends this analysis to a distinct dynamical regime in which the measured pitch angles trace the curvature of bar-driven gas flows rather than spiral arms.

Accordingly, we calculated the pitch angle of the main arms and substructures in seven spiral galaxies, using NIR and MIR images from the PHANGS–JWST project, with a total of eight filters per galaxy. Similar to the approach of Puerari et al. (2014), this study concentrates on feathers, since we use the emission of gas and dust to trace these substructures. In particular, MIR filters are especially sensitive to dust emission, making feathers more easily detectable and providing clearer constraints on their morphology.

Our primary objective is to investigate how local pitch angles differ between the main spiral arms and smaller interarm substructures, and how these measurements depend on the adopted window scale and infrared wavelength. By combining global and local measurements, we aim to characterize the spatial variations in spiral morphology without imposing a single representative pitch angle on an entire galaxy.

In Section 2, we describe the data used (Section 2.1) and the preprocessing steps applied to the images prior to analysis (Section 2.2). We then present the methods employed, including the two-dimensional Fourier Transform (Section 2.3) and the circular-window correlation method in the $(\ln(R), \theta)$ space, as proposed by Puerari et al. (2014) (Section 2.4). This approach enables the determination of the local pitch angle at different scales, depending on the window size. In Section 3, we present a representative full analysis together with the sample-wide results; the detailed diagnostic plots for the remaining galaxies are provided in Appendix A. Finally, in Section 4, we summarize the conclusions of this study.

## 2 DATA PROCESSING AND PITCH ANGLE ANALYSIS

Our analysis follows a multi-step approach designed to characterize spiral structure in the selected galaxies. In this section, we describe the procedure adopted for calculating the pitch angle. We first present the data (Section 2.1) and the preprocessing steps required to prepare the images for analysis (Section 2.2). This preprocessing includes the deprojection of the images to correct for inclination, the construction of $\ln(R)$ vs. $\theta$ diagrams, and the definition of the minimum and maximum radii that delimit the regions of interest. We then apply the two-dimensional Fourier Transform (Section 2.3) to separate the arm and interarm regions in each galaxy. Finally, we employ the circular-window correlation method in the $(\ln(R), \theta)$ space (Section 2.4), which allows the determination of the local pitch angle at different scales. This sequence ensures a structured and rigorous processing of the images for the analysis of spiral patterns.

This methodology is general and can be applied to different morphological regimes, including both spiral arm-dominated and bar-dominated systems, as explored in the companion paper (Faria-Costa et al. 2026).

### 2.1 Data

The images analysed in this work were retrieved from the publicly available PHANGS–JWST high-level science products (HLSP) hosted by the Mikulski Archive for Space Telescopes (MAST). We used the reduced images provided by the PHANGS–JWST collaboration and did not perform an independent reduction of the raw JWST observations. The procedures adopted for data reduction, calibration, and image processing are described by Williams et al. (2024).

Our subsequent analysis, including image deprojection, coordinate transformations, and pitch-angle measurements, was performed using these publicly released images.

Out of the 19 galaxies observed by the PHANGS–JWST project, we selected 13 for the structural analysis presented across this two-paper study (see Figure 1, which summarizes the selection process). The galaxies IC 5332, NGC 1087, NGC 1365, NGC 1385, NGC 1672, and NGC 5068 were excluded from the analysis due to the limited field of view in the available images, which capture only the central regions without significant spiral arm coverage. In the case of IC 5332, the exclusion was motivated by its multiple-arm morphology in the MIR, which complicates the approach adopted in this study.

The selected galaxies were organized into two groups, based on the presence of visible bars in the images. The first group includes galaxies of types SA or SAB, with less prominent bars and a more pronounced presence of spiral arms (Figure 2). The second group includes SB galaxies with well-defined bars that dominate the observed region and exhibit little or no detectable spiral arm structure. In this paper, we focus on the analysis of seven SA and SAB spiral galaxies (Table 1), since their spiral structure is sufficiently well defined to allow a reliable determination of the pitch angle and associated substructures. The analysis of the second group, composed of strongly barred galaxies, is presented in the companion Paper II (Faria-Costa et al. 2026).

Each galaxy was observed in eight different filters, four from the Near-Infrared Camera (NIRCam) and four from the Mid-Infrared Instrument (MIRI). The NIRCam filters are $F200W$ (2 $\mu$m), $F300M$ (3 $\mu$m), $F335M$ (3.35 $\mu$m), and $F360M$ (3.6 $\mu$m), while the MIRI filters are $F770W$ (7.7 $\mu$m), $F1000W$ (10 $\mu$m), $F1130W$ (11.3 $\mu$m), and $F2100W$ (21 $\mu$m). This combination of filters enables a detailed analysis of galactic structures across different wavelengths.

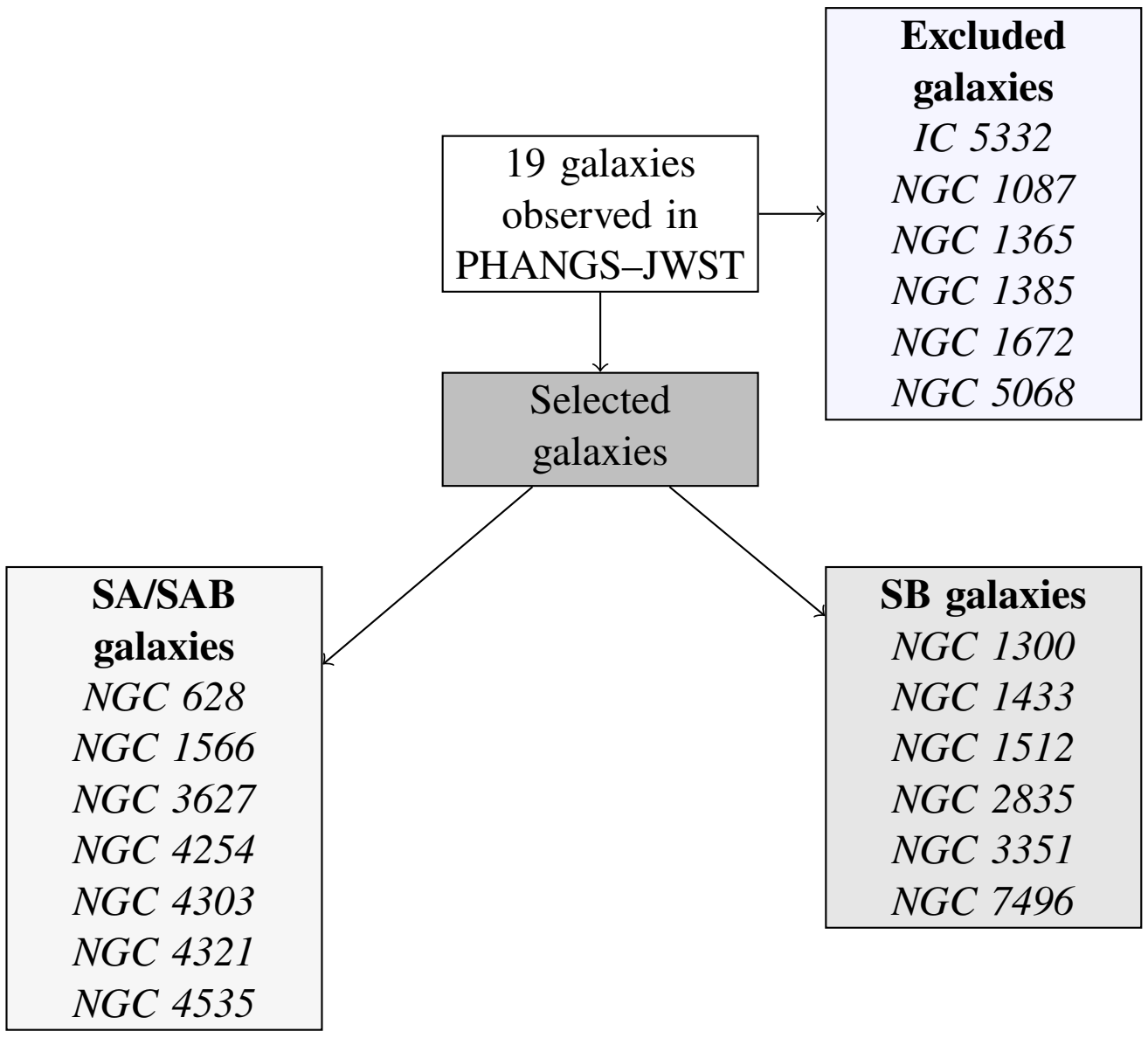

**Figure 1.** Flowchart of the selection and organization of galaxies analyzed from PHANGS–JWST images. In this work, we focus on the SA/SAB group, while the SB group is analysed in the companion Paper II.

**Table 1.** General properties of the galaxy sample. Columns: (1) Galaxy name; (2) morphological classification; (3) inclination ($i$); (4) position angle (PA); (5) distance ($d$); and (6) physical scale, calculated from the adopted distance.

| Galaxy | Morphology[a] | $i$[b] (deg) | PA[b] (deg) | $d$[c] (Mpc) | Scale (pc/arcsec) |
|---|---|---|---|---|---|
| NGC 628 | SA(s)c | 9 | 21 | 9.84 | 47.70 |
| NGC 1566 | SAB(s)bc | 30 | 215 | 17.69 | 85.76 |
| NGC 3627 | SAB(s)b | 57 | 173 | 11.32 | 54.88 |
| NGC 4254 | SA(s)c | 34 | 68 | 13.1 | 63.51 |
| NGC 4303 | SAB(rs)bc | 24 | 312 | 16.99 | 82.37 |
| NGC 4321 | SAB(s)bc | 39 | 156 | 15.21 | 73.74 |
| NGC 4535 | SAB(s)c | 45 | 180 | 15.77 | 76.45 |

[a] Morphological classification from the NASA/IPAC Extragalactic Database (NED).
[b] Inclination and position angle adopted from Lang et al. (2020).
[c] Distances adopted from Anand et al. (2021).

The $F200W$, $F300M$, and $F360M$ filters trace photospheric stellar emission with low extinction, in addition to contributions from nebular emission and hot dust. The $F335M$, $F770W$, and $F1130W$ filters detect polycyclic aromatic hydrocarbon (PAH) emission, tracing properties such as the size and charge of these molecules. The $F1000W$ and $F2100W$ filters trace the dust continuum, with a possible contribution from silicate absorption in the $F1000W$ band (Lee et al. 2023). NIRCam images have a pixel scale of $0.0630''$ pixel$^{-1}$, while MIRI images have $0.1109''$ pixel$^{-1}$.

### 2.2 Image Preprocessing

#### 2.2.1 *Deprojection*

Galaxy disks are inclined with respect to the plane of the sky, which can distort the appearance of their structures, including spiral arms. For an accurate analysis of spiral morphology, it is preferable to study galaxies in a face-on orientation. To correct for projection effects, the images were deprojected using the position angle (PA) and inclination angle ($i$) values listed in Table 1.

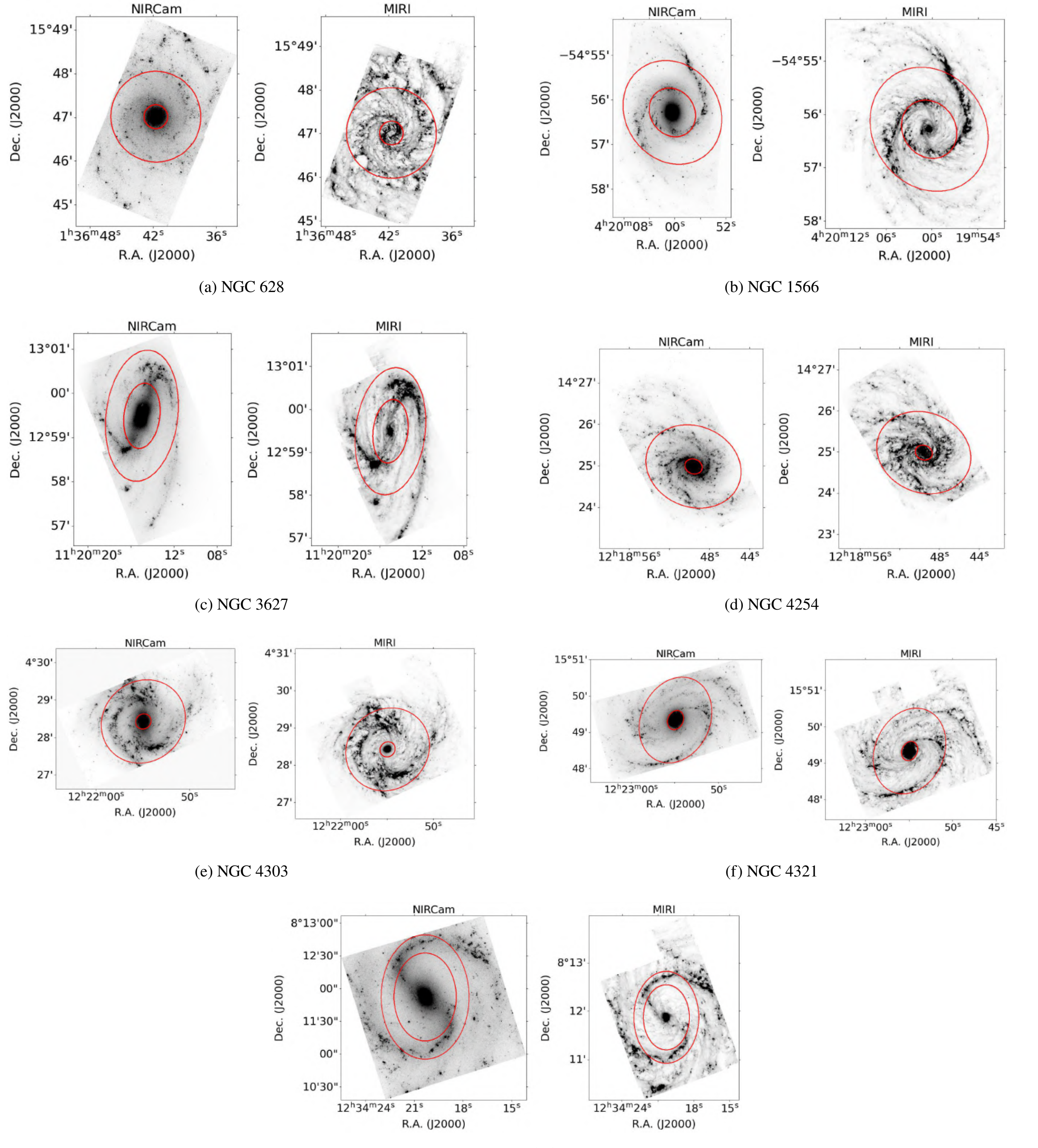

(a) NGC 628

(b) NGC 1566

(c) NGC 3627

(d) NGC 4254

(e) NGC 4303

(f) NGC 4321

(g) NGC 4535

**Figure 2.** Observed $F335M$ (NIRCam) and $F1130W$ (MIRI) images of the selected galaxies. The solid red lines indicate the radial limits $R_{min}$ and $R_{max}$ considered in our analysis.

### 2.2.2 *The* $\ln(R/\mathrm{arcsec})$ *vs.* $\theta$ *diagram*

A common approach to studying spiral arms is to transform the luminosity distribution of each galaxy image in order to straighten the spiral arms. This is achieved by representing the galaxy images in an alternative coordinate system, with axes $\ln(R/\mathrm{arcsec})$ and $\theta$ (Figures 3, A1, A6, A11, A16, A21 and A26). In this diagram, the degree of symmetry and continuity of the spiral arms becomes easier to visualize. Logarithmic spirals, in particular, appear as straight lines in the $\ln(R/\mathrm{arcsec})$ vs. $\theta$ space, which provides a convenient framework for measuring pitch angles.

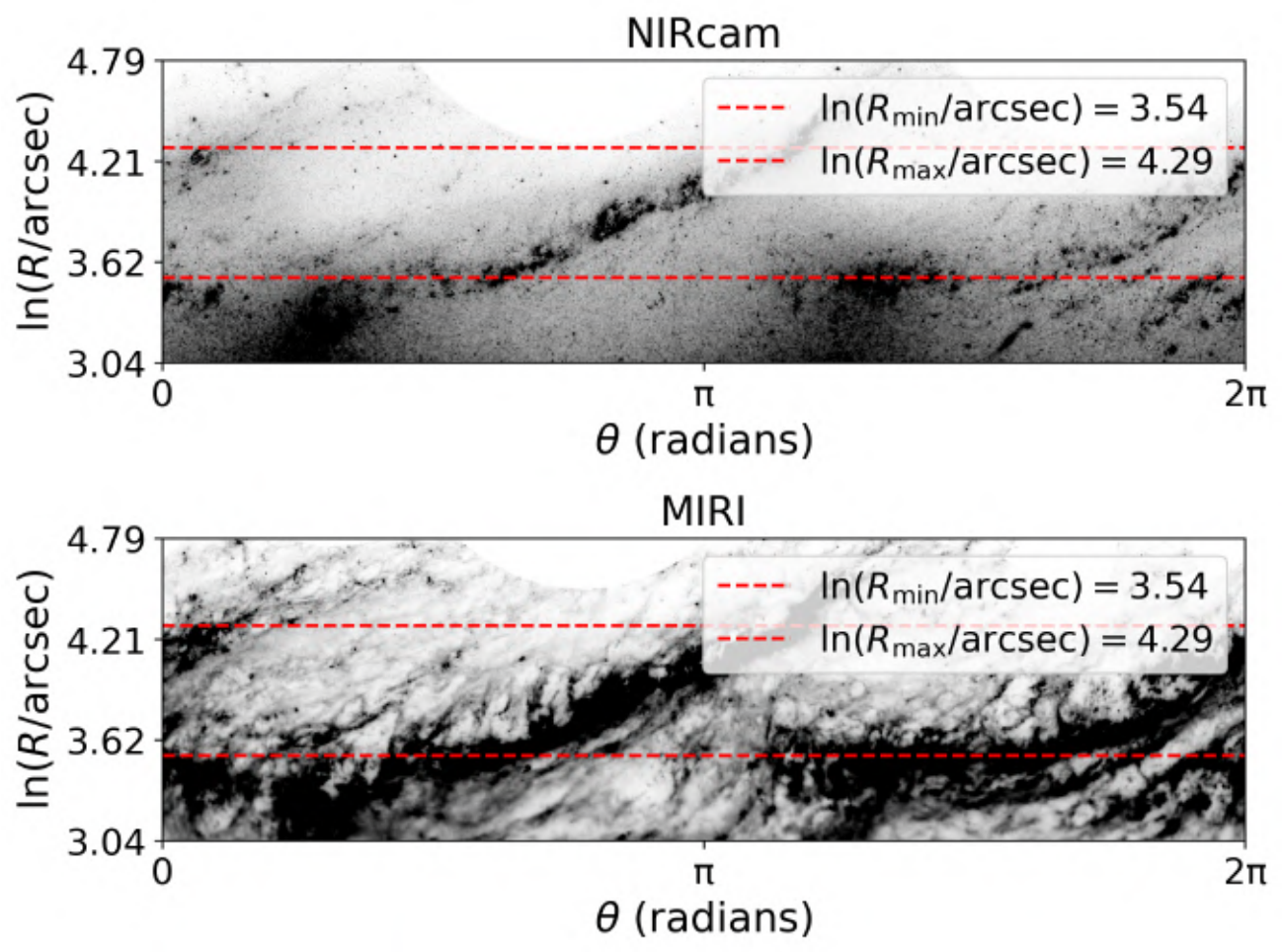

**Figure 3.** Diagrams $\ln(R/\mathrm{arcsec})$ vs. $\theta$ for the $F335M$ (NIRCam) and $F1130W$ (MIRI) images of NGC 1566. The dashed red lines indicate the values of $\ln(R_{min}/\mathrm{arcsec})$ and $\ln(R_{max}/\mathrm{arcsec})$ adopted in the analysis.

Throughout this work, logarithmic radii are expressed as $\ln(R/\mathrm{arcsec})$, corresponding to the natural logarithm of the numerical value of $R$ when the radius is measured in arcseconds. For compactness, $\ln R$ is used throughout the remaining text and equations as shorthand for this same dimensionless quantity.

### 2.2.3 *Minimum and maximum radii*

For each galaxy, a minimum radius ($R_{\min}$), corresponding to the beginning of the spiral arms, and a maximum radius ($R_{\max}$), defined as the outermost point where the spiral arms remain fully visible in the images, were determined, as illustrated in Figure 2. This procedure establishes a specific radial range for the analysis of the structures. The choice of $R_{\min}$ also avoids contamination from the central bulge, while the definition of $R_{\max}$ prevents the inclusion of noisy regions in the outskirts and excludes areas near the frame borders.

## 2.3 Two-Dimensional Fourier Transform Method

To isolate the spiral structure in the image, we express the intensity distribution in $(\ln(R), \theta)$ space and apply a 2D Fourier transform to identify the dominant spiral modes.

Let $f : \mathbb{R} \times \mathbb{R} \to \mathbb{C}$ be a function that satisfies the conditions of integrability. The two-dimensional Fourier Transform (2D-FT) of $f(x, y)$ is defined as:

$$\mathscr{F}[f(x,y)] = F(k,l) = \int_{-\infty}^{+\infty} \int_{-\infty}^{+\infty} f(x,y)\, e^{-i(kx+ly)}\, dx\, dy, \quad (1)$$

and the two-dimensional Inverse Fourier Transform (2D-IFT):

$$\mathscr{F}^{-1}[F(k,l)] = f(x,y) = \frac{1}{4\pi^2} \int_{-\infty}^{+\infty} \int_{-\infty}^{+\infty} F(k,l)\, e^{i(kx+ly)}\, dk\, dl. \quad (2)$$

In our application, $f(x, y)$ corresponds to the deprojected galaxy image, and its Fourier transform $F(k, l)$ provides a decomposition of the spiral structure into modes that can be analyzed to estimate the pitch angle and identify substructures.

The continuous density distribution $I(u, \theta)$, corresponding to the discrete matrix $I(u_k, \theta_l)$, can be expanded into a Fourier series in the azimuthal angle $\theta$ and an integral over the logarithm of the radius $u$ (e.g. Considère & Athanassoula 1982; Puerari & Dottori 1992). Therefore, we have:

$$I(u,\theta) = \frac{1}{2\pi} \sum_{m=0}^{+\infty} \int_{-\infty}^{+\infty} A(p,m) e^{i(pu+m\theta)}\, dp \quad (3)$$

where the Fourier coefficient $A(p, m)$ can be obtained by performing the Fourier Transform of $I(u, \theta)$, such that:

$$A(p,m) = \frac{1}{\sum_{k=1}^{K}\sum_{l=1}^{L} w_{kl}} \sum_{k=1}^{K}\sum_{l=1}^{L} w_{kl} e^{-i(pu_{kl}+m\theta_{kl})}, \quad (4)$$

in which $w_{kl}$ is the weight of the point $(k, l)$ located at coordinates $(r_{kl}, \theta_{kl})$, $u_{kl} = \ln(r_{kl}/\mathrm{arcsec})$, $m$ is the azimuthal wave number (the number of spiral arms), and $p$ is the variable associated with the pitch angle $\alpha$, defined as $\alpha = \arctan\left(-\frac{m}{p_{\max}}\right)$, with $p_{\max}$ being the value of $p$ corresponding to the highest amplitude for a given component. Positive values of $p$ indicate an S-type spiral pattern, in which the arms wind clockwise, with $\alpha < 0°$. Negative values of $p$ correspond to a Z-type spiral pattern, winding counterclockwise, with $\alpha > 0°$. For $p = 0$, the pattern is not spiral but bar-like, with $\alpha = \pm 90°$. This formalism therefore provides a direct connection between the Fourier modes of the density distribution and the observed morphology of spiral arms.

To obtain the matrix $A(p, m)$, the Fast Fourier Transform (FFT) algorithm was used[1]. We employed a basis of six logarithmic spirals, with the azimuthal wave number (or number of arms) in the range $0 \le m \le 6$, and the logarithmic radial wave number in the range $-50 \le p \le 50$, with $\Delta m = 1$ and $\Delta p = 0.25$.

The galaxy image can then be reconstructed by performing the inverse transform of the complex functions $A(p, m)$. The surface density is calculated using the expression:

$$S(u,\theta) = \sum_m S_m(u) e^{im\theta} \quad (5)$$

where

$$S_m(u) = \frac{\sum_{k=1}^{K}\sum_{l=1}^{L} w_{kl}}{e^{2u} 4\pi^2} \int_{-\infty}^{+\infty} G_m(p) A(p,m) e^{ipu}\, dp, \quad (6)$$

and

$$G_m(p) = \exp\left[-\frac{1}{2}\left(\frac{p - p_{\max}}{10}\right)^2\right], \quad (7)$$

where $p_{\max}$ is the value of $p$ at which the spectrum reaches its maximum amplitude for a given $m$.

Equation (7) applies a Gaussian filter in the $p$-space that suppresses high-frequency components, thereby smoothing the spectra $A(p, m)$ toward the edges of the interval $-50 \le p \le 50$. This filtering emphasizes the central frequency of interest, $p_{\max}$, while attenuating the contributions from extreme frequencies, resulting in a cleaner reconstruction of the dominant spiral modes.

The 2D-FT and 2D-IFT were applied individually to the images of each galaxy in the $F1130W$ filter, with the goal of obtaining a mask that systematically and automatically defines the arm and interarm regions. First, from Equation (4), the value of $p_{\max}$ and the dominant $m$ mode were determined in each case. With these parameters, Equation (5) was used to compute the 2D-IFT and reconstruct an image that highlights the spiral pattern. From this reconstruction, a specific mask was generated for each galaxy. Figure 4 shows, as an example, the mask obtained for the galaxy NGC 1566. In the image, the region of the main arms is highlighted in gray, the interarm region in white, and the separation between these regions is represented in black.

[1] This routine is part of a series of algorithms presented in the book *Numerical Recipes in C* (Press et al. 1992).

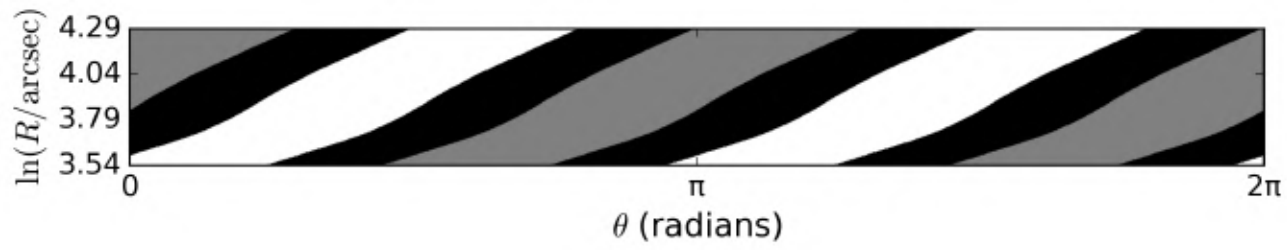

**Figure 4.** Mask generated from the 2D-FT and 2D-IFT used to define the arm and interarm regions of the galaxy NGC 1566. The gray area corresponds to the spiral arm region, the white area represents the interarm region, and the black area indicate the boundaries between them.

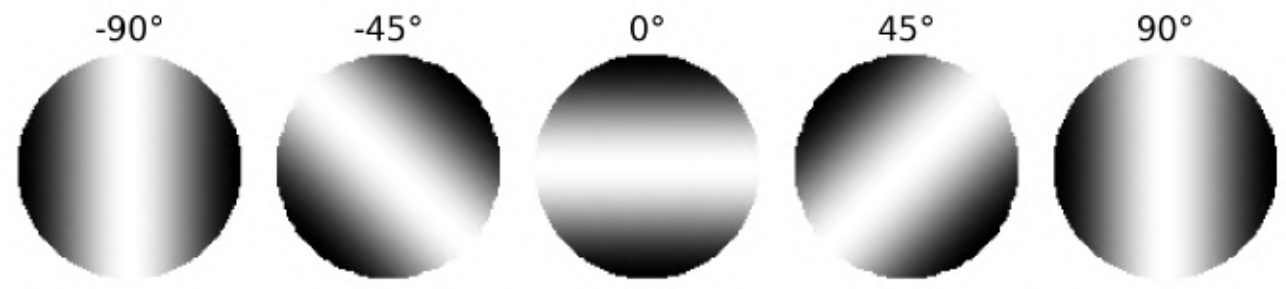

**Figure 5.** Cosine filters on circular windows with $P_w = -90°, -45°, 0°, 45°$, and $90°$. Filters of this type were applied with $P_w$ varying from $-90°$ to $+90°$ in order to calculate the correlation between the filter and the galaxy image at a given position $(\ln(R), \theta)$.

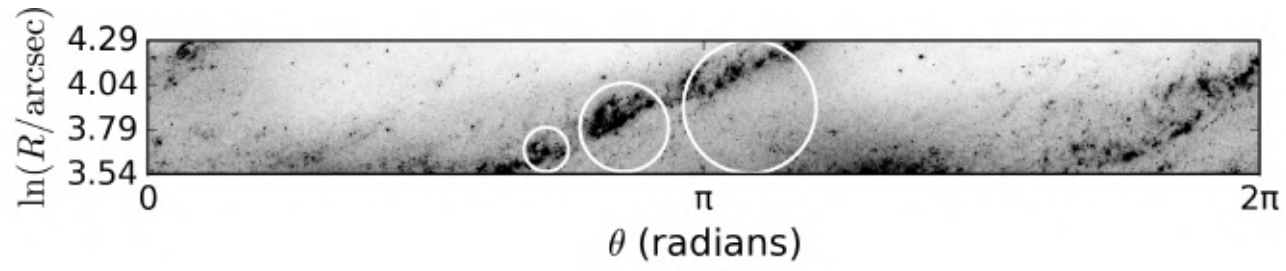

**Figure 6.** Image of NGC 1566 in the $F335M$ filter, presented in the $\ln(R/\mathrm{arcsec})$ vs. $\theta$ diagram, with three circles representing the dimensionless window diameters used in the analysis: $D_w = 0.25, 0.5$, and $0.75$, shown from left to right.

### 2.4 Circular-Window Correlation Method

To estimate the pitch angle, we apply a correlation method similar to the one proposed by Puerari et al. (2014), using circular windows in the $(\ln(R), \theta)$ space. For each window, a filter is defined with the function $-\cos(x)$ varying from 0 to $2\pi$, meaning that it reaches its maximum value at the center of the window and gradually decreases toward the edges (Figure 5). The filter is expressed as:

$$F = F(\ln(R_c), \theta_c, P_w, D_w) \quad (8)$$

where $F$ is a circular filter centered at $(\ln(R_c), \theta_c)$, with pitch angle $P_w$ and diameter $D_w$. The parameter $P_w$ varies between $-90°$ and $90°$, while $D_w$ takes the dimensionless values 0.25, 0.50, and 0.75 in the $(\ln R, \theta)$ plane. The size of these windows compared to the structures being measured is illustrated in Figure 6, which allows the analysis of spiral features at different spatial scales.

At each position in the $(\ln(R), \theta)$ space, we compute the correlation of the galaxy image with the filter, varying both the pitch angle and the window diameter. Higher correlation values are obtained when the structure has the same pitch angle as the filter at that position $(\ln(R), \theta)$.

Thus, the correlation can be described as:

$$C(\ln(R_c), \theta_c, P_w, D_w) = \sum_{i=1}^{N} F(\ln(R_c), \theta_c, P_w, D_w)\, I(\ln(R), \theta), \quad (9)$$

where $I(\ln(R), \theta)$ is the galaxy image deprojected into the $(\ln(R), \theta)$ space. The correlation reaches its maximum value when the pitch angle of a structure centered at $(\ln(R_c), \theta_c)$ coincides with the pitch angle of the filter $F$. The pitch angle associated with the maximum correlation at each position corresponds to the local orientation of the spiral pattern.

The pitch angle $P_w$ was sampled from $-90°$ to $+90°$ in steps of $1°$. The three window diameters were adopted to probe different structural scales. Smaller windows are more sensitive to localized features, whereas larger windows preferentially trace coherent large-scale structures; the intermediate value $D_w = 0.50$ provides sensitivity to both regimes. This scale-dependent behaviour is consistent with the original application of the circular-window method (Puerari et al. 2014) and was also verified directly from the spatial pitch-angle maps in our sample.

Representative pitch angles were determined by considering the summed-correlation spectra together with the spatial distribution of the corresponding local pitch angles. The arm and interarm regions were analysed separately. Within each instrument, the four filters generally showed very similar spectral behaviour. Main-arm pitch angles were therefore identified primarily from recurrent peaks in the NIRCam spectra, whereas pitch angles associated with feathers and other interarm structures were identified primarily from recurrent peaks in the MIRI spectra. Agreement between NIRCam and MIRI was not required, since the two instruments emphasize different morphological components.

Candidate peaks were subsequently cross-checked against the local pitch-angle maps. A peak was considered representative when the corresponding pitch-angle values were spatially concentrated along a recognizable morphological structure. When several nearby peaks were associated with the same structure, small intervals of pitch angle were inspected in the maps to determine whether they represented distinct features or local variations of a single structure. In the latter case, an interval or a representative value was reported. This association was performed by visual inspection of the spectra and spatial maps; no parametric fitting of the summed-correlation peaks or formal statistical significance criterion was applied.

## 3 RESULTS

Each galaxy has eight images, four from NIRCam and four from MIRI, and the arm and interarm regions are analysed separately for the three window diameters, $D_w = 0.25, 0.50$, and $0.75$. Following the procedure described in Section 2.4, the representative pitch angles discussed below are identified by considering recurrent features in the summed-correlation spectra together with their spatial distribution in the local pitch-angle maps.

Because the two instruments emphasize different morphological components, the main-arm geometry is constrained primarily from the NIRCam results, whereas feathers and other dusty interarm structures are identified primarily from MIRI. Within each instrument, the four filters generally show similar spectral behaviour. Agreement between NIRCam and MIRI is therefore considered complementary

**Table 2.** Summary of the analysis for the seven SA/SAB galaxies. The logarithmic radial limits are expressed as $\ln(R/\mathrm{arcsec})$. The column "$D_w$ emphasized" lists the window sizes highlighted in the analysis of the indicated structures; all three window sizes (0.25, 0.50, and 0.75) were calculated for every galaxy.

| Galaxy | Winding | $\ln(R_{\mathrm{min}}/\mathrm{arcsec})$ | $\ln(R_{\mathrm{max}}/\mathrm{arcsec})$ | Main arms (°) | Interarm/substructures (°) | $D_w$ emphasized |
|---|---|---|---|---|---|---|
| NGC 628 | Z | 2.80 | 4.14 | $\sim 17$ | $\sim 50, \sim 58$ | 0.75 (arms); 0.25–0.50 (interarm) |
| NGC 1566 | Z | 3.54 | 4.29 | $\sim 24$ | $\sim 32, \sim 76$ | 0.50–0.75 (arms); 0.25 (interarm) |
| NGC 3627 | S | 3.80 | 4.50 | $\sim -32, \sim -45, \sim -8$ | $\sim -15, \sim -32$ | 0.50–0.75 (arms); 0.25–0.50 (interarm) |
| NGC 4254 | Z | 2.55 | 4.26 | $\sim 24, \sim 17$ | $\sim 58$ | 0.50–0.75 (arms); 0.25–0.50 (interarm) |
| NGC 4303 | Z | 2.55 | 4.25 | $\sim 56, \sim 10, \sim 17$ | $\sim 31, \sim 40$ | 0.50–0.75 (arms); 0.25 (interarm) |
| NGC 4321 | S | 2.80 | 4.30 | $\sim -55$ | $\sim -76, \sim -46, \sim -29, \sim -8$ | 0.50 (main-arm onset) |
| NGC 4535 | S | 3.70 | 4.05 | $\sim -14, \sim -2$ | $\sim -59, \sim -76$ | 0.50 (arms); 0.25 (interarm) |

rather than a requirement for retaining a representative pitch-angle component.

### 3.1 Overview of the sample

The seven galaxies span a range of spiral morphologies. According to Elmegreen & Elmegreen (1987), NGC 628, NGC 4254, NGC 4303, and NGC 4535 show "two symmetric inner arms; multiple long and continuous outer arms", while NGC 1566 and NGC 4321 show "two long symmetric arms dominating the optical disk". NGC 3627 is described as having "two symmetric, long outer arms; feathery or irregular inner arms". The sample contains both S- and Z-type winding patterns, as summarized in Table 2.

In the $\ln(R/\mathrm{arcsec})$ vs. $\theta$ diagrams, where the spiral arms are rectified and $\theta$ ranges from 0 to $2\pi$, we define the first main arm as the one that appears first from left to right and the second main arm as the subsequent one. Table 2 also lists the radial interval analysed for each galaxy, the representative pitch angles reported for the main arms and interarm structures, and the window scales emphasized in their interpretation. All three values of $D_w$ were calculated for every galaxy. The window scales listed in the final column reflect the expected sensitivity to structures of different sizes, together with the spatial coherence of the corresponding pitch-angle values in the local maps; they do not result from a formal statistical optimization. Detailed correlation spectra and local pitch-angle maps for all galaxies other than the representative case NGC 1566 are presented in Appendix A.

### 3.2 Representative case: NGC 1566

We use NGC 1566 as a representative example to illustrate the complete analysis sequence in the main text. This galaxy is classified as SAB(s)bc, has Z-type spiral winding, and was analysed over the radial interval $\ln(R_{\mathrm{min}}/\mathrm{arcsec}) = 3.54$ to $\ln(R_{\mathrm{max}}/\mathrm{arcsec}) = 4.29$ (Figure 3). The arm and interarm regions defined from the 2D-FT/2D-IFT mask are shown in Figures 8a and 9a.

Figure 7 shows the summed correlation as a function of pitch angle for the arm and interarm regions and for the three window diameters, $D_w = 0.25, 0.5,$ and $0.75$. The corresponding local pitch-angle maps for the $F335M$ and $F1130W$ images are shown in Figures 8 and 9.

For $D_w = 0.50$ (Figure 7b), the NIRCam arm-region spectra show recurrent components near $\alpha \approx 19°$ and $\alpha \approx 28°$. The MIRI spectra show a broader component over approximately $21° \lesssim \alpha \lesssim 28°$, with a maximum near 24°. At the larger scale, $D_w = 0.75$ (Figure 7c), the four NIRCam filters show a dominant component near $\alpha \approx 23°$, while $F200W$ and $F300M$ also show a component near 17°. The corresponding MIRI spectra contain a component near 24°. Inspection of the local pitch-angle maps shows that values around 24° trace the main spiral-arm geometry coherently, and we therefore adopt $\alpha \approx 24°$ as the representative value for the main arms.

At the smaller scale, $D_w = 0.25$ (Figure 7a), the interarm spectra contain components near $\alpha \approx 32°$ and $\alpha \approx 76°$. Their morphological association is most clearly established from the MIRI pitch-angle maps, where the dusty interarm structures are better defined. Examining the spatial distribution of values around these spectral components shows that $\alpha \approx 76°$ traces feathers projecting outward from the main spiral arms, whereas values near $\alpha \approx 32°$ trace a more extended interarm structure.

At larger window sizes these localized structures become less distinct, as expected from the scale dependence of the method. The adopted values are illustrated in Figure 10, while the spatial distributions for the smaller windows are shown in Figures 9b and 9c.

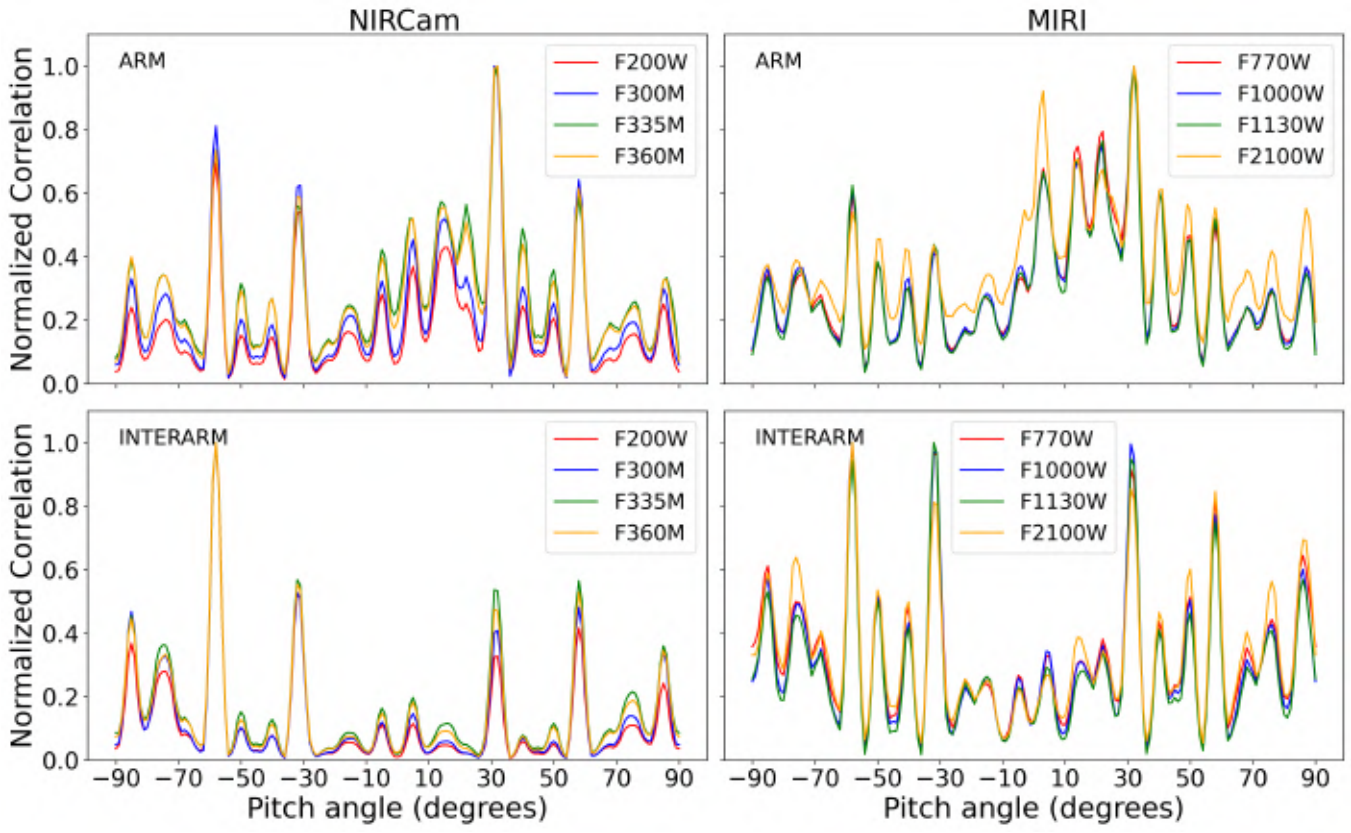

(a) Sum of the correlations calculated with $D_w = 0.25$ for the arm and interarm regions of NGC 1566.

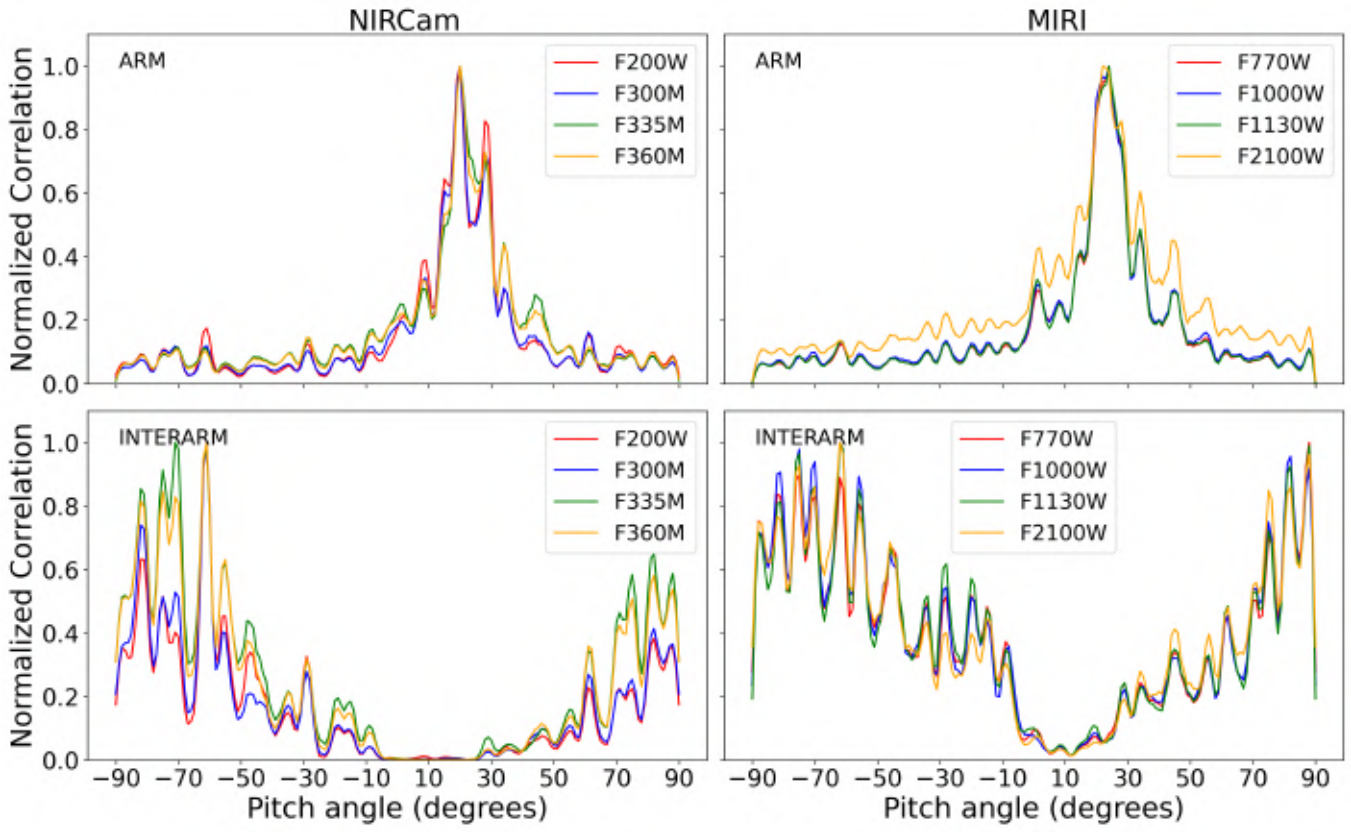

(b) Sum of the correlations calculated with $D_w = 0.5$ for the arm and interarm regions of NGC 1566.

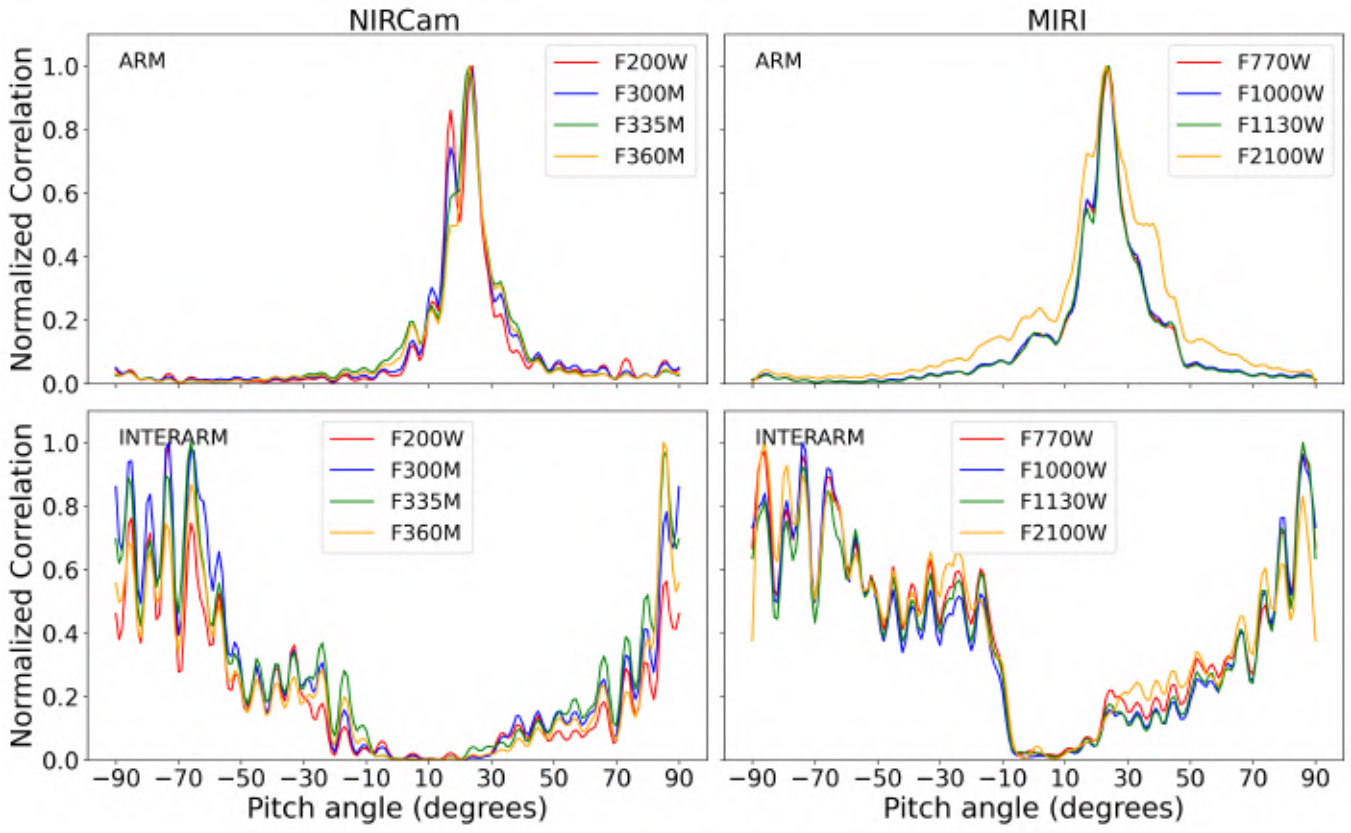

(c) Sum of the correlations calculated with $D_w = 0.75$ for the arm and interarm regions of NGC 1566.

**Figure 7.** Results for the sum of the correlations in the analyzed regions of NGC 1566. On the left, plots from NIRCam images with filters $F200W$ (red), $F300M$ (blue), $F335M$ (green), and $F360M$ (yellow), where the upper panel corresponds to the arm region and the lower panel to the interarm region. On the right, plots from MIRI images with filters $F770W$ (red), $F1000W$ (blue), $F1130W$ (green), and $F2100W$ (yellow), following the same division between arm (upper) and interarm (lower) regions.

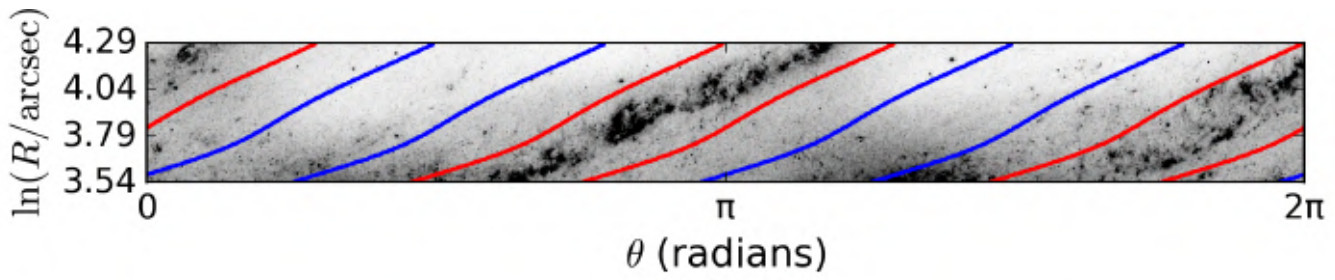

(a) Regions delimited in the $F335M$ image of NGC 1566 from the mask (Figure 4). The spiral arm regions are located between the red lines, while the interarm regions are between the blue lines.

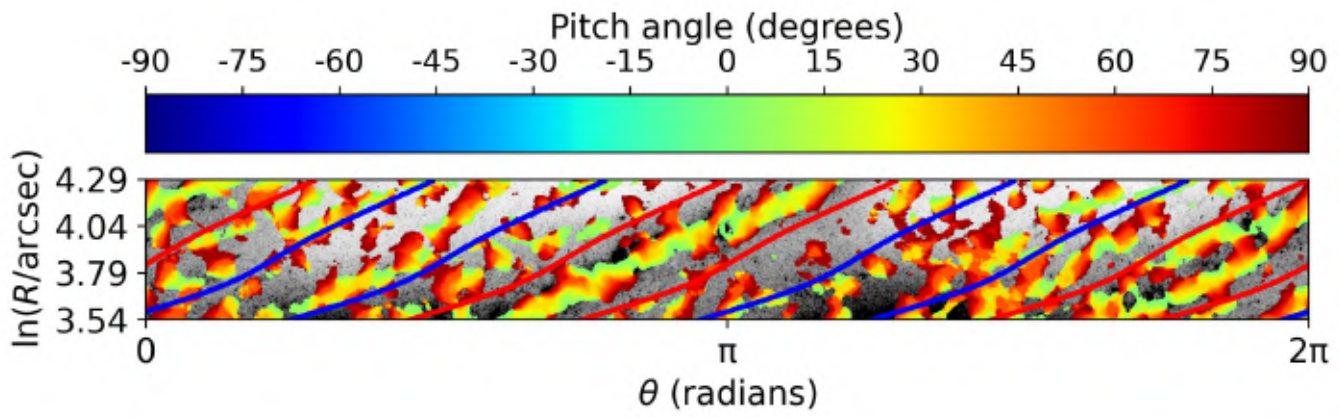

(b) Distribution of pitch angles obtained with the $D_w = 0.25$ filter.

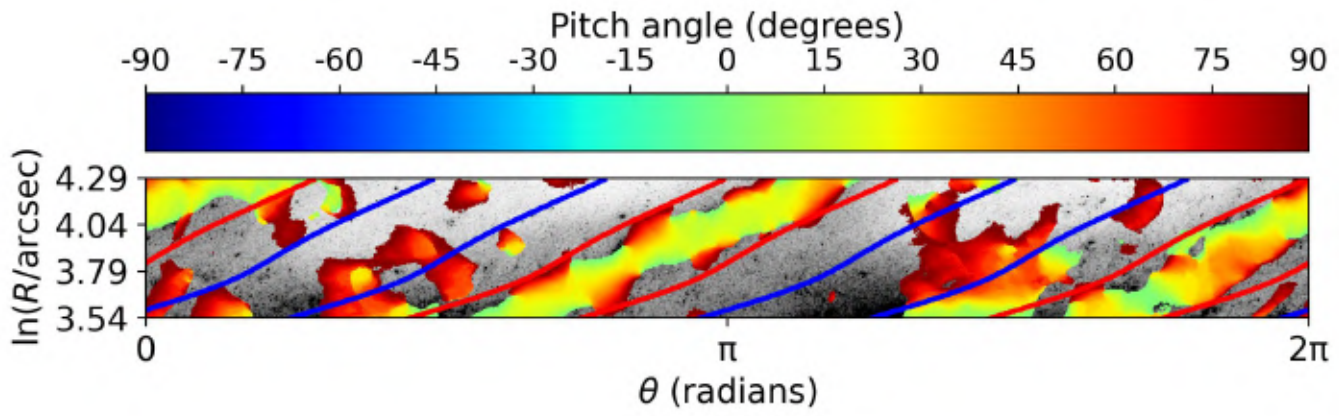

(c) Distribution of pitch angles obtained with the $D_w = 0.5$ filter.

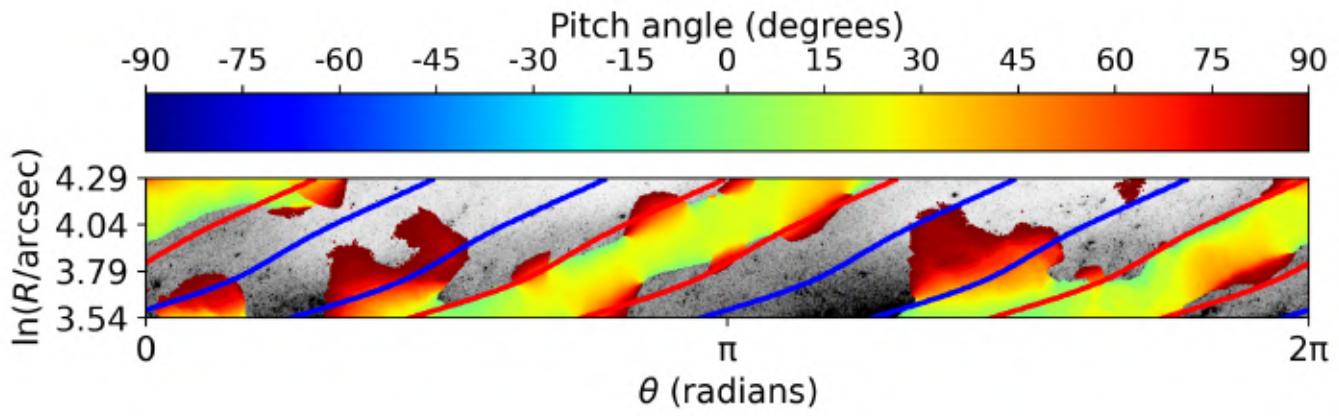

(d) Distribution of pitch angles obtained with the $D_w = 0.75$ filter.

**Figure 8.** Distribution of pitch angles obtained from the highest correlations at each position $(\ln(R/\mathrm{arcsec}), \theta)$ in the $F335M$ image of NGC 1566. The colorbar represents pitch angle values, ranging from $-90^\circ$ (dark blue) to $90^\circ$ (dark red). The colors of the points indicate the pitch angle associated with the maximum correlation, based on the idea that the strongest correlations occur when the structure has the same pitch angle as the filter at the corresponding position.

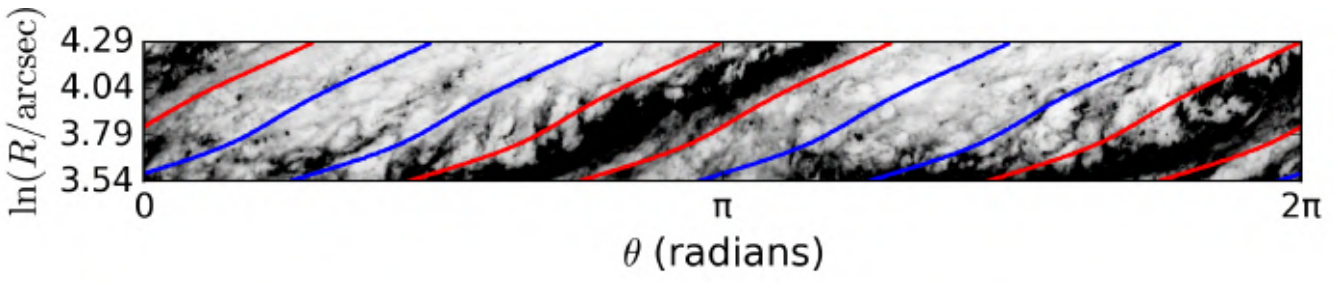

(a) Regions delimited in the $F1130W$ image of NGC 1566 from the mask (Figure 4). The spiral arm regions are located between the red lines, while the interarm regions are between the blue lines.

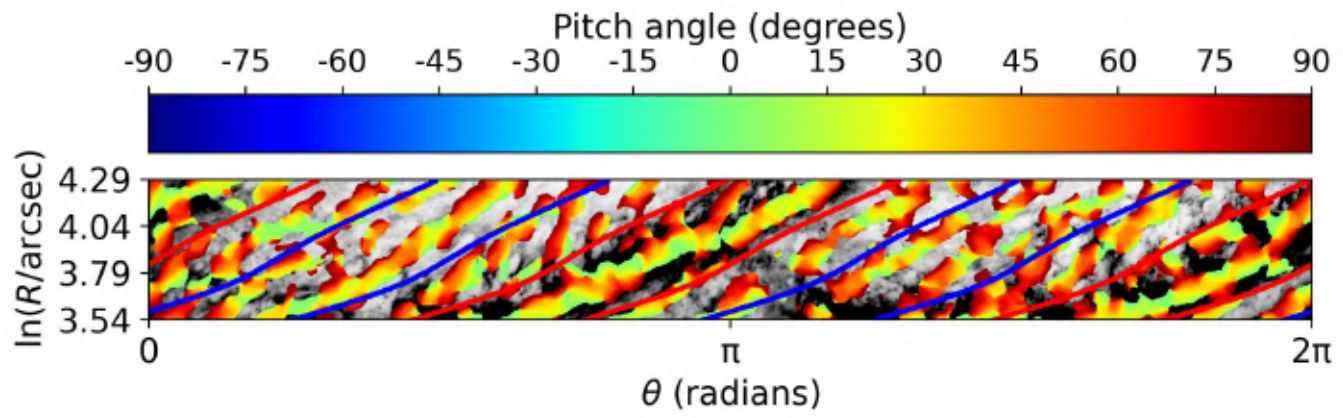

(b) Distribution of pitch angles obtained with the $D_w = 0.25$ filter.

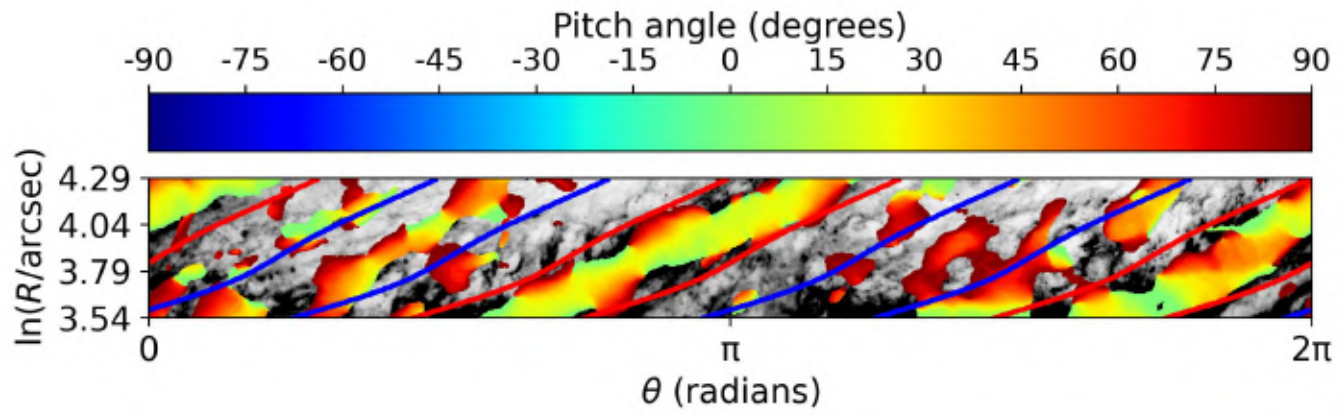

(c) Distribution of pitch angles obtained with the $D_w = 0.5$ filter.

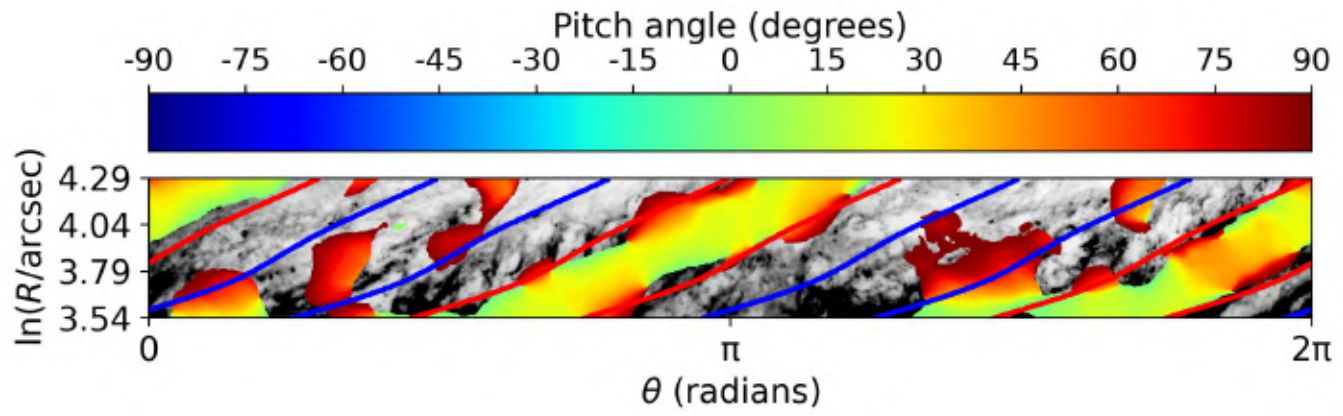

(d) Distribution of pitch angles obtained with the $D_w = 0.75$ filter.

**Figure 9.** Distribution of pitch angles obtained from the highest correlations at each position $(\ln(R/\text{arcsec}), \theta)$ in the $F1130W$ image of NGC 1566. The colorbar represents pitch angle values, ranging from $-90°$ (dark blue) to $90°$ (dark red). The colors of the points indicate the pitch angle associated with the maximum correlation, based on the idea that the strongest correlations occur when the structure has the same pitch angle as the filter at the corresponding position.

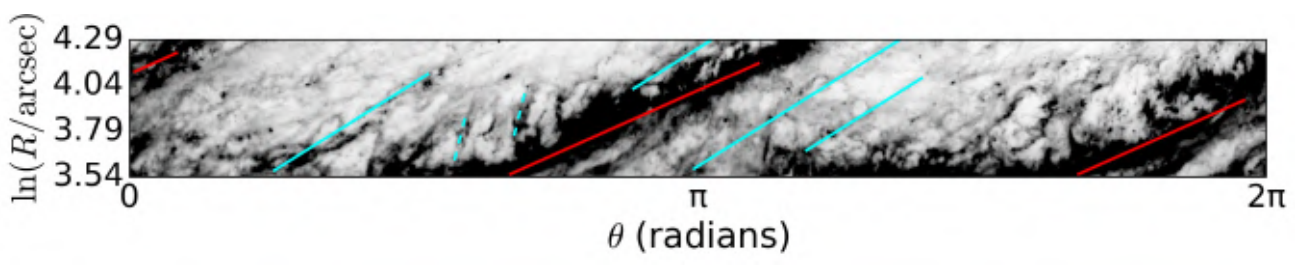

**Figure 10.** Lines overlaid on the $F1130W$ image of NGC 1566, indicating the pitch angles of the spiral structures. The solid red line corresponds to 24°, the solid blue line to 32°, and the dashed blue line to 76°.

### 3.3 Main spiral-arm behaviour

The main spiral arms generally occupy lower absolute pitch-angle ranges than the most open interarm substructures, although several galaxies show clear radial variations. NGC 628 is well represented by a main-arm pitch angle of approximately 17°, while NGC 1566 is characterized by a value near 24°. NGC 4254 shows a radial change from approximately 24° in the inner portions of the arms to approximately 17° farther out.

The clearest departures from a single characteristic arm pitch angle occur in galaxies where the analysed region includes a transition between the bar and the spiral pattern. In NGC 3627, the first main arm changes from $|\alpha| \approx 40°$–$45°$ near its inner portion to approximately $|\alpha| \approx 32°$ farther out, while the second arm emerges with a substantially smaller absolute pitch angle, near 8°. NGC 4303 similarly changes from a relatively open inner configuration, with $\alpha \sim 56°$, to values of approximately 10° and 17° along the outer portions of the two main arms. In NGC 4321, the analysed region predominantly samples the bar–spiral transition, where the beginning of the main arm reaches $\alpha \sim -55°$. NGC 4535, whose analysed radial interval is comparatively narrow, shows values near $-14°$ and $-2°$ for the two main arms. The full diagnostic plots supporting these cases are shown in Appendix A.

### 3.4 Interarm substructures, spatial scale, and wavelength

The spatial-map analysis confirms that interarm structures are preferentially recovered at smaller window sizes and frequently show larger absolute pitch angles than the main spiral pattern. Examples include the features near 50° and 58° in NGC 628, 76° in NGC 1566, 58° in NGC 4254, and $-76°$ in NGC 4321. Other interarm or branching structures occupy intermediate pitch-angle ranges, such as the values near 31°–40° in NGC 4303 and the $-15°$ feature in NGC 3627. NGC 4535 contains interarm structures near $-59°$ and $-76°$, although no feather is identified within the analysed region.

The behaviour across the sample confirms the expected scale dependence of the circular-window analysis. The larger windows, particularly $D_w = 0.50$–$0.75$, emphasize the coherent geometry of extended spiral arms, whereas $D_w = 0.25$ is more sensitive to localized interarm features. The two instruments provide complementary information: the four NIRCam filters generally give a consistent description of the main spiral pattern, while the four MIRI filters preferentially reveal dusty interarm structures and feathers. The representative values quoted above are those for which the corresponding pitch-angle distributions are spatially associated with the relevant morphological structures.

### 3.5 Summary of the results

Taken together, the sample shows that a single global pitch angle does not capture the full geometry of a spiral galaxy. The main arms, radial transitions, and smaller interarm structures can occupy different pitch-angle ranges within the same system, and the structures recovered depend on the spatial scale probed by the correlation window. The distinction between the large-scale spiral pattern and localized dusty substructures is particularly clear when the NIRCam and MIRI results are considered jointly.

## 4 CONCLUSIONS

We have presented a spatially resolved analysis of spiral arms and interarm substructures in seven nearby SA/SAB galaxies observed by PHANGS–JWST. By combining the 2D-FT/2D-IFT decomposition with the circular-window correlation analysis, we characterize local pitch angles over multiple spatial scales rather than assigning a single pitch angle to an entire galaxy.

The main spiral pattern is generally recovered most clearly with intermediate or large windows, whereas the smallest window enhances localized interarm structures. Across the sample, the main arms commonly occupy smaller absolute pitch-angle ranges than feathers and other dusty substructures, although strong radial variations occur in systems where the analysed region includes transitions between bars and spiral arms. The quantitative results and the window scales emphasized for each galaxy are summarized in Table 2.

NIRCam and MIRI provide complementary views of these structures. NIRCam images yield robust measurements of the large-scale spiral pattern with reduced influence from fine dusty features, while MIRI images are particularly effective at revealing feathers and other arm and interarm substructures (Williams et al. 2022). Within each instrument, the principal pitch-angle features are broadly consistent across the four filters, while their relative prominence differs between NIRCam and MIRI because the two instruments emphasize different morphological components.

A limitation of the present analysis is the finite spatial coverage of the PHANGS–JWST images, which in some galaxies does not include the full radial extent of the spiral pattern. Nevertheless, the local analysis demonstrates that pitch angle is scale dependent within individual galaxies and can distinguish large-scale spiral arms from smaller-scale interarm features.

In the companion Paper II (Faria-Costa et al. 2026), the same framework is extended to strongly barred galaxies, where the local angles characterize the orientation and curvature of dust lanes rather than classical spiral arms. Considered together, the two studies provide a common geometrical framework for comparing structures that arise in different morphological and dynamical environments.

## ACKNOWLEDGEMENTS

Stella Faria Costa acknowledges the support received from Coordenação de Aperfeiçoamento de Pessoal de Nível Superior - Brasil (CAPES) under process numbers 88887.823252/2023-00 and 88881.935622/2024-01; Irapuan Rodrigues acknowledges support from Conselho Nacional de Desenvolvimento Científico e Tecnológico (CNPq/Brazil, Grant No. 314267/2021-2). We thank the anonymous referee for the constructive comments and suggestions, which were particularly helpful in restructuring the manuscript and improving the presentation of the results. The JWST imaging data analysed in this work are publicly available as part of the PHANGS–JWST high-level science products hosted by the Mikulski Archive for Space Telescopes. The data can be accessed at https://doi.org/10.17909/ew88-jt15. The derived products generated in this study are available from the corresponding author upon reasonable request.

## DATA AVAILABILITY

The images used in this study were obtained from the Mikulski Archive for Space Telescopes (MAST), available at https://mast.stsci.edu/portal/Mashup/Clients/Mast/Portal.html.

## APPENDIX A: DETAILED RESULTS FOR INDIVIDUAL GALAXIES

This appendix presents the complete correlation spectra and local pitch-angle maps for the six galaxies whose detailed diagnostic figure sets are not shown in the main text. NGC 1566 is presented as the representative case in Section 3.2. The same analysis procedure and the same three window diameters were applied to all galaxies.

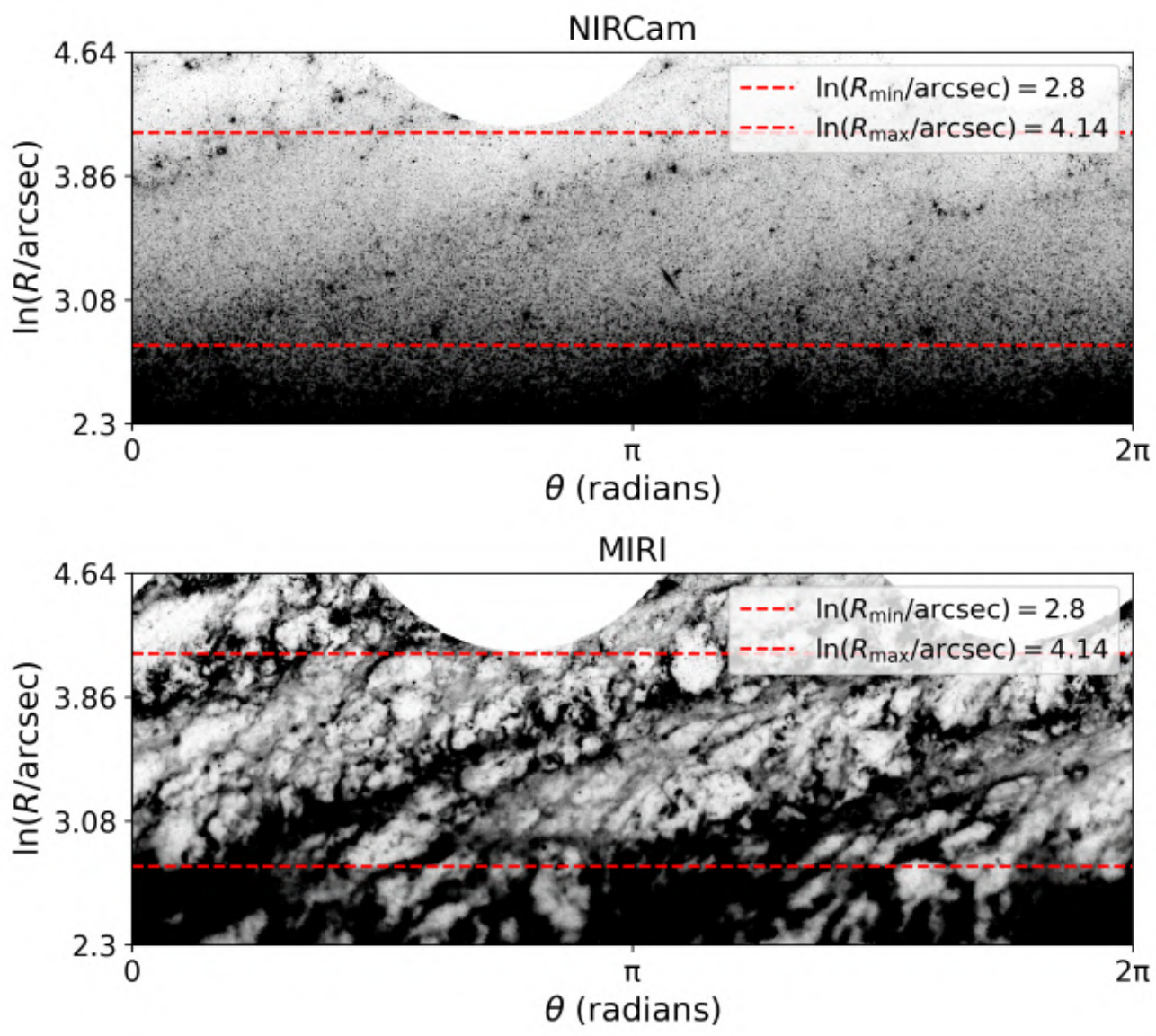

**Figure A1.** Diagrams $\ln(R/\text{arcsec})$ vs. $\theta$ for the $F335M$ (NIRCam) and $F1130W$ (MIRI) images of NGC 628. The dashed red lines indicate the values of $\ln(R_{min}/\text{arcsec})$ and $\ln(R_{max}/\text{arcsec})$ adopted in the analysis.

## A1 NGC 628

The galaxy NGC 628 is morphologically classified as SA(s)c and, according to Elmegreen & Elmegreen (1987), presents "two symmetric inner arms; multiple long and continuous outer arms". The analysis of this galaxy was performed between $\ln(R_{\rm min}/\text{arcsec}) = 2.8$ and $\ln(R_{\rm max}/\text{arcsec}) = 4.14$, defined in the $(\ln(R/\text{arcsec}), \theta)$ space (Figure A1). Within this range, we investigated the regions corresponding to the main arm and the interarm, based on the mask obtained through the 2D-FT and 2D-IFT (Figures A3a and A4a).

Figure A2 shows the sum of the correlations as a function of pitch angle for the arm and interarm regions of NGC 628, for the three different window diameters, $D_w = 0.25$, 0.5, and 0.75 (Figures A2a, A2b, and A2c, respectively). Since this galaxy has a Z-type spiral winding, we focus on the positive peaks, i.e., $0^\circ \leq \alpha \leq 90^\circ$. At $D_w = 0.75$ (Figure A2c), the arm-region spectra show recurrent components near $\alpha \approx 17^\circ$ and $\alpha \approx 23^\circ$ across the filters. Inspection of the corresponding local pitch-angle maps shows that values near $17^\circ$ provide a coherent description of the main spiral arms, and this value is therefore adopted as representative. For the interarm region, the smaller windows reveal additional components. With $D_w = 0.25$ (Figure A2a), components occur near $40^\circ$, $50^\circ$, and $58^\circ$, whereas at $D_w = 0.50$ (Figure A2b) the distribution is concentrated near $60^\circ$. The spatial maps show that values near $50^\circ$ and $58^\circ$ trace the prominent interarm structures retained in our final interpretation.

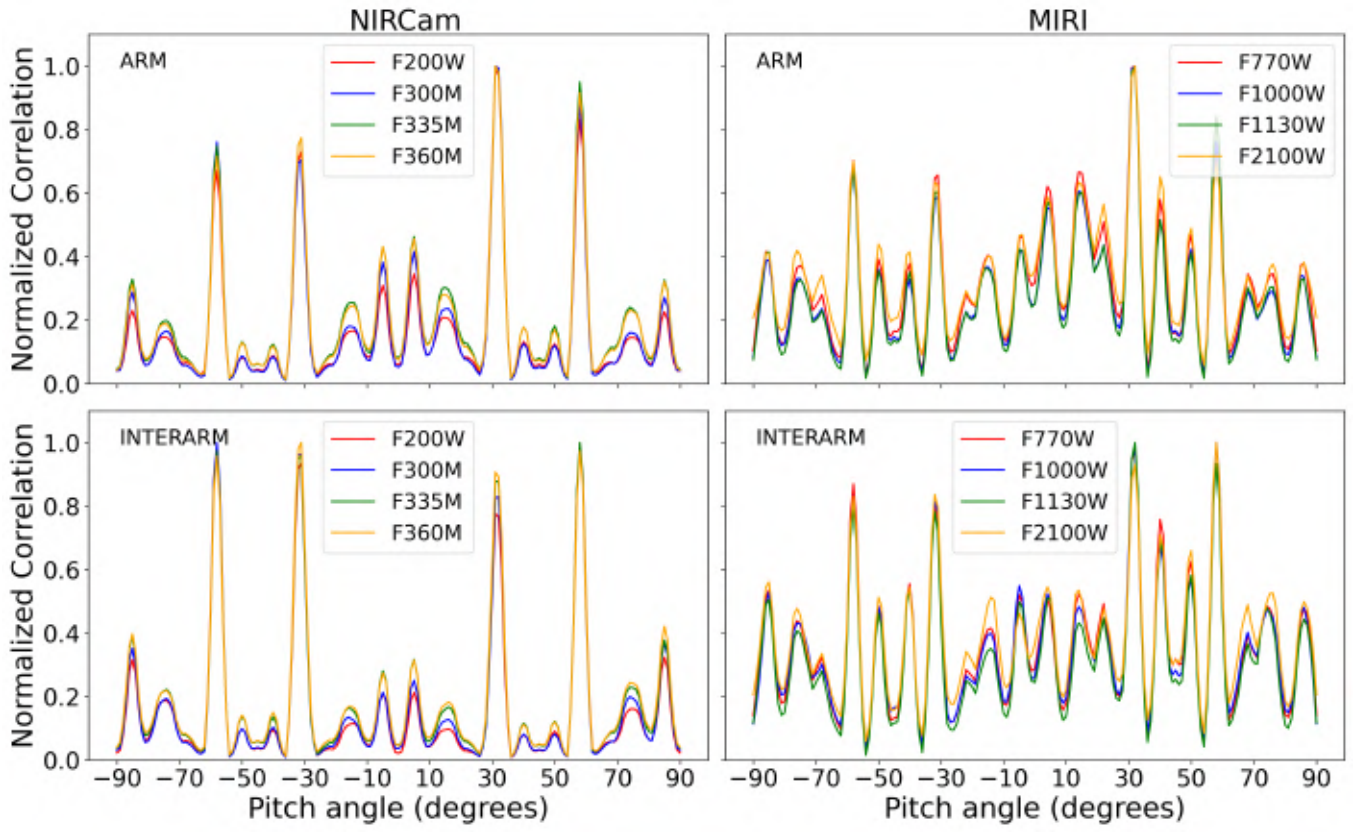

(a) Sum of the correlations calculated with $D_w = 0.25$ for the arm and interarm regions of NGC 628.

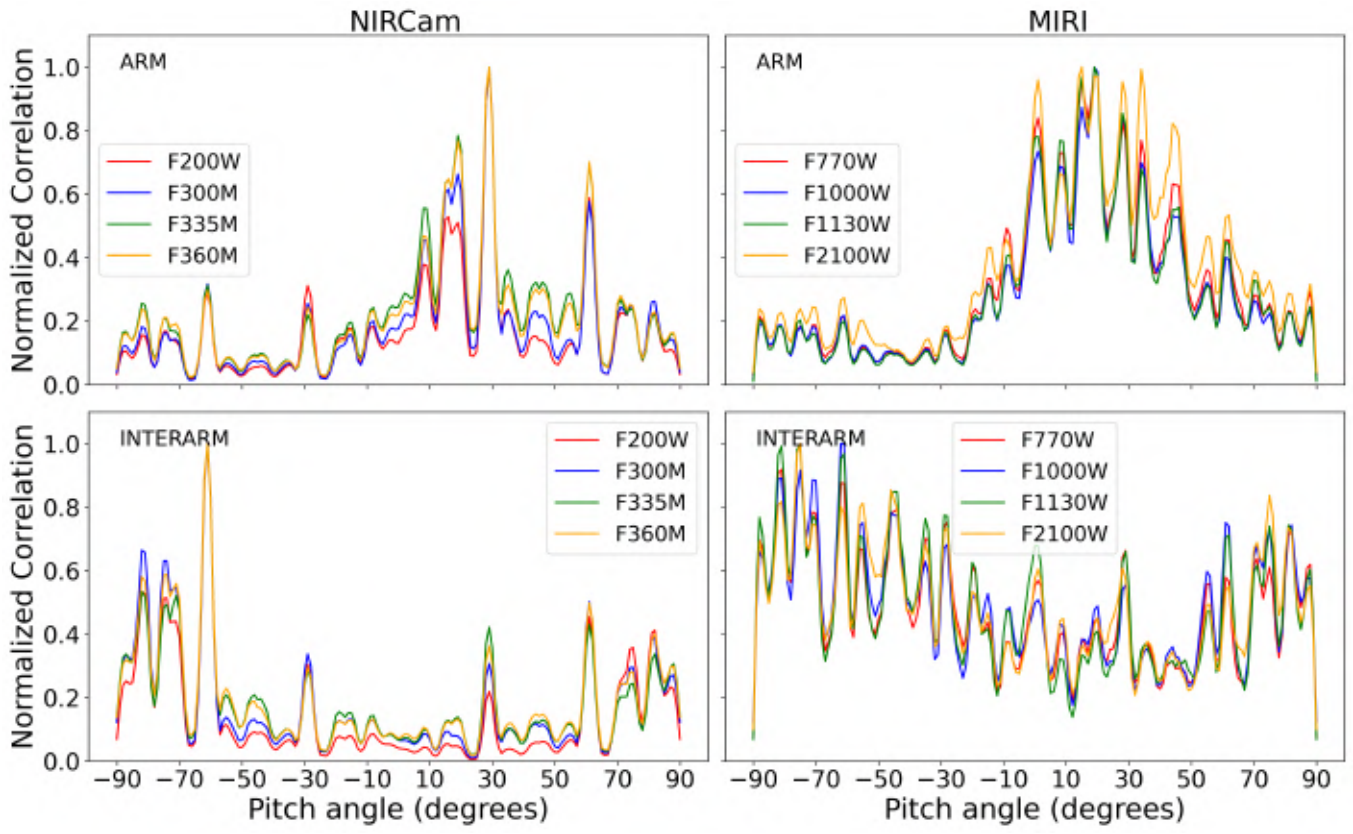

(b) Sum of the correlations calculated with $D_w = 0.5$ for the arm and interarm regions of NGC 628.

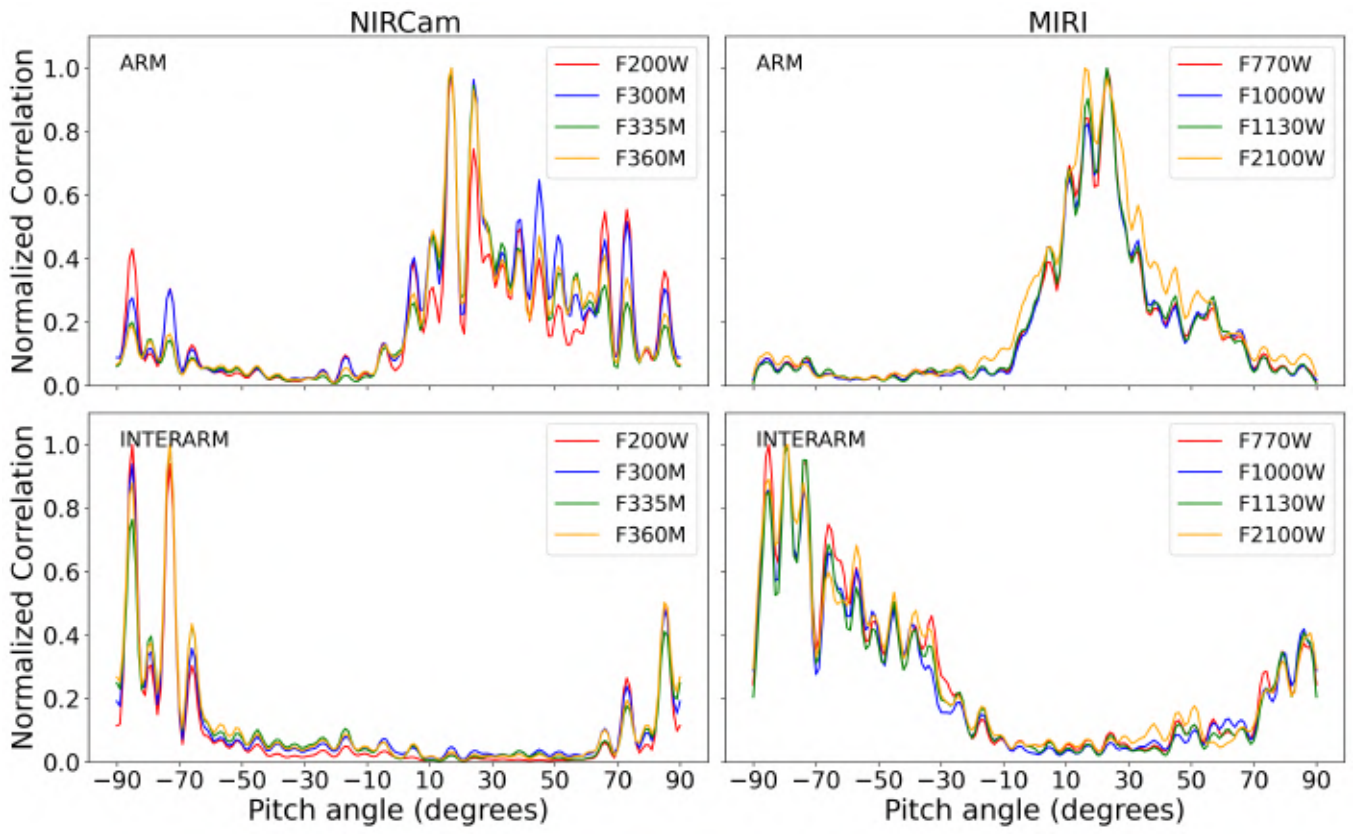

(c) Sum of the correlations calculated with $D_w = 0.75$ for the arm and interarm regions of NGC 628.

**Figure A2.** Results for the sum of the correlations in the analyzed regions of NGC 628. On the left, plots from NIRCam images with filters $F200W$ (red), $F300M$ (blue), $F335M$ (green), and $F360M$ (yellow), where the upper panel corresponds to the arm region and the lower panel to the interarm region. On the right, plots from MIRI images with filters $F770W$ (red), $F1000W$ (blue), $F1130W$ (green), and $F2100W$ (yellow), following the same division between arm (upper) and interarm (lower) regions.

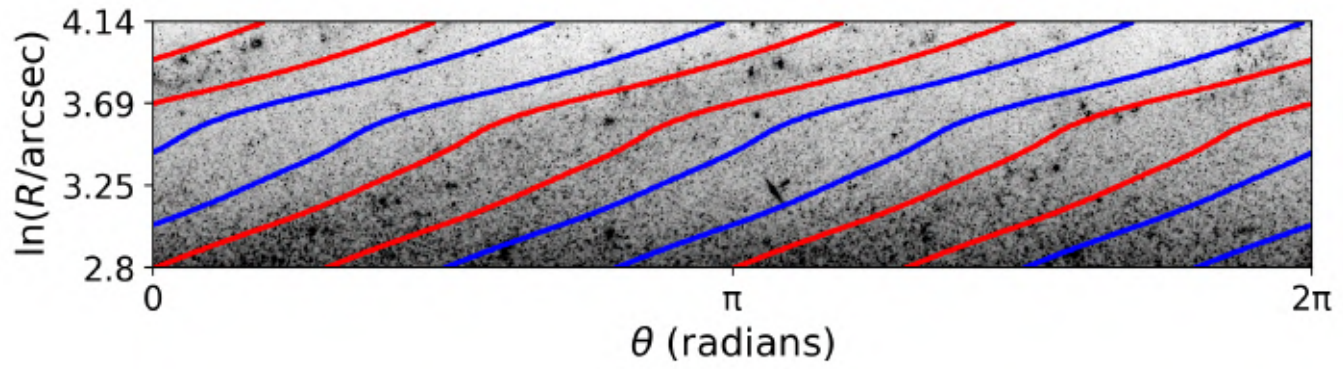

(a) Regions delimited in the $F335M$ image of the galaxy NGC 628. The regions defined as the arm are located between the red lines, and the interarm regions are between the blue lines.

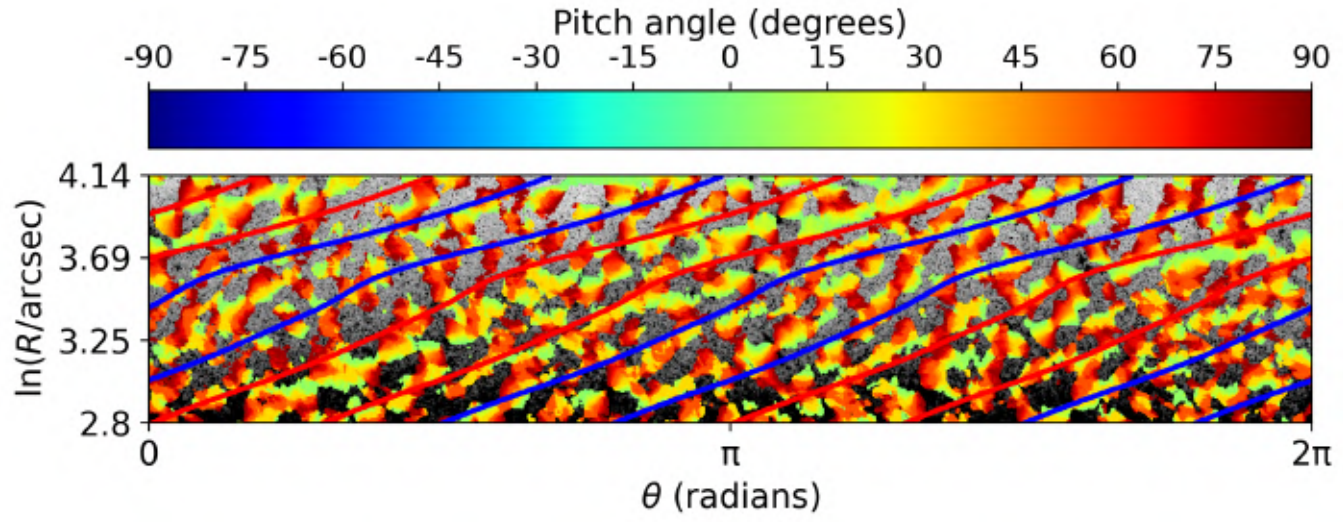

(b) Distribution of pitch angles obtained with the $D_w = 0.25$ filter.

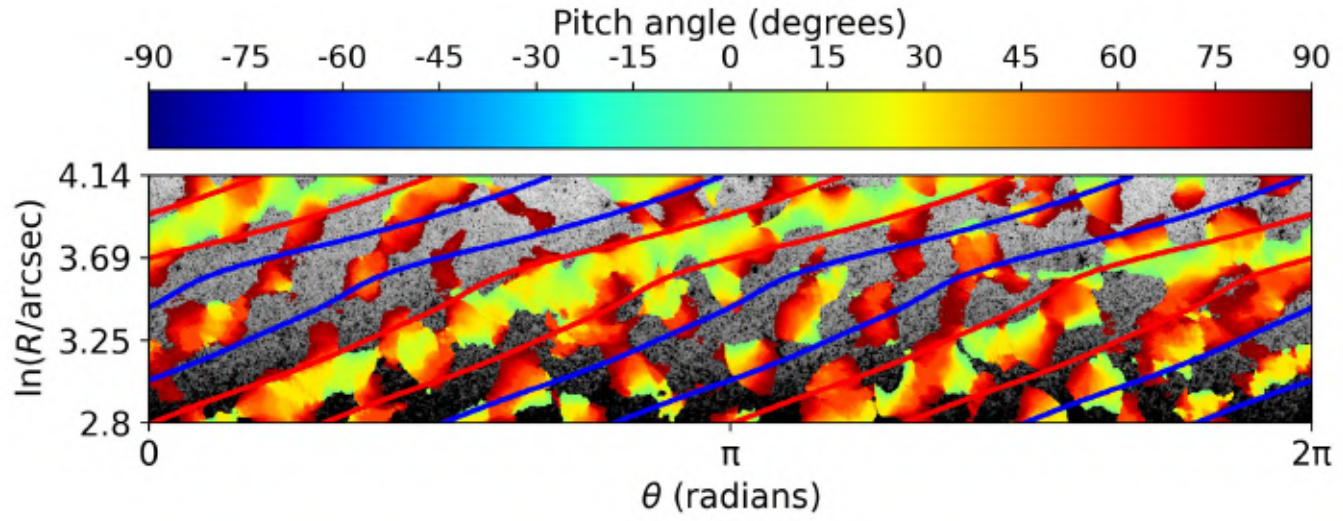

(c) Distribution of pitch angles obtained with the $D_w = 0.5$ filter.

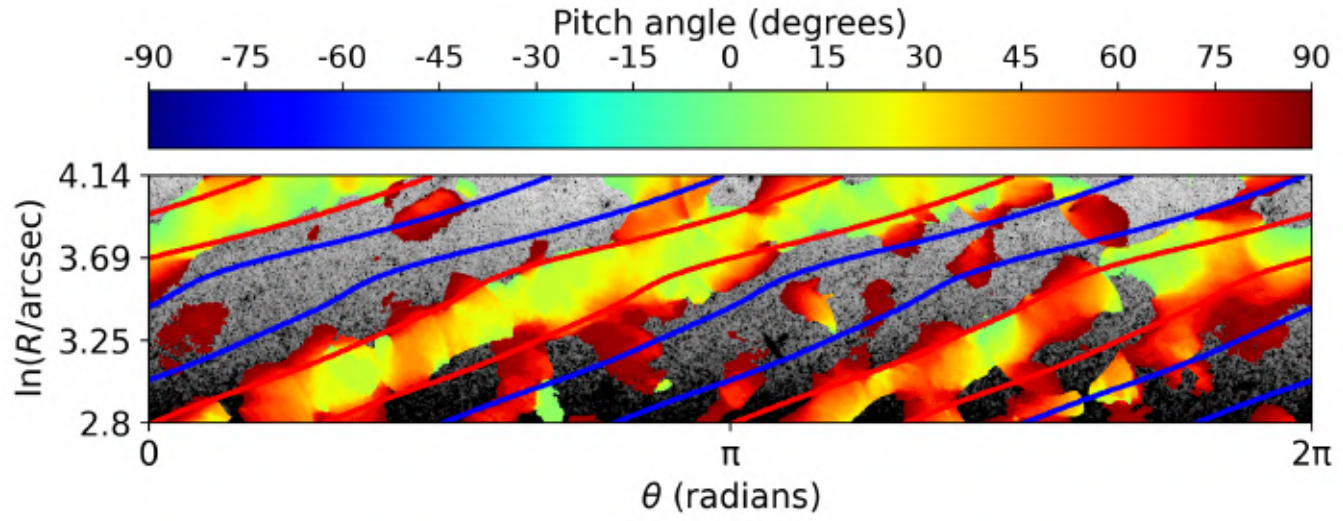

(d) Distribution of pitch angles obtained with the $D_w = 0.75$ filter.

**Figure A3.** Distribution of pitch angles obtained from the highest correlations at each position $(\ln(R/\mathrm{arcsec}), \theta)$ in the $F335M$ image of NGC 628. The colorbar represents pitch angle values, ranging from $-90°$ (dark blue) to $90°$ (dark red). The colors of the points indicate the pitch angle associated with the maximum correlation, based on the idea that the strongest correlations occur when the structure has the same pitch angle as the filter at the corresponding position.

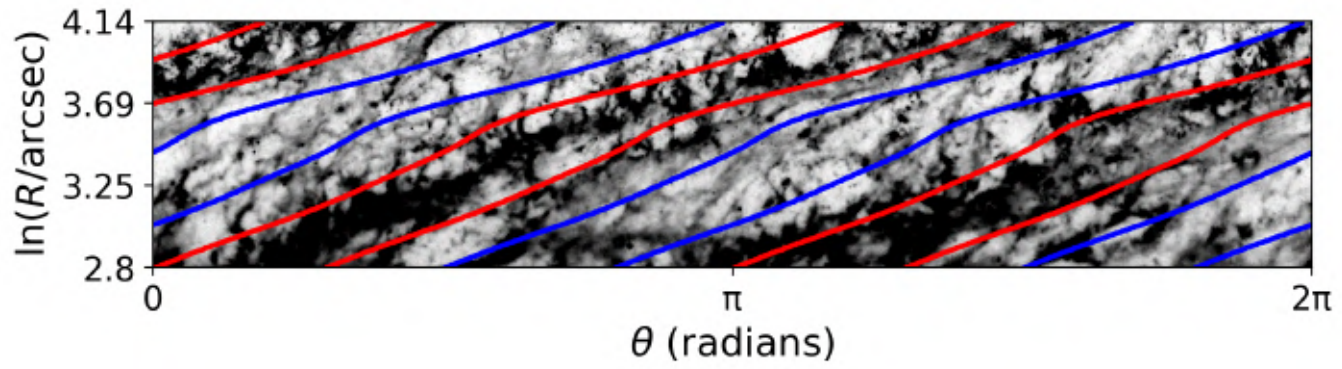

(a) Regions delimited in the $F1130W$ image of the galaxy NGC 628. The spiral arm regions are located between the red lines, while the interarm regions are between the blue lines.

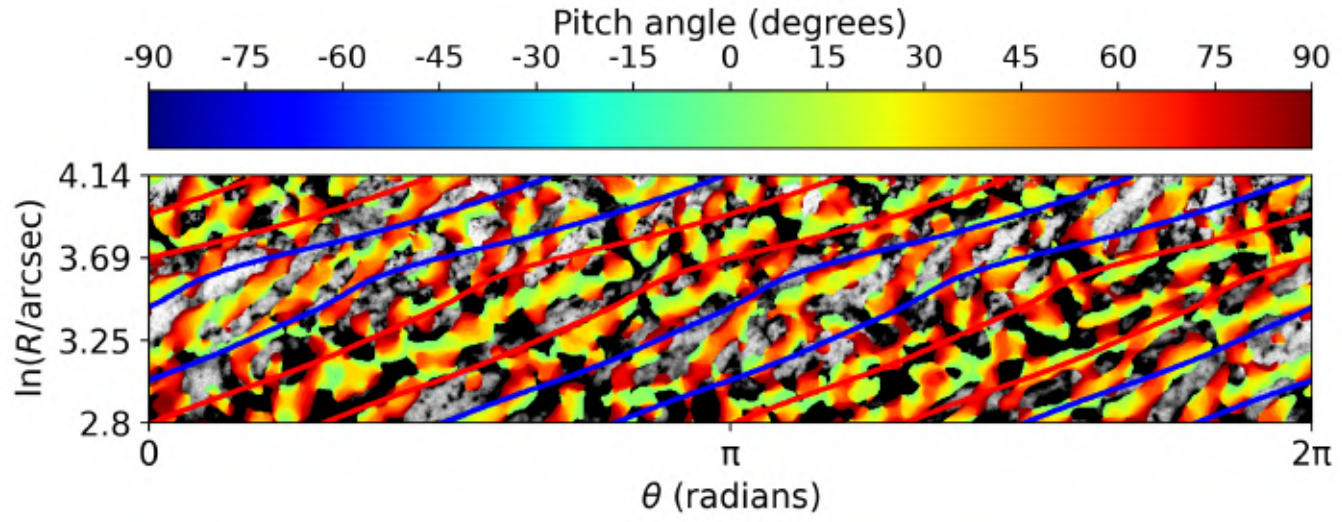

(b) Distribution of pitch angles obtained with the $D_w = 0.25$ filter.

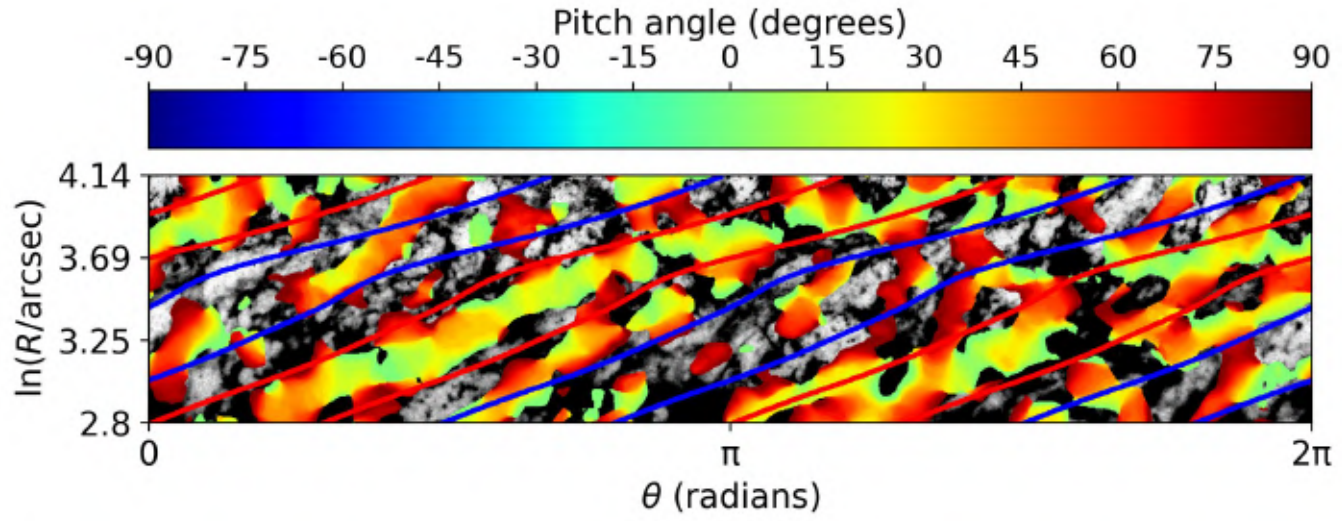

(c) Distribution of pitch angles obtained with the $D_w = 0.5$ filter.

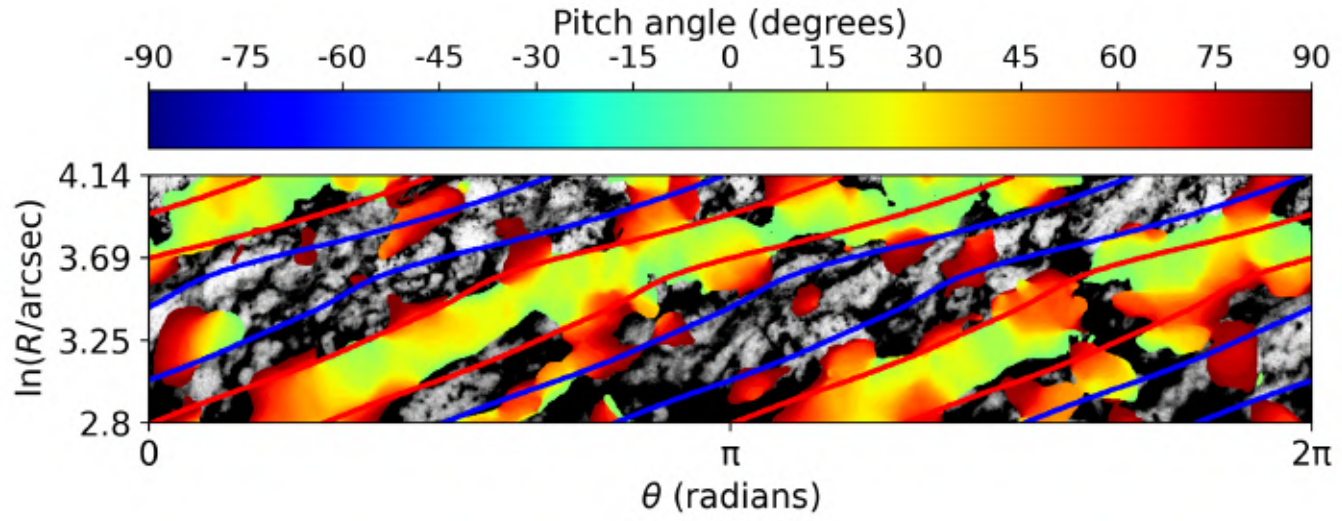

(d) Distribution of pitch angles obtained with the $D_w = 0.75$ filter.

**Figure A4.** Distribution of pitch angles obtained from the highest correlations at each position $(\ln(R/\mathrm{arcsec}), \theta)$ in the $F1130W$ image of NGC 628. The colorbar represents pitch angle values, ranging from $-90°$ (dark blue) to $90°$ (dark red). The colors of the points indicate the pitch angle associated with the maximum correlation, based on the idea that the strongest correlations occur when the structure has the same pitch angle as the filter at the corresponding position.

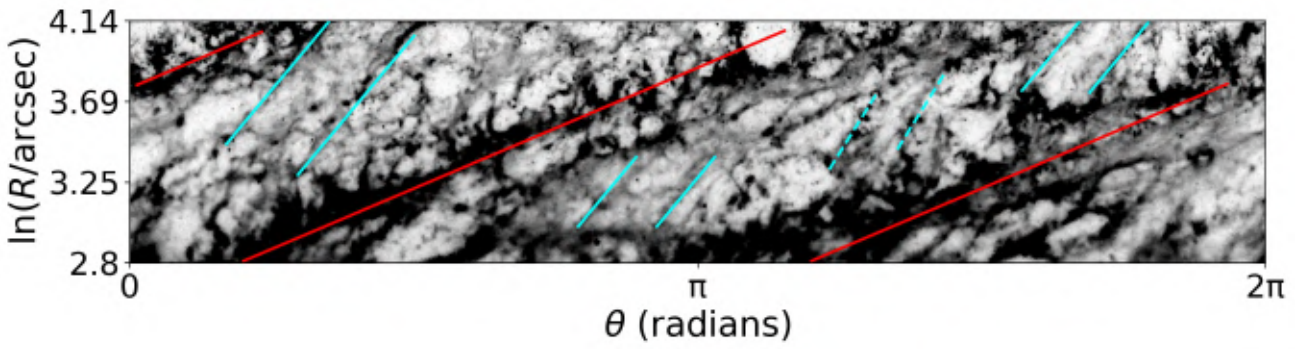

**Figure A5.** Lines overlaid on the $F1130W$ image of NGC 628, indicating the pitch angles of the spiral structures. The solid red line corresponds to $17°$, the solid blue line to $50°$, and the dashed blue line to $58°$.

## A2 NGC 3627

The analysis of NGC 3627 was performed between $\ln(R_{\rm min}/\rm arcsec) = 3.8$ and $\ln(R_{\rm max}/\rm arcsec) = 4.5$, defined in the $(\ln(R/\rm arcsec), \theta)$ space (Figure A6). Within this range, we investigated the regions corresponding to the main arms and interarm regions, based on the mask obtained with the 2D-FT (Figures A8a and A9a).

Figure A7 shows the sum of the correlations as a function of pitch angle for the arm and interarm regions of NGC 3627, for the three different window diameters $D_w = 0.25$, 0.5, and 0.75 (Figures A7a, A7b, and A7c, respectively). Since this galaxy has S-type spiral winding, we analyze the negative peaks, where $-90^\circ \leq \alpha \leq 0^\circ$.

NGC 3627 is classified as SAB(s)b with a central bar, but the field of view also captures spiral arms and analyzable substructures. According to Elmegreen & Elmegreen (1987), this galaxy presents "two symmetric, long outer arms; feathery or irregular inner arms". In the selected analysis range, $R_{\rm min}$ was chosen to avoid the bar region and mark the beginning of the main spiral arms, as described in Section 2.2.3. However, Figure A6 shows that both arms maintain a certain pitch angle up to $\ln(R/\rm arcsec) \approx 4.27$. Beyond this point, the winding decreases — especially in the second main arm, where the structure becomes almost horizontal. This change can be attributed to the transition between the end of the bar and the onset of the spiral arm. The transition region still shows a distinct inclination, resulting in larger pitch angle values up to $\ln(R/\rm arcsec) \approx 4.27$. After this point, with the bar's influence no longer dominant, a reduction in pitch angle is observed.

For the spiral arms, considering $D_w = 0.5$ (Figure A7b), peaks appear at approximately $\alpha \approx -29^\circ$, $\alpha \approx -34^\circ$, and $\alpha \approx -8^\circ$ in both the NIRCam and MIRI images. The value $\alpha \approx -32^\circ$ provides a good description of the first main arm beyond $\ln(R/\rm arcsec) \approx 4.27$. At the beginning of this arm, however, larger pitch angle values are observed, in the range $-45^\circ \leq \alpha \leq -40^\circ$. Between the two main arms, in the interarm region, the MIRI images reveal a substructure resembling a branch, with $-40^\circ \leq \alpha \leq -30^\circ$. The second main arm emerges from the bar with a tightly wound configuration, $-10^\circ \leq \alpha \leq -5^\circ$, which explains the peak around $\alpha \approx -8^\circ$.

For $D_w = 0.75$ (Figure A7c), peaks are observed at $\alpha \approx -25^\circ$, $\alpha \approx -29^\circ$, and $\alpha \approx -32^\circ$ in the NIRCam images, and at $\alpha \approx -23^\circ$ and $\alpha \approx -34^\circ$ in the MIRI images. At smaller scales, $D_w = 0.25$ (Figure A7a) and $D_w = 0.5$ (Figure A7b), focused on the interarm regions, a peak at $\alpha \approx -15^\circ$ is highlighted. This peak corresponds to a substructure branching from the first main arm and extending to its left, most clearly visible in the MIRI images.

Thus, for the main spiral arms we estimate $\alpha \approx -32^\circ$, while in the interarm regions we identify a substructure with $\alpha \approx -15^\circ$, along with another feature showing a winding similar to that of the main arm.

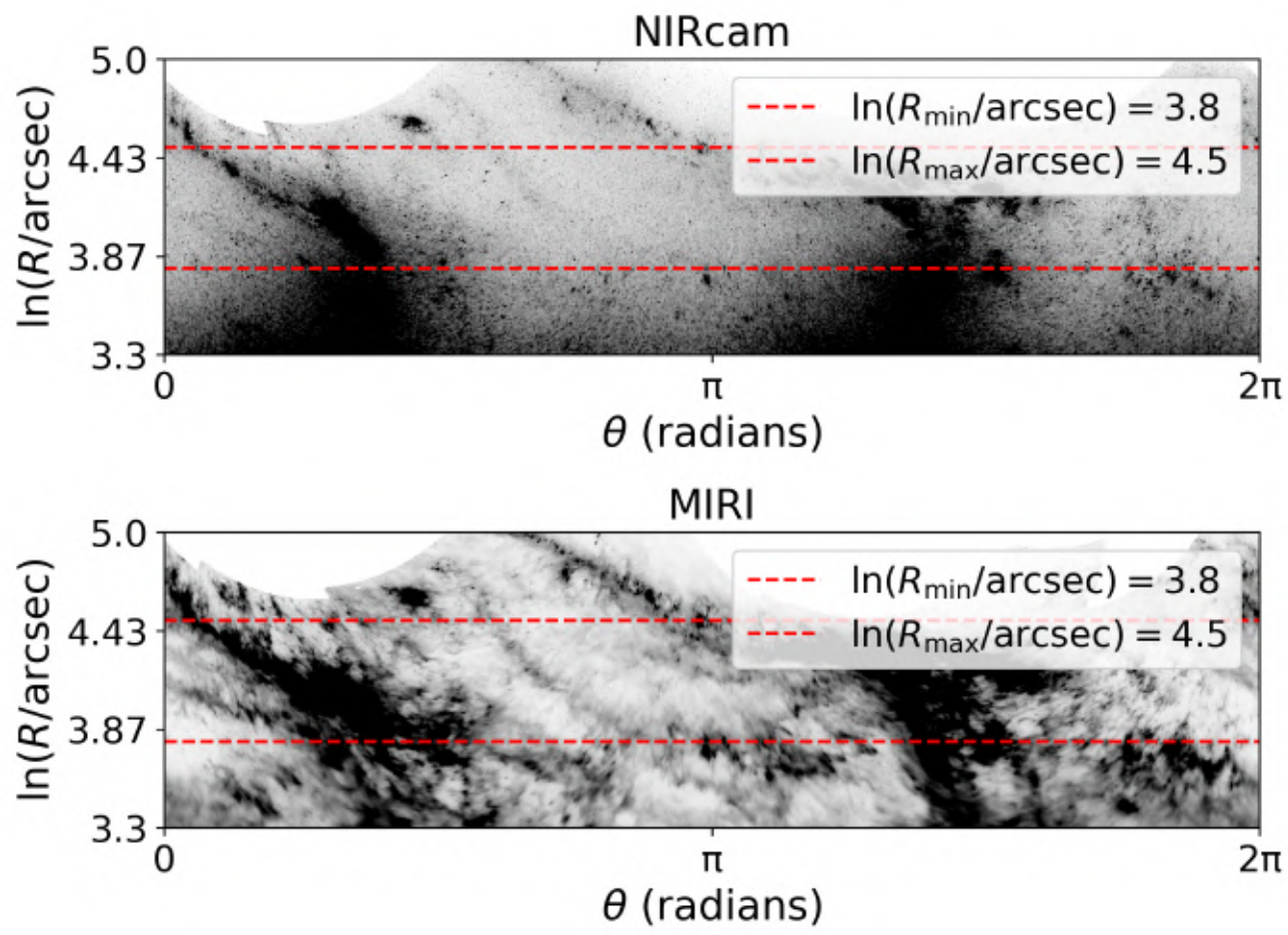

**Figure A6.** Diagrams $\ln(R/\rm arcsec)$ vs. $\theta$ for the $F335M$ (NIRCam) and $F1130W$ (MIRI) images of NGC 3627. The dashed red lines indicate the values of $\ln(R_{min}/\rm arcsec)$ and $\ln(R_{max}/\rm arcsec)$ adopted in the analysis.

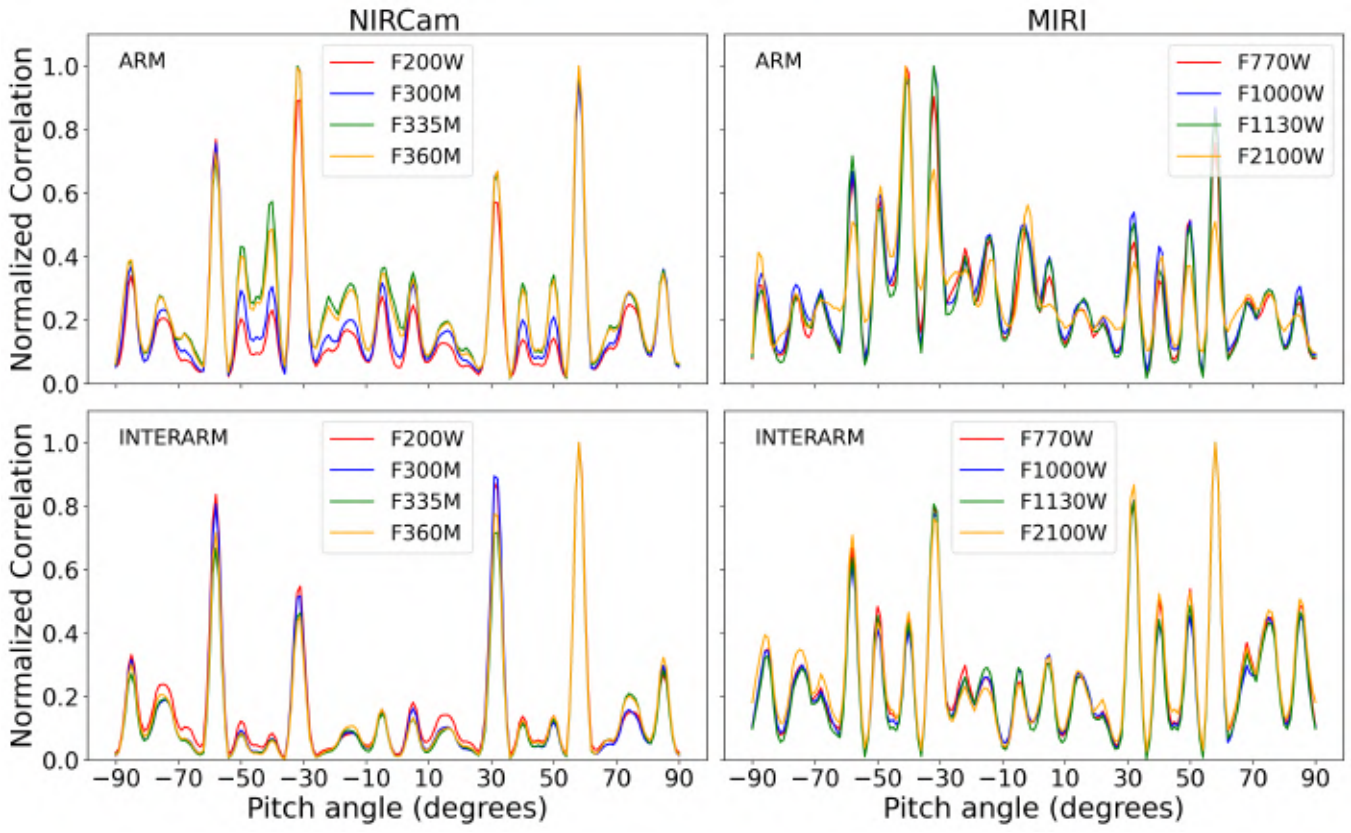

(a) Sum of the correlations calculated with $D_w = 0.25$ for the arm and interarm regions of NGC 3627.

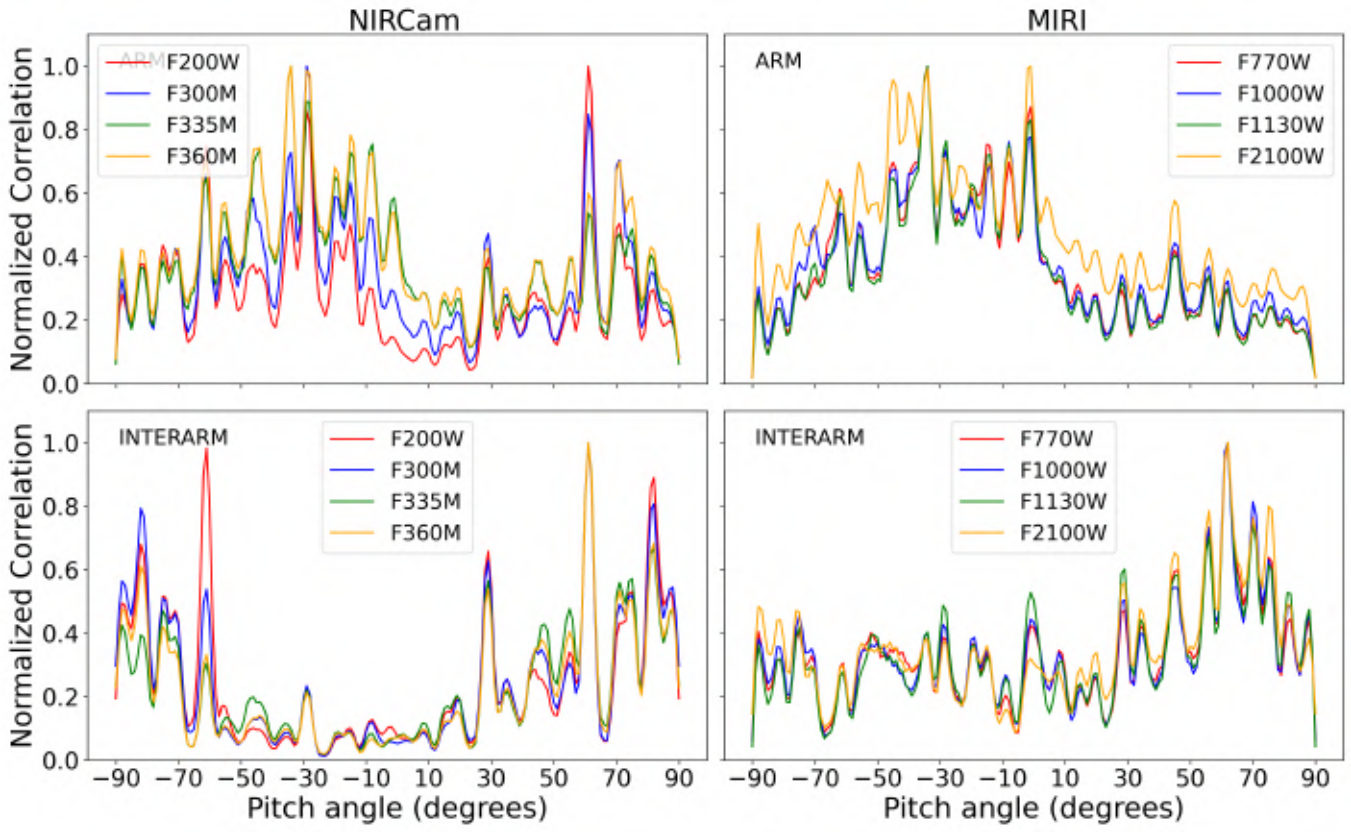

(b) Sum of the correlations calculated with $D_w = 0.5$ for the arm and interarm regions of NGC 3627.

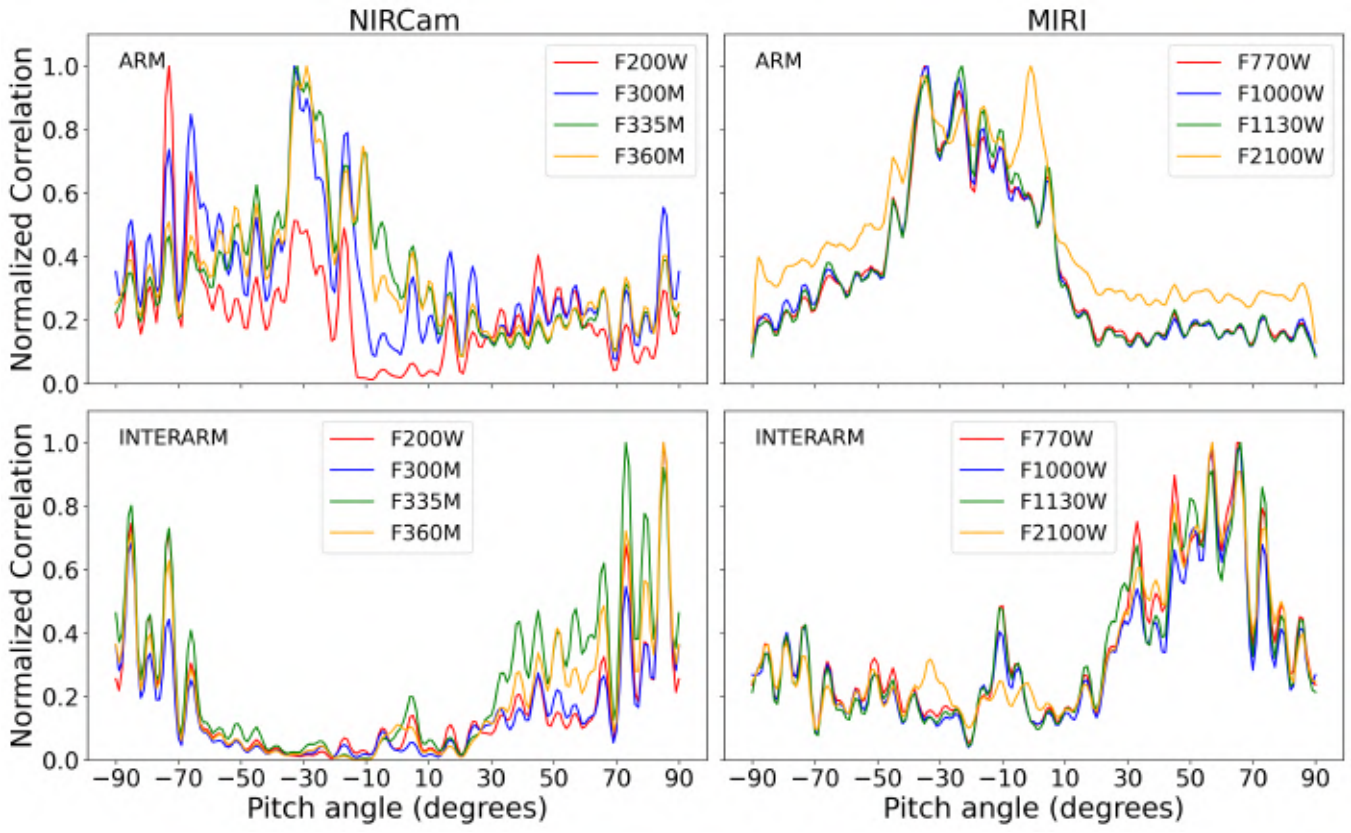

(c) Sum of the correlations calculated with $D_w = 0.75$ for the arm and interarm regions of NGC 3627.

**Figure A7.** Results for the sum of the correlations in the analyzed regions of NGC 3627. On the left, plots from NIRCam images with filters $F200W$ (red), $F300M$ (blue), $F335M$ (green), and $F360M$ (yellow), where the upper panel corresponds to the arm region and the lower panel to the interarm region. On the right, plots from MIRI images with filters $F770W$ (red), $F1000W$ (blue), $F1130W$ (green), and $F2100W$ (yellow), following the same division between arm (upper) and interarm (lower) regions.

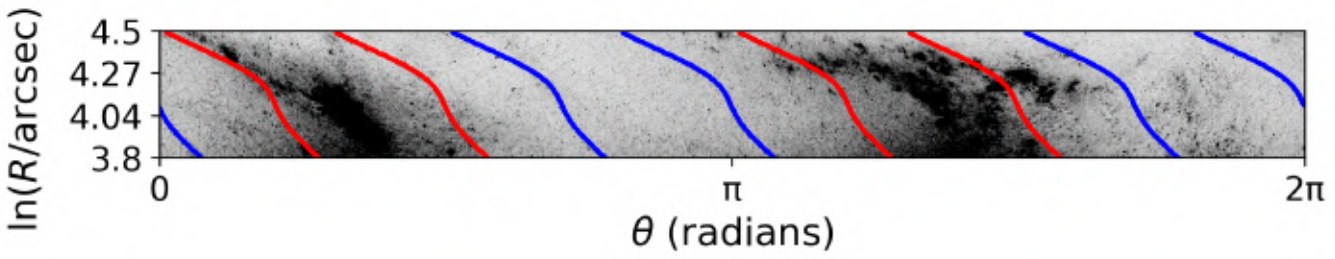

(a) Regions delimited in the $F335M$ image of NGC 3627. The spiral arm regions are located between the red lines, while the interarm regions are between the blue lines.

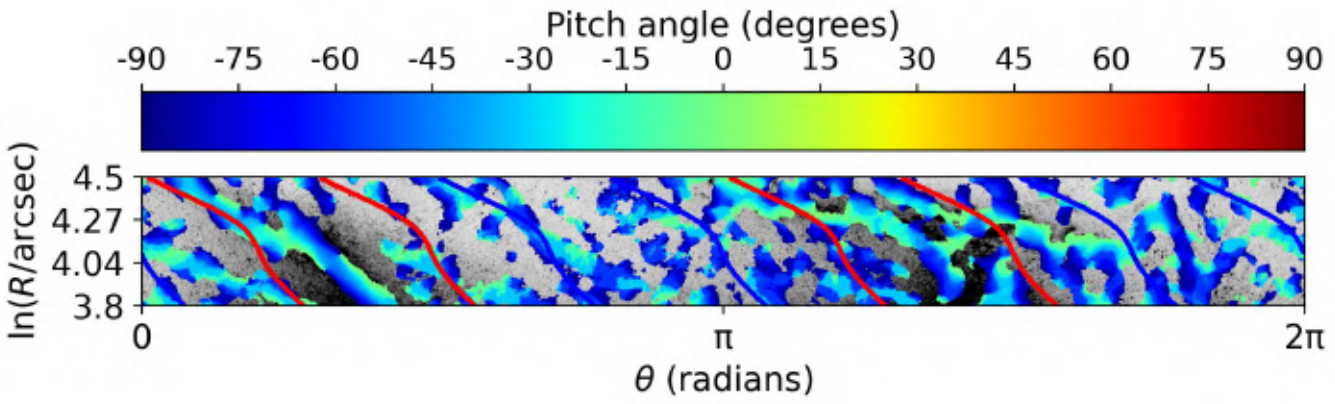

(b) Distribution of pitch angles obtained with the $D_w = 0.25$ filter.

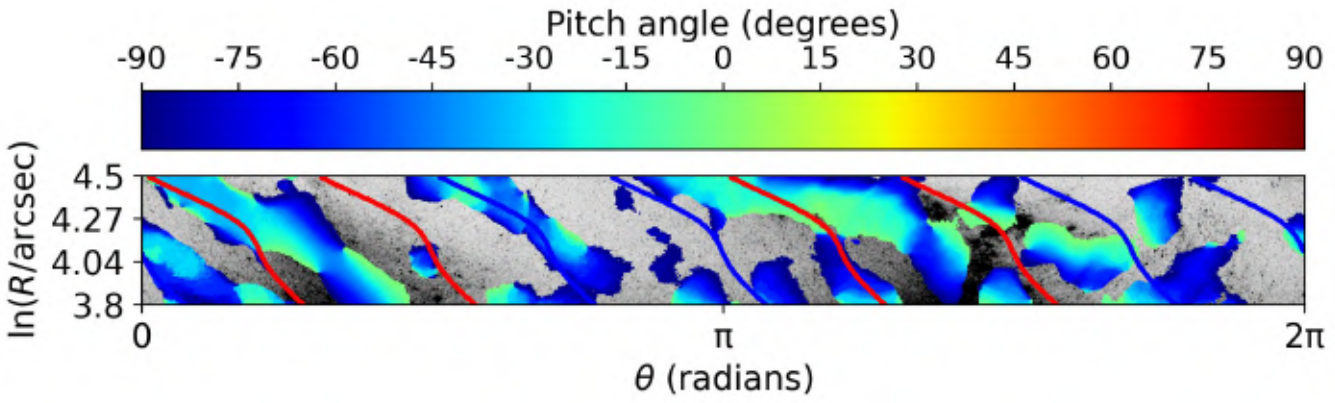

(c) Distribution of pitch angles obtained with the $D_w = 0.5$ filter.

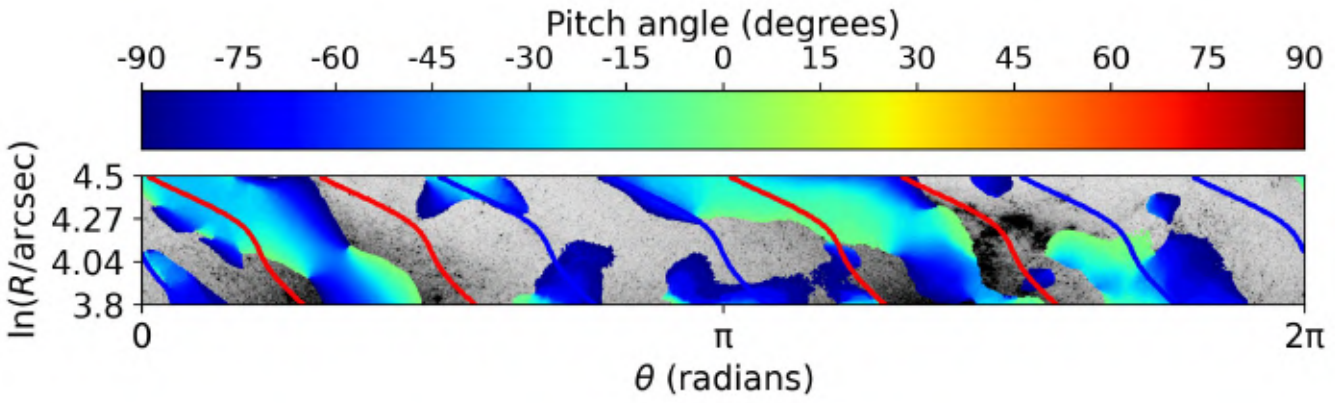

(d) Distribution of pitch angles obtained with the $D_w = 0.75$ filter.

**Figure A8.** Distribution of pitch angles obtained from the highest correlations at each position $(\ln(R/\mathrm{arcsec}), \theta)$ in the $F335M$ image of NGC 3627. The colorbar represents pitch angle values, ranging from $-90°$ (dark blue) to $90°$ (dark red). The colors of the points indicate the pitch angle associated with the maximum correlation, based on the idea that the strongest correlations occur when the structure has the same pitch angle as the filter at the corresponding position.

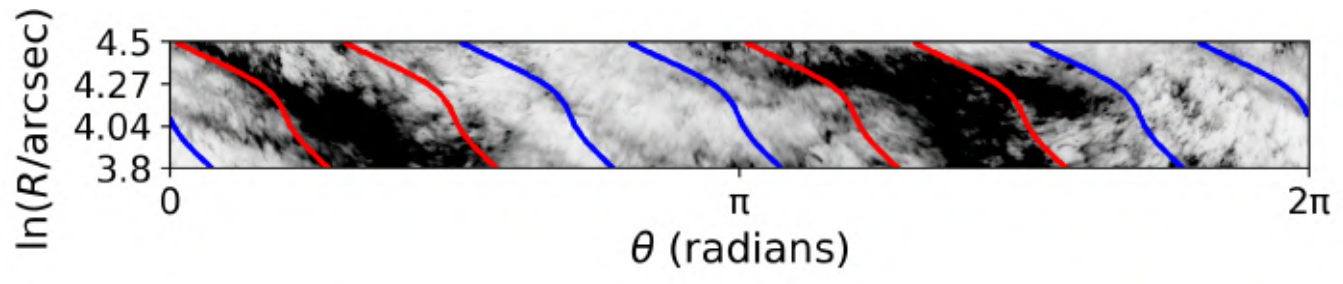

(a) Regions delimited in the $F1130W$ image of NGC 3627. The spiral arm regions are located between the red lines, while the interarm regions are between the blue lines.

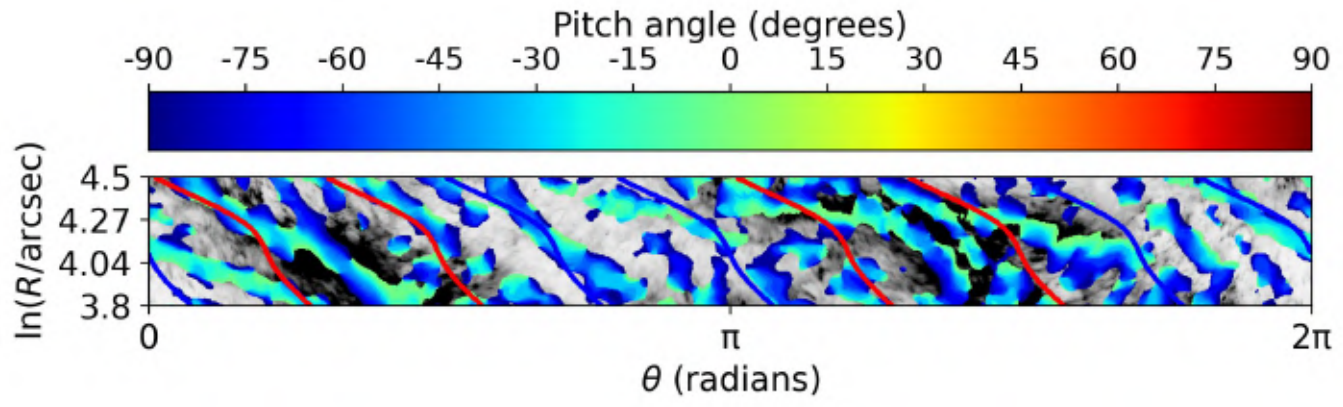

(b) Distribution of pitch angles obtained with the $D_w$ = 0.25 filter.

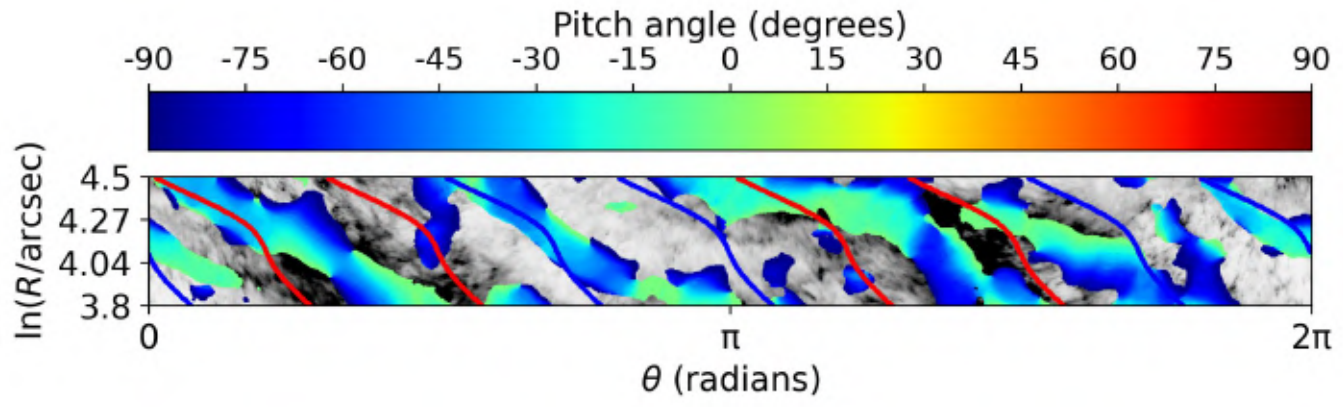

(c) Distribution of pitch angles obtained with the $D_w$ = 0.5 filter.

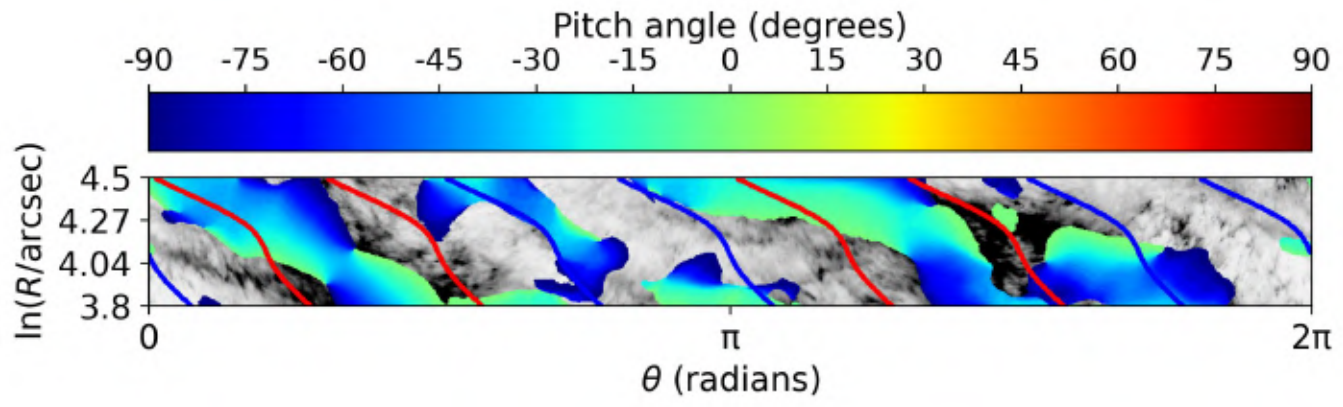

(d) Distribution of pitch angles obtained with the $D_w$ = 0.75 filter.

**Figure A9.** Distribution of pitch angles obtained from the highest correlations at each position $(\ln(R/\text{arcsec}), \theta)$ in the $F1130W$ image of NGC 3627. The colorbar represents pitch angle values, ranging from $-90°$ (dark blue) to 90° (dark red). The colors of the points indicate the pitch angle associated with the maximum correlation, based on the idea that the strongest correlations occur when the structure has the same pitch angle as the filter at the corresponding position.

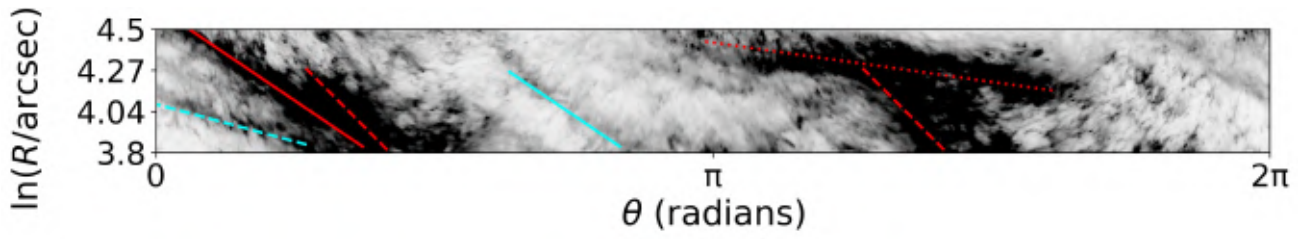

**Figure A10.** Lines overlaid on the $F1130W$ image of NGC 3627, indicating the pitch angles of the spiral structures. The solid red line corresponds to $-32°$, the dashed red line to $-45°$, the dotted red line to $-8°$, the solid blue line to $-32°$, and the dashed blue line to $-15°$.

## A3 NGC 4254

NGC 4254 is morphologically classified as SA(s)c and exhibits Z-type spiral arms. According to Elmegreen & Elmegreen (1987), this galaxy presents "two symmetric inner arms; multiple long and continuous outer arms". The analysis was performed between $\ln(R_{\rm min}/\text{arcsec}) = 2.55$ and $\ln(R_{\rm max}/\text{arcsec}) = 4.26$, defined in the $(\ln(R/\text{arcsec}), \theta)$ space (Figure A11). Figure A12 shows the sum of the correlations as a function of pitch angle for the arm and inter-arm regions of NGC 4254, for the three different window diameters $D_w$ = 0.25, 0.5, and 0.75 (Figures A12a, A12b, and A12c, respectively).

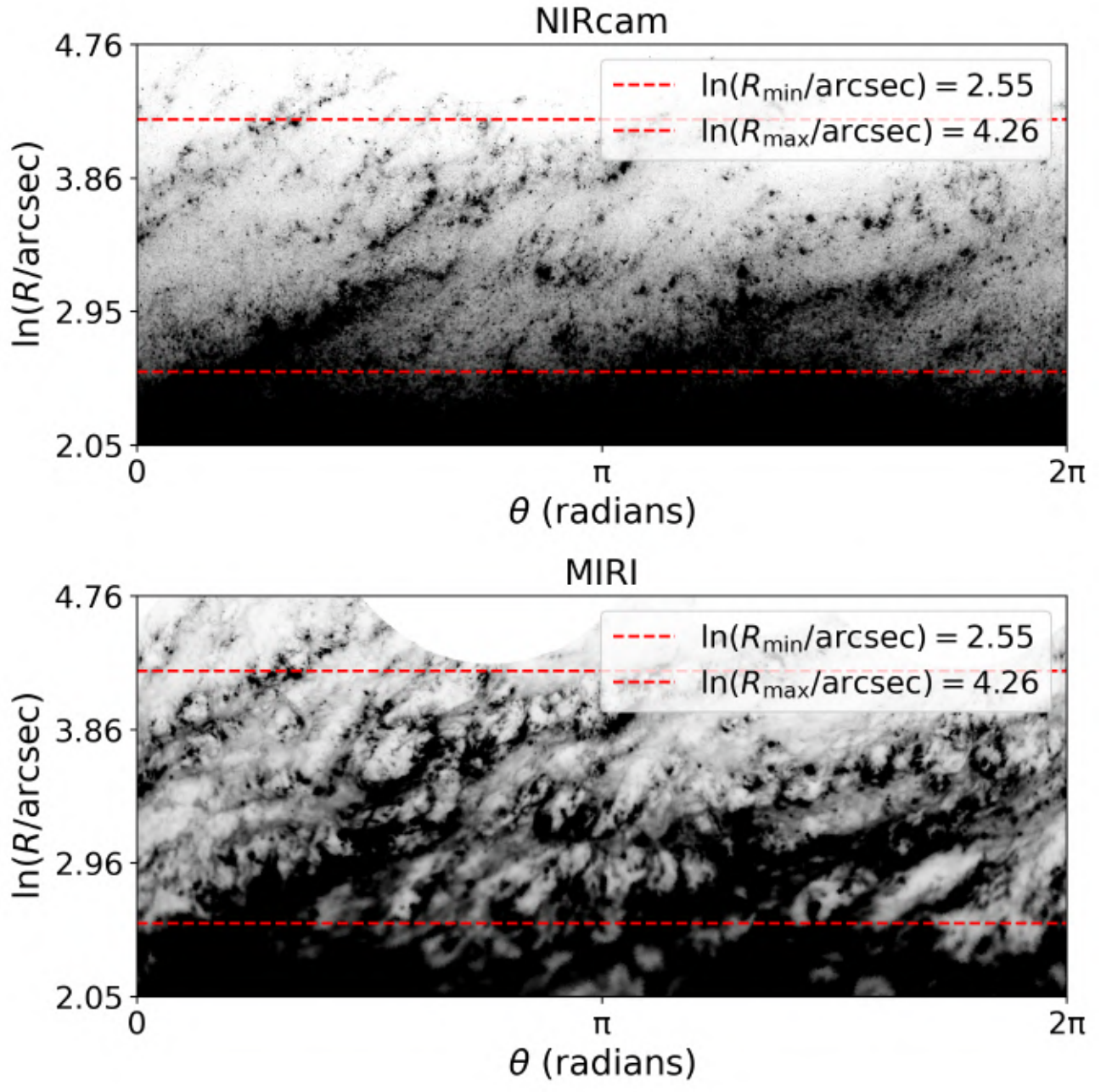

**Figure A11.** Diagrams $\ln(R/\text{arcsec})$ vs. $\theta$ for the $F335M$ (NIRCam) and $F1130W$ (MIRI) images of NGC 4254. The dashed red lines indicate the values of $\ln(R_{min}/\text{arcsec})$ and $\ln(R_{max}/\text{arcsec})$ adopted in the analysis.

In the NIRCam images, an apparent "break" in the main arms is observed at $\ln(R/\text{arcsec}) \approx 3.4$ (see Figure A13a), indicating that up to this radius the main arms are symmetric and follow a given pitch angle. Beyond this point, the arms change, exhibiting smaller pitch angles.

In the correlation analysis for the arm regions, with $D_w$ = 0.5 (Figure A12b), peaks appear at $\alpha \approx 15°$, 19°, and 29° in the NIRCam images, and at $\alpha \approx 15°$, 19°, and 28° in the MIRI images. For $D_w$ = 0.75 (Figure A12c), the predominant peaks are $\alpha \approx 17°$, 24°, and 28° in both NIRCam and MIRI.

From Figures A13d and A14d, it can be seen that the range $24° \leq \alpha \leq 28°$ describes the inner parts of the main arms well. However, beyond $\ln(R/\text{arcsec}) \approx 3.4$, smaller values of $15° \leq \alpha \leq 20°$ are observed, reflecting a lower degree of arm opening in the outer regions. This occurs because two inner arms are observed to be symmetric, with larger pitch angles.

For the interarm regions, with $D_w$ = 0.25 (Figure A12a) and $D_w$ = 0.5 (Figure A12b), a peak at $\alpha \approx 58°$ is prominent, representing the substructures that emerge from the main arms. These feathers are visible in the MIRI images and are best captured at smaller spatial scales (Figures A13b and A14b).

In summary, for the main spiral arms we find a pitch angle of $\alpha \approx 24°$ in the inner radii and $\alpha \approx 17°$ in the outer regions, while the feathers are characterized by substantially larger pitch angles, around $\alpha \approx 58°$.

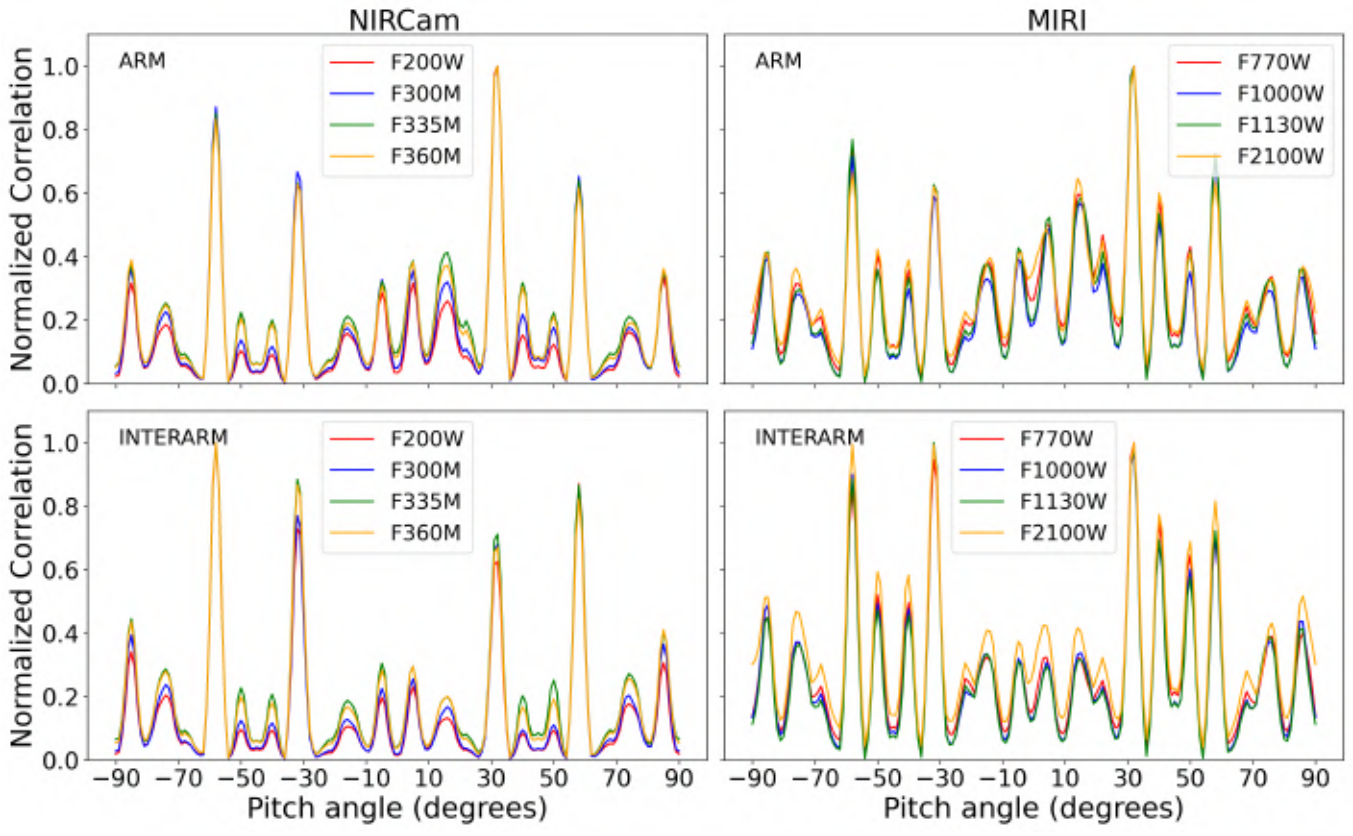

(a) Sum of the correlations calculated with $D_w = 0.25$ for the arm and interarm regions of NGC 4254.

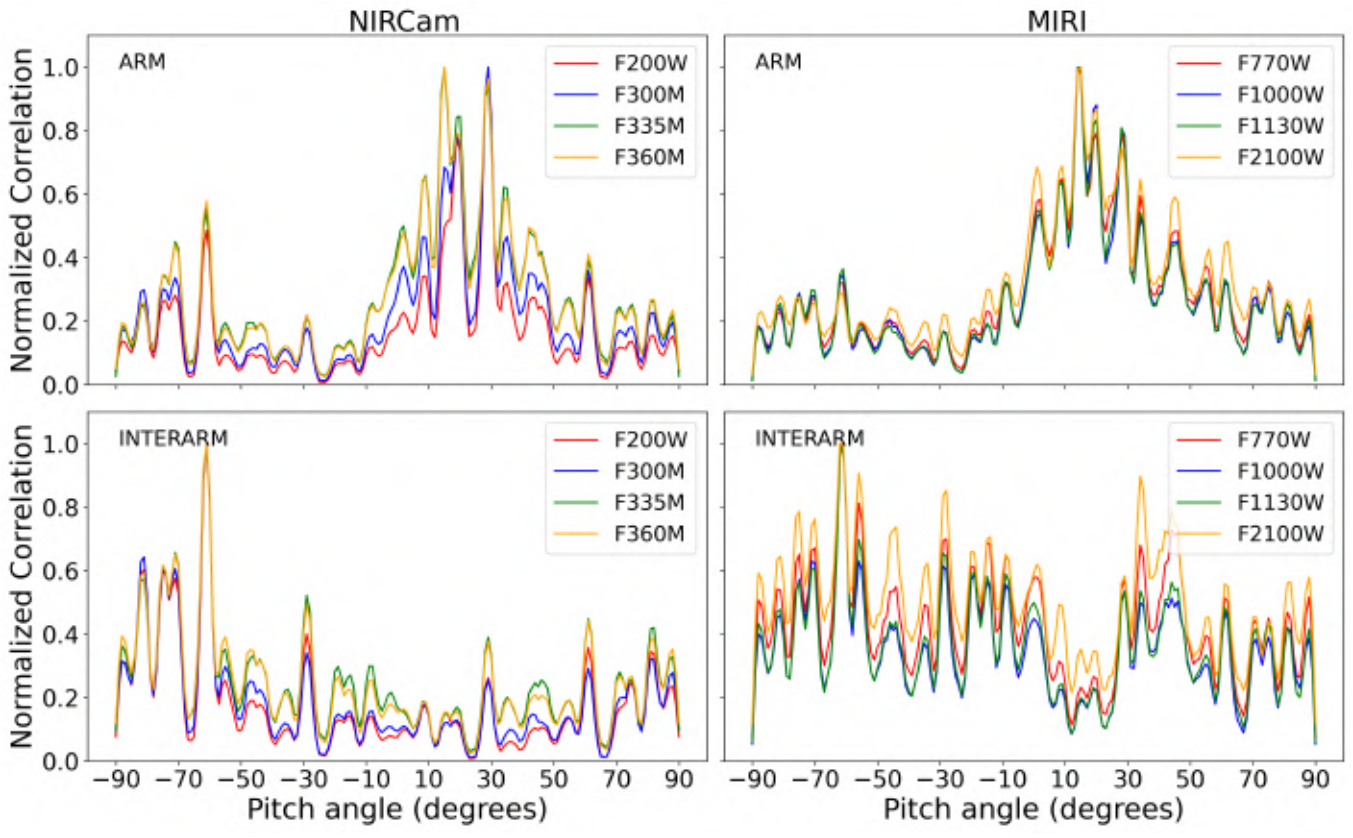

(b) Sum of the correlations calculated with $D_w = 0.5$ for the arm and interarm regions of NGC 4254.

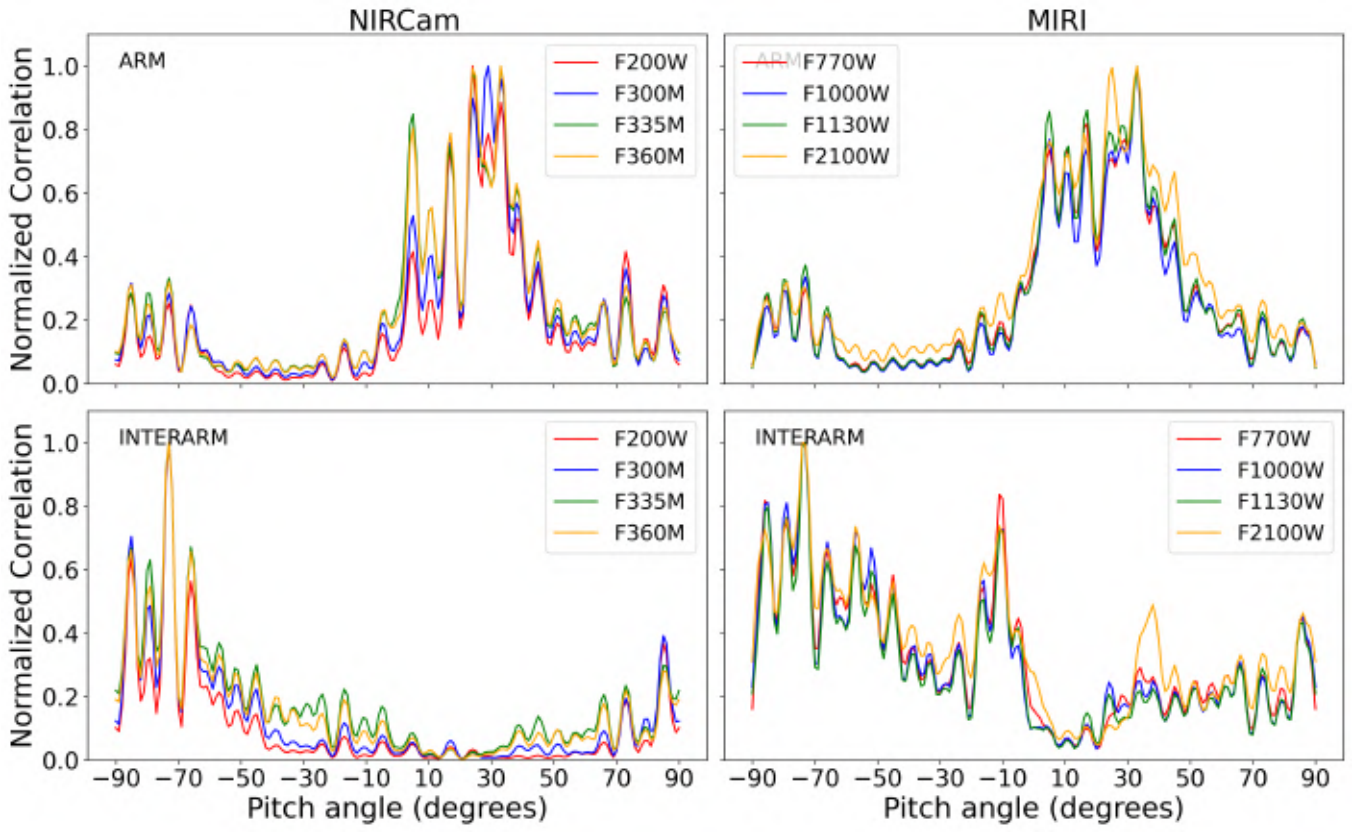

(c) Sum of the correlations calculated with $D_w = 0.75$ for the arm and interarm regions of NGC 4254.

**Figure A12.** Results for the sum of the correlations in the analyzed regions of NGC 4254. On the left, plots from NIRCam images with filters $F200W$ (red), $F300M$ (blue), $F335M$ (green), and $F360M$ (yellow), where the upper panel corresponds to the arm region and the lower panel to the interarm region. On the right, plots from MIRI images with filters $F770W$ (red), $F1000W$ (blue), $F1130W$ (green), and $F2100W$ (yellow), following the same division between arm (upper) and interarm (lower) regions.

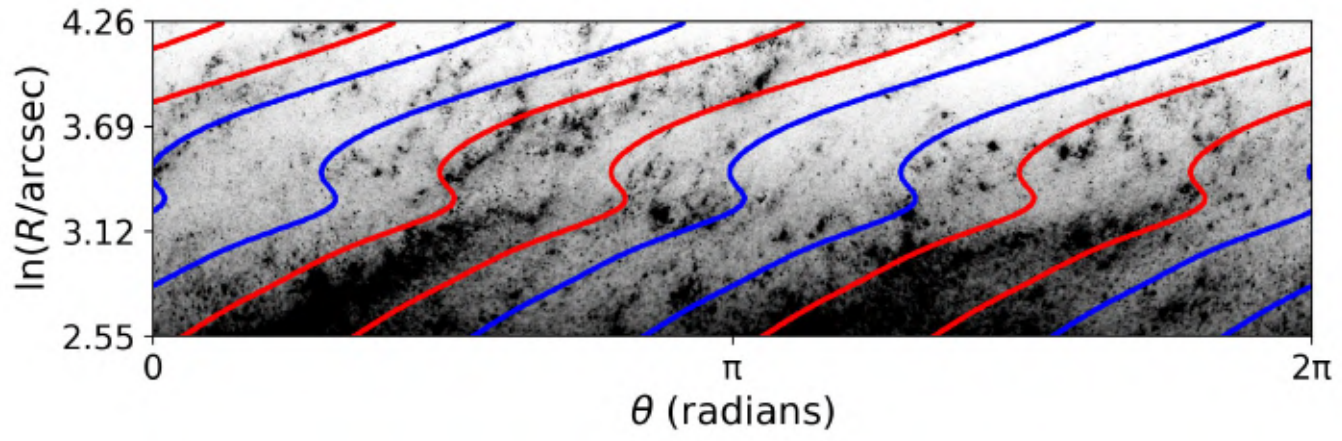

(a) Regions delimited in the $F335M$ image of NGC 4254. The spiral arm regions are located between the red lines, while the interarm regions are between the blue lines.

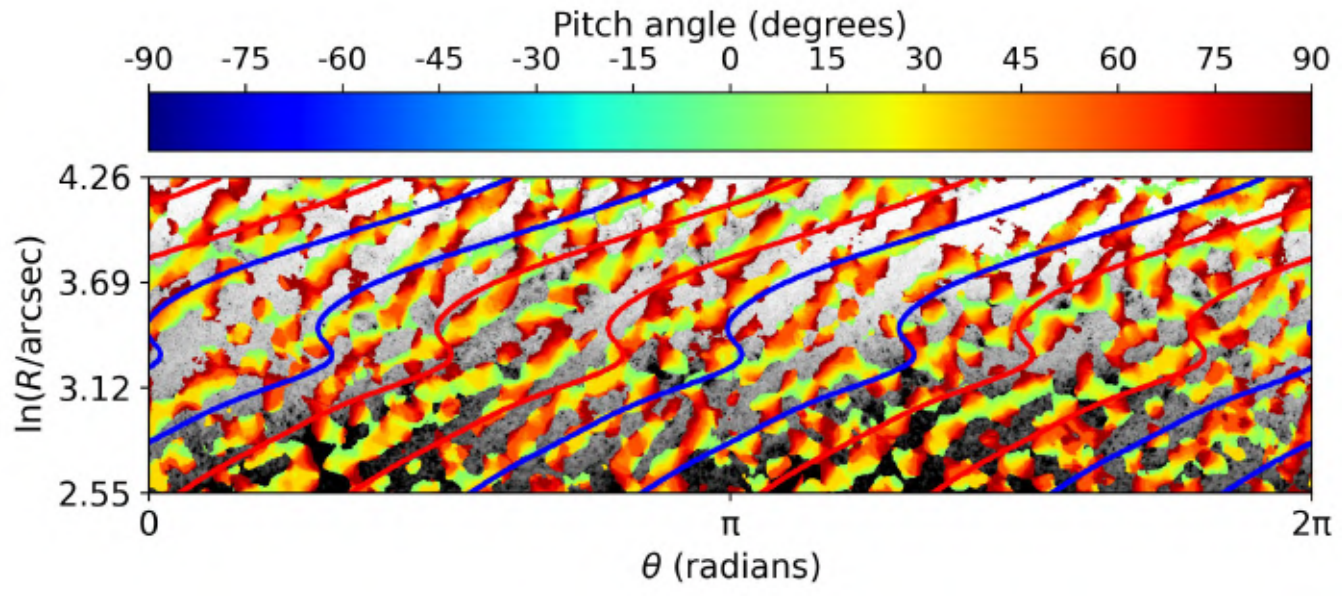

(b) Distribution of pitch angles obtained with the $D_w = 0.25$ filter.

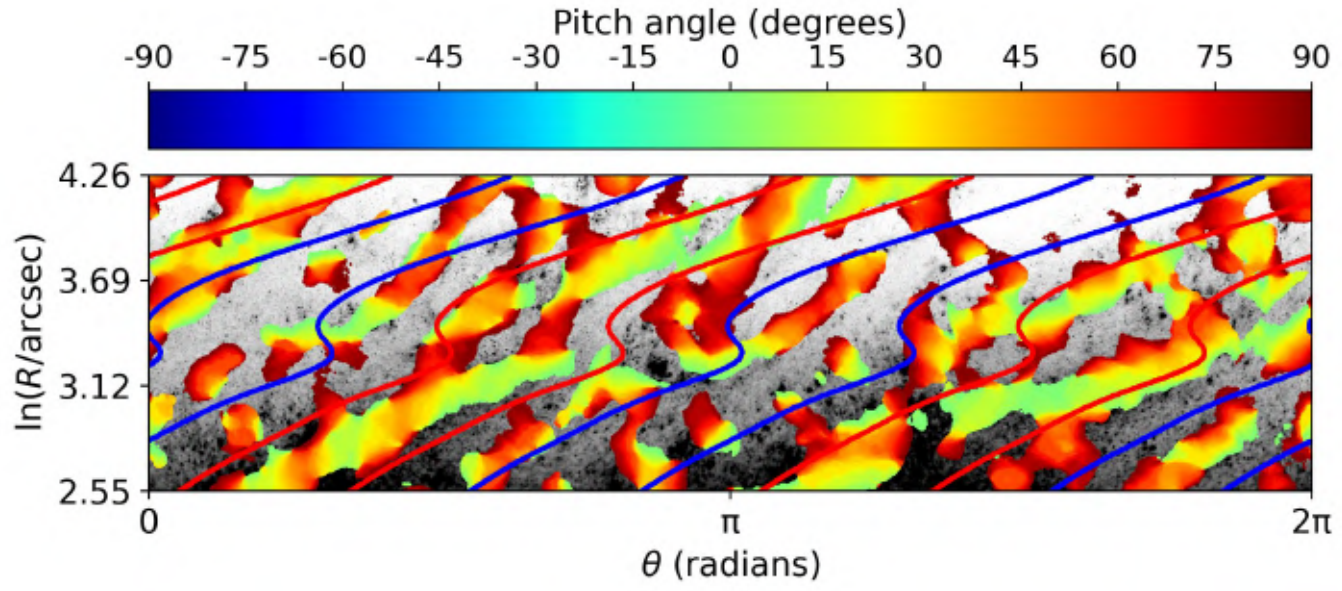

(c) Distribution of pitch angles obtained with the $D_w = 0.5$ filter.

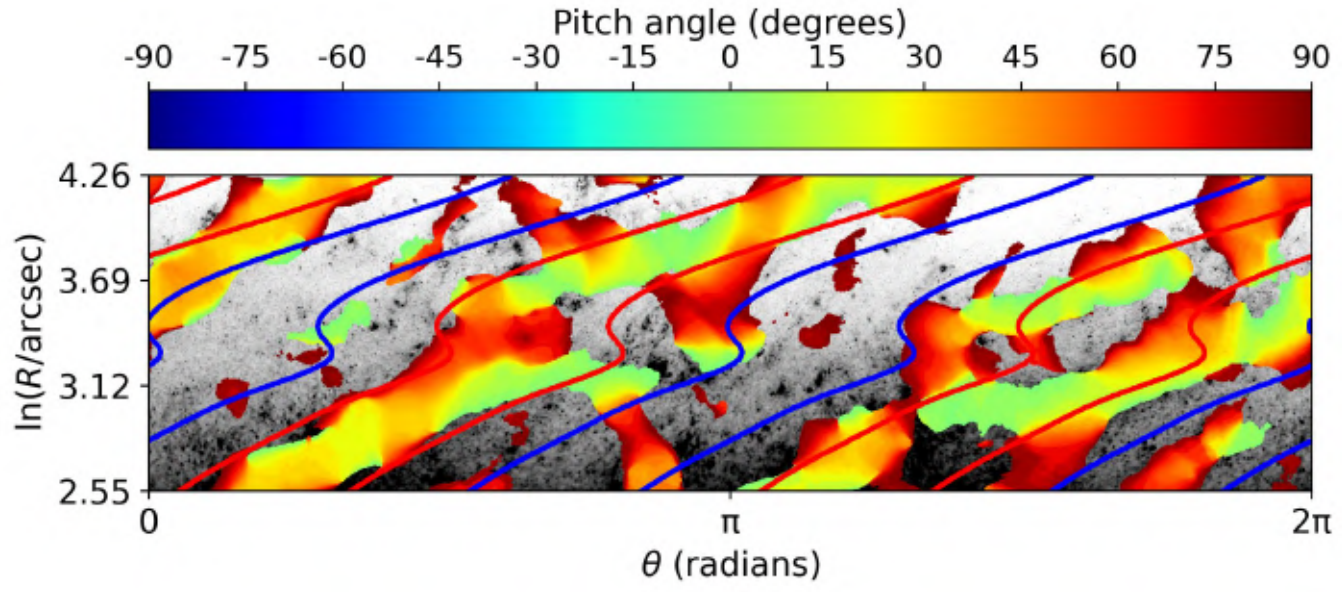

(d) Distribution of pitch angles obtained with the $D_w = 0.75$ filter.

**Figure A13.** Distribution of pitch angles obtained from the highest correlations at each position $(\ln(R/\mathrm{arcsec}), \theta)$ in the $F335M$ image of NGC 4254. The colorbar represents pitch angle values, ranging from $-90°$ (dark blue) to $90°$ (dark red). The colors of the points indicate the pitch angle associated with the maximum correlation, based on the idea that the strongest correlations occur when the structure has the same pitch angle as the filter at the corresponding position.

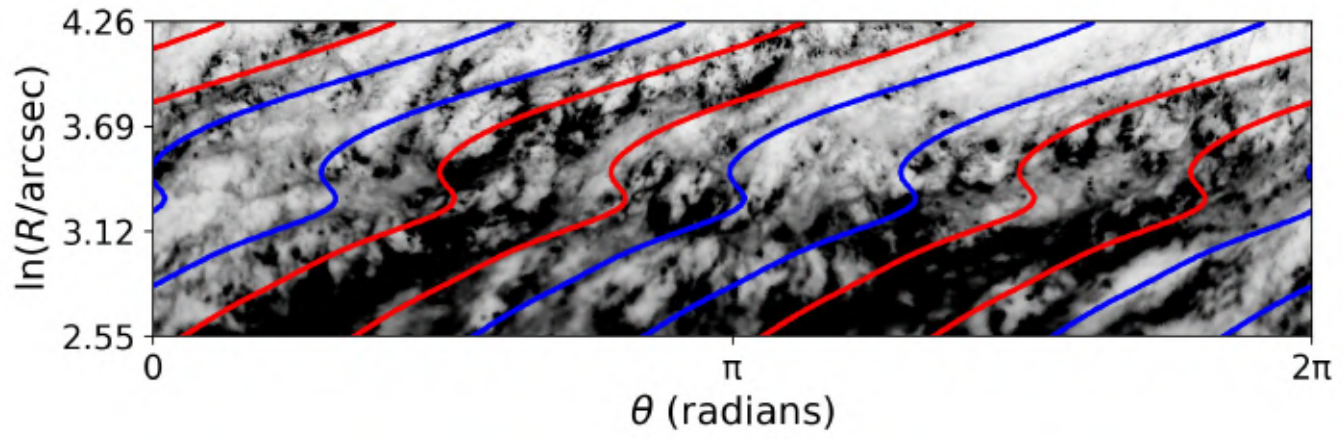

(a) Regions delimited in the $F1130W$ image of NGC 4254. The spiral arm regions are located between the red lines, while the interarm regions are between the blue lines.

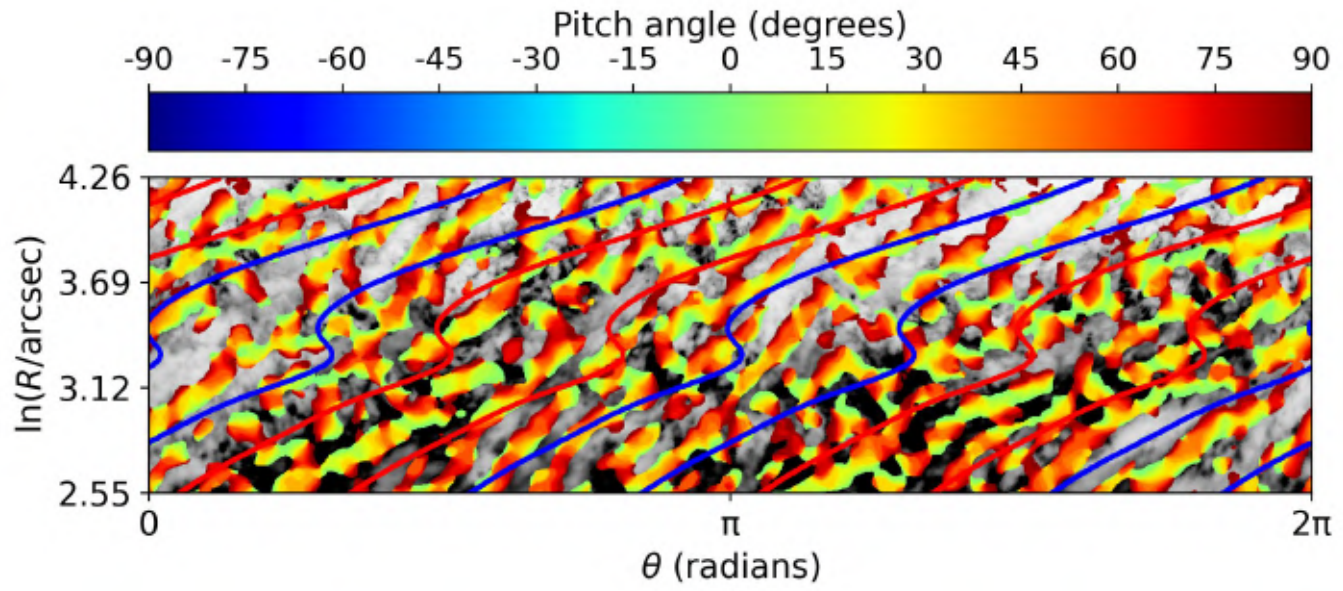

(b) Distribution of pitch angles obtained with the $D_w = 0.25$ filter.

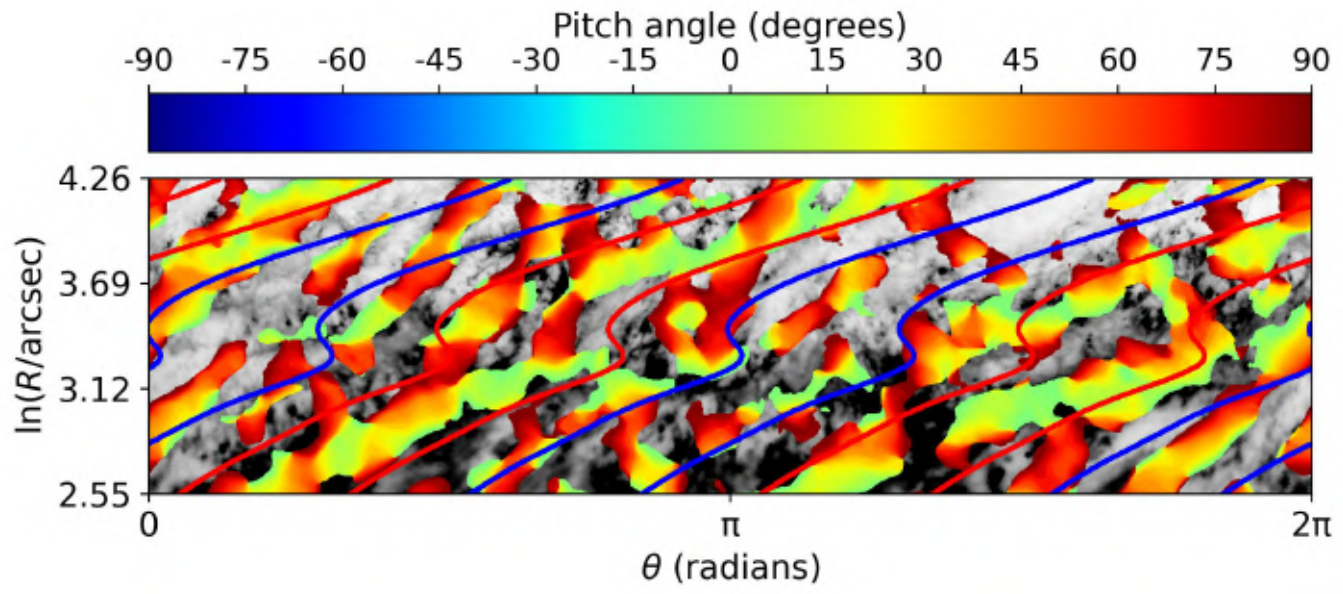

(c) Distribution of pitch angles obtained with the $D_w = 0.5$ filter.

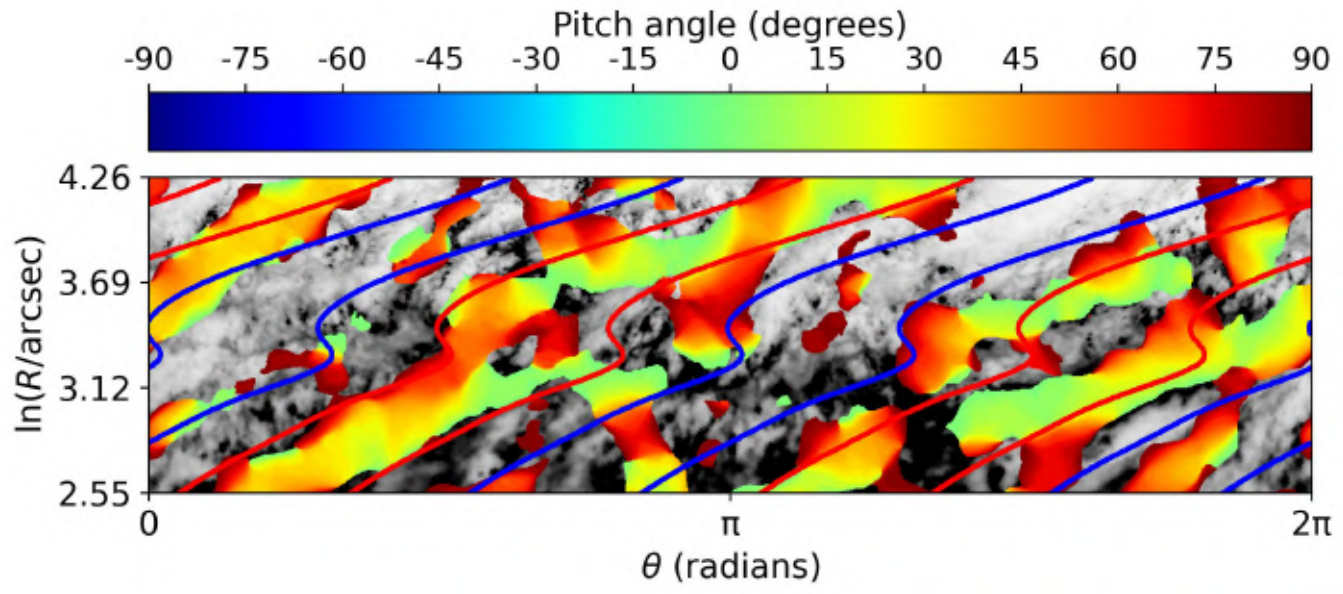

(d) Distribution of pitch angles obtained with the $D_w = 0.75$ filter.

**Figure A14.** Distribution of pitch angles obtained from the highest correlations at each position $(\ln(R/\mathrm{arcsec}), \theta)$ in the $F1130W$ image of NGC 4254. The colorbar represents pitch angle values, ranging from $-90°$ (dark blue) to $90°$ (dark red). The colors of the points indicate the pitch angle associated with the maximum correlation, based on the idea that the strongest correlations occur when the structure has the same pitch angle as the filter at the corresponding position.

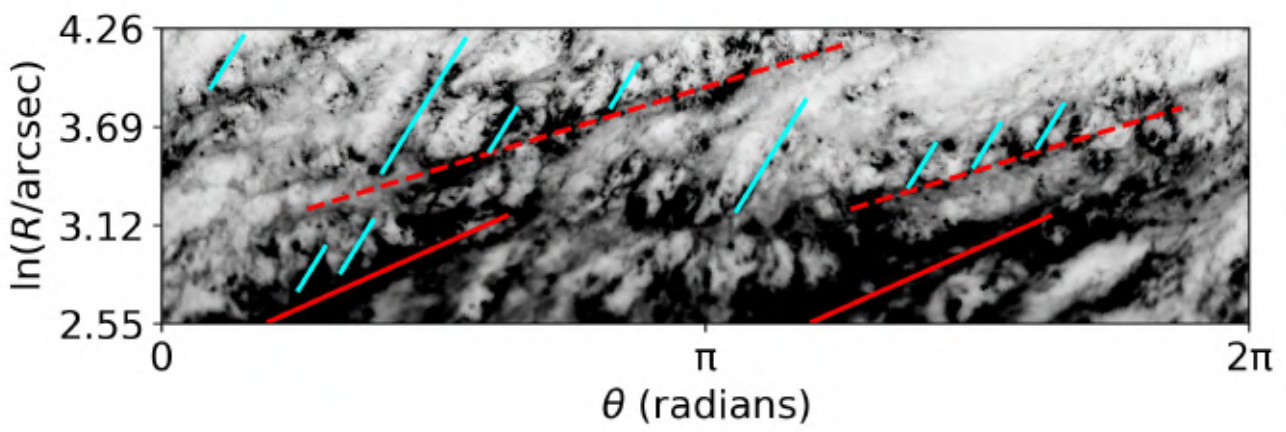

**Figure A15.** Lines overlaid on the $F1130W$ image of NGC 4254, indicating the pitch angles of the spiral structures. The solid red line corresponds to $24°$, the dashed red line to $17°$, and the solid blue line to $58°$.

### A4 NGC 4303

NGC 4303 is morphologically classified as SAB(rs)bc, with Z-type spiral arms, and according to Elmegreen & Elmegreen (1987), it presents "two symmetric inner arms; multiple long and continuous outer arms". The analysis was carried out between $\ln(R_{\rm min}/{\rm arcsec}) = 2.55$ and $\ln(R_{\rm max}/{\rm arcsec}) = 4.25$ in the $(\ln(R/{\rm arcsec}), \theta)$ space (Figure A16), with the arm and interarm regions for this radial range highlighted in Figures A18a and A19a, based on the mask obtained from the 2D-FT and 2D-IFT.

Figure A17 shows the sum of the correlations as a function of pitch angle for the arm and interarm regions of NGC 4303, for the three different window diameters $D_w = 0.25, 0.5$, and $0.75$ (Figures A17a, A17b, and A17c, respectively).

The inner parts of the main arms exhibit a smaller degree of winding, with relatively larger pitch angles compared to the outer radii of the galaxy. This feature is evident up to $\ln(R/{\rm arcsec}) \approx 3.4$, where the arms remain relatively open. Beyond this point, the arms become more tightly wound—especially the first arm, which appears almost horizontal in the diagram (Figure A16), indicating a small pitch angle. It is also worth noting that part of the first main arm falls into the region defined as interarm, a consequence of this abrupt change in winding.

Figures A18 and A19 illustrate this variation and the decrease in pitch angle values along the main arms. Since the beginnings of the arms are not clearly visible in the NIRCam images, the pitch angle estimates for the inner radii of the galaxy are better constrained by the MIRI results.

Considering the arm regions at smaller radii, the MIRI images suggest that $56° \lesssim \alpha \lesssim 61°$ provide a good description of the initial winding, particularly at larger window scales ($D_w = 0.5$, Figure A17b, and $D_w = 0.75$, Figure A17c). For the remaining part of the first arm, we find $5° \lesssim \alpha \lesssim 10°$, while for the second main arm $\alpha \approx 17°$.

In the interarm region, feathers are identified with $\alpha \approx 31°$ and $\alpha \approx 40°$, appearing as clear peaks in the MIRI images obtained with $D_w = 0.25$ (Figure A17a).

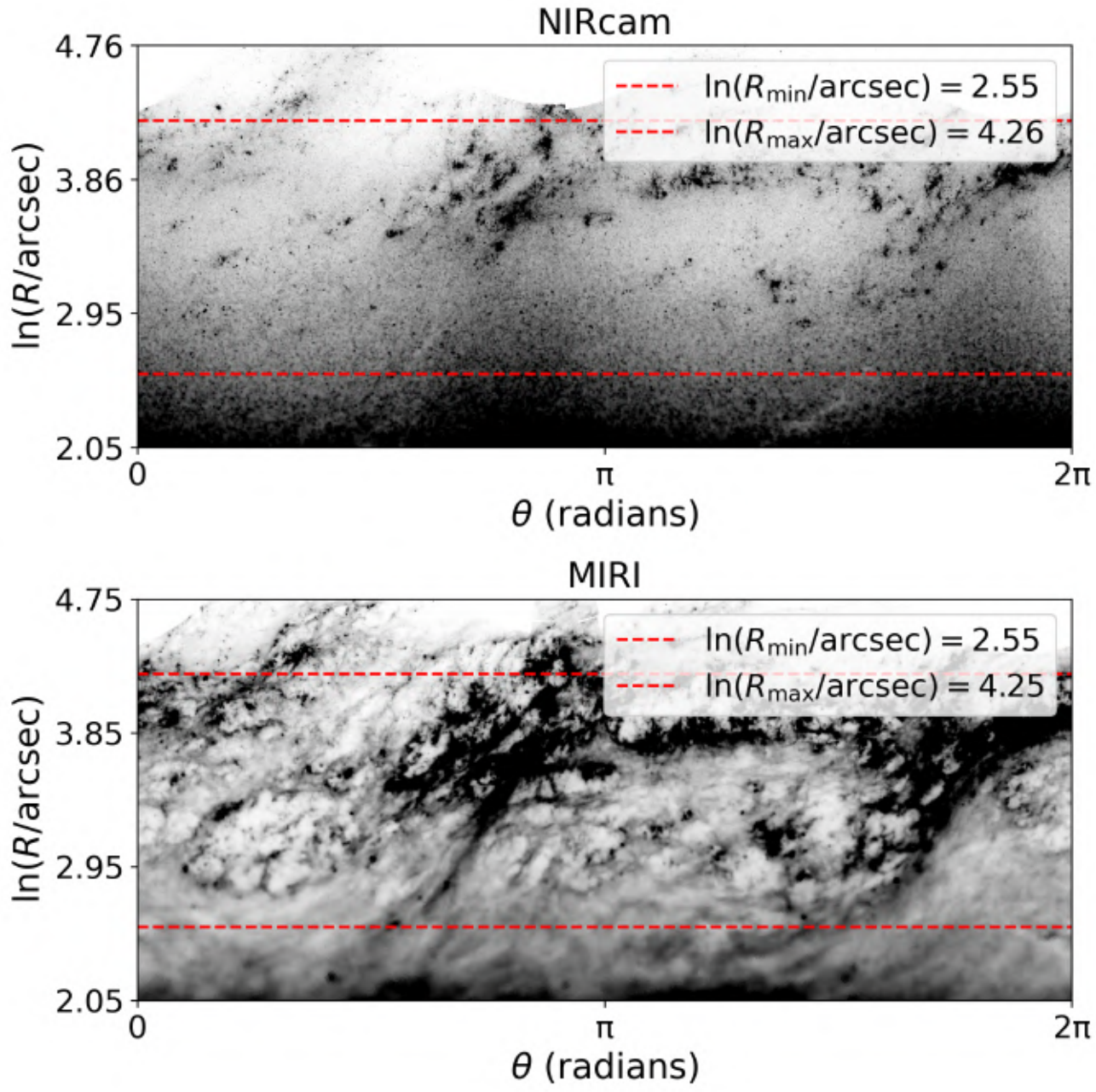

**Figure A16.** Diagrams $\ln(R/{\rm arcsec})$ vs. $\theta$ for the $F335M$ (NIRCam) and $F1130W$ (MIRI) images of NGC 4303. The dashed red lines indicate the values of $\ln(R_{min}/{\rm arcsec})$ and $\ln(R_{max}/{\rm arcsec})$ adopted in the analysis.

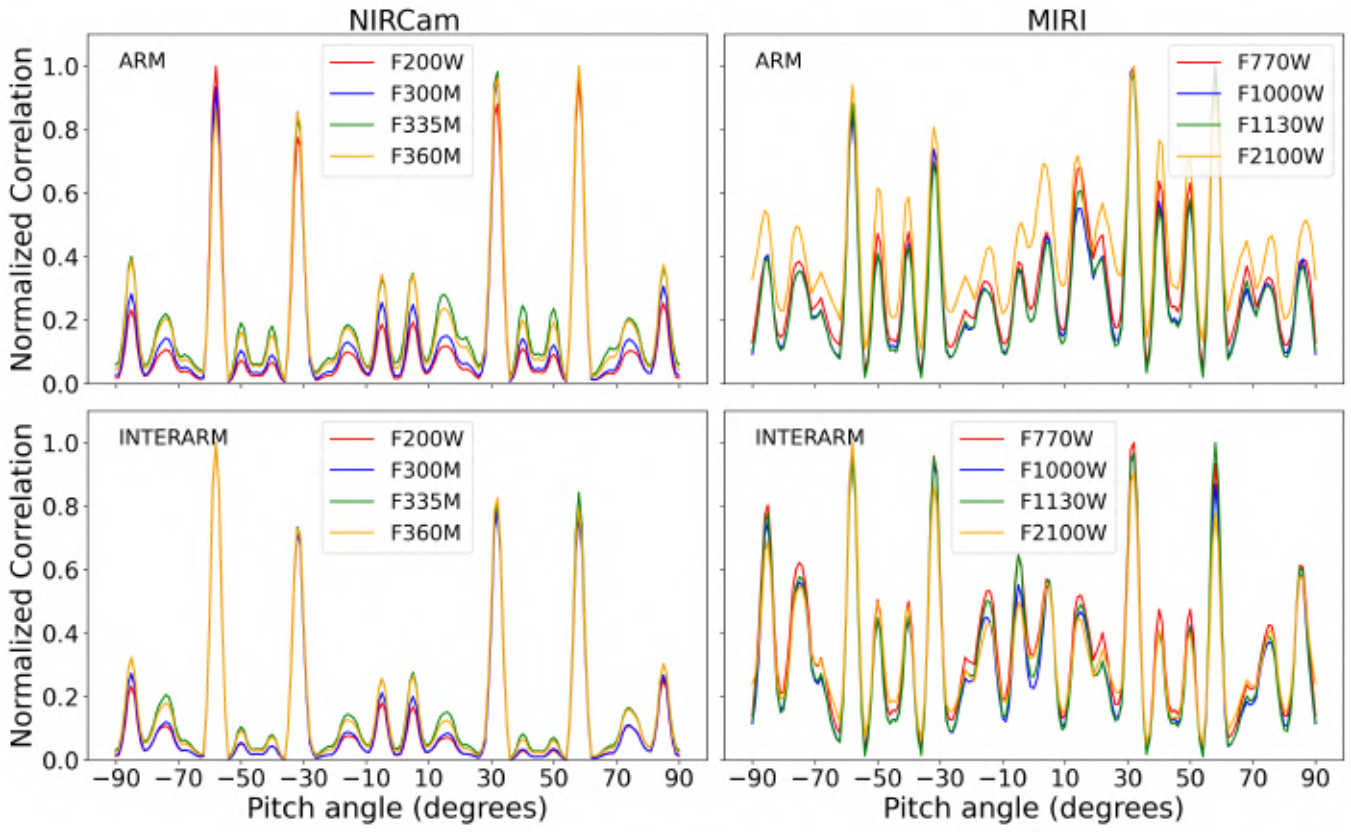

(a) Sum of the correlations calculated with $D_w = 0.25$ for the arm and interarm regions of NGC 4303.

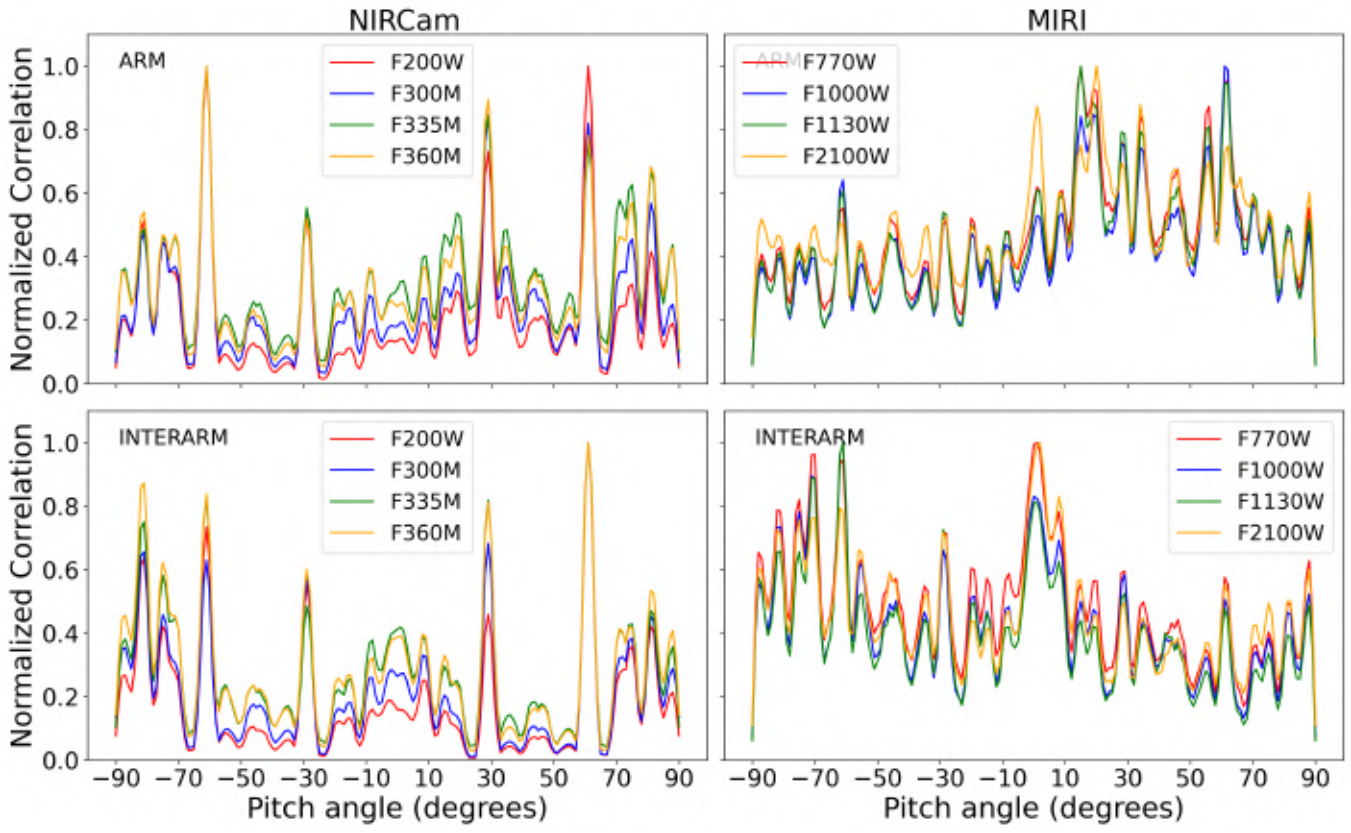

(b) Sum of the correlations calculated with $D_w = 0.5$ for the arm and interarm regions of NGC 4303.

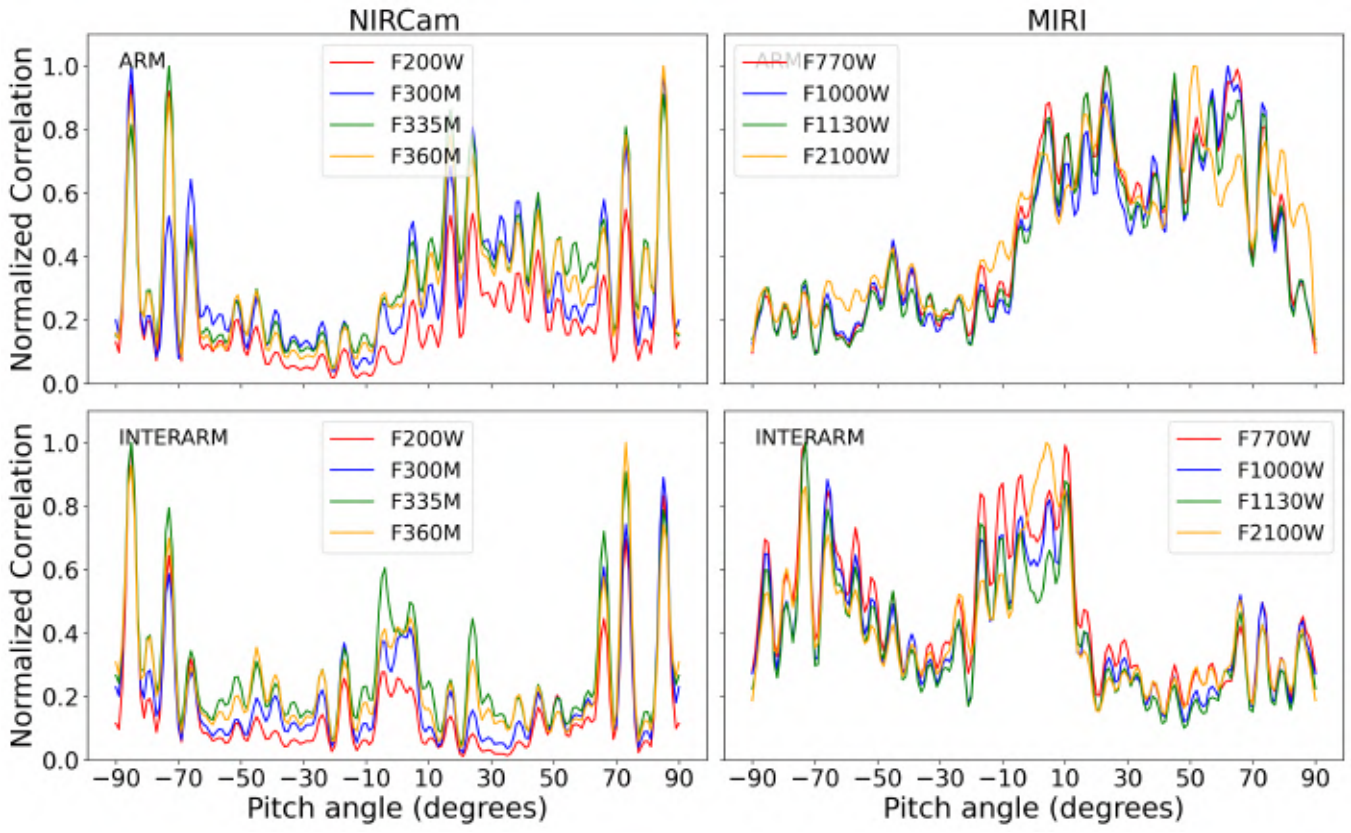

(c) Sum of the correlations calculated with $D_w = 0.75$ for the arm and interarm regions of NGC 4303.

**Figure A17.** Results for the sum of the correlations in the analyzed regions of NGC 4303. On the left, plots from NIRCam images with filters $F200W$ (red), $F300M$ (blue), $F335M$ (green), and $F360M$ (yellow), where the upper panel corresponds to the arm region and the lower panel to the interarm region. On the right, plots from MIRI images with filters $F770W$ (red), $F1000W$ (blue), $F1130W$ (green), and $F2100W$ (yellow), following the same division between arm (upper) and interarm (lower) regions.

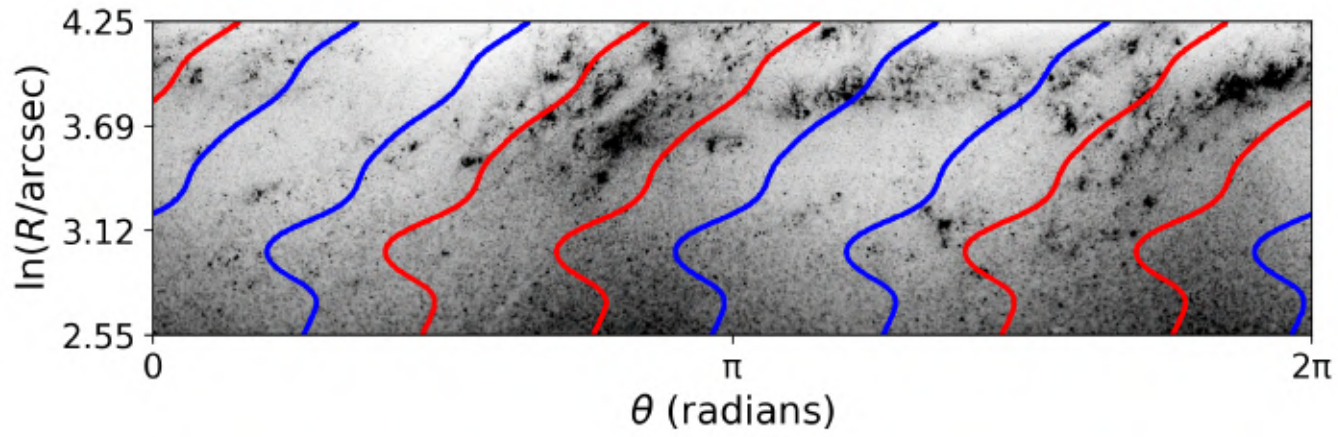

(a) Regions delimited in the $F335M$ image of NGC 4303. The spiral arm regions are located between the red lines, while the interarm regions are between the blue lines.

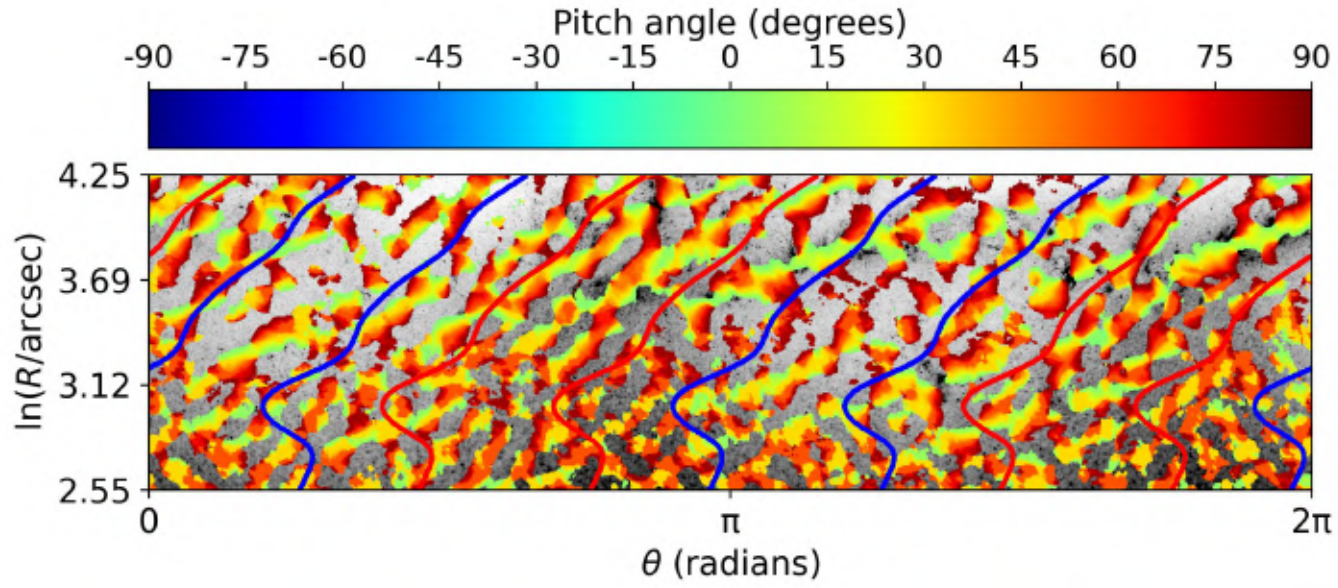

(b) Distribution of pitch angles obtained with the $D_w = 0.25$ filter.

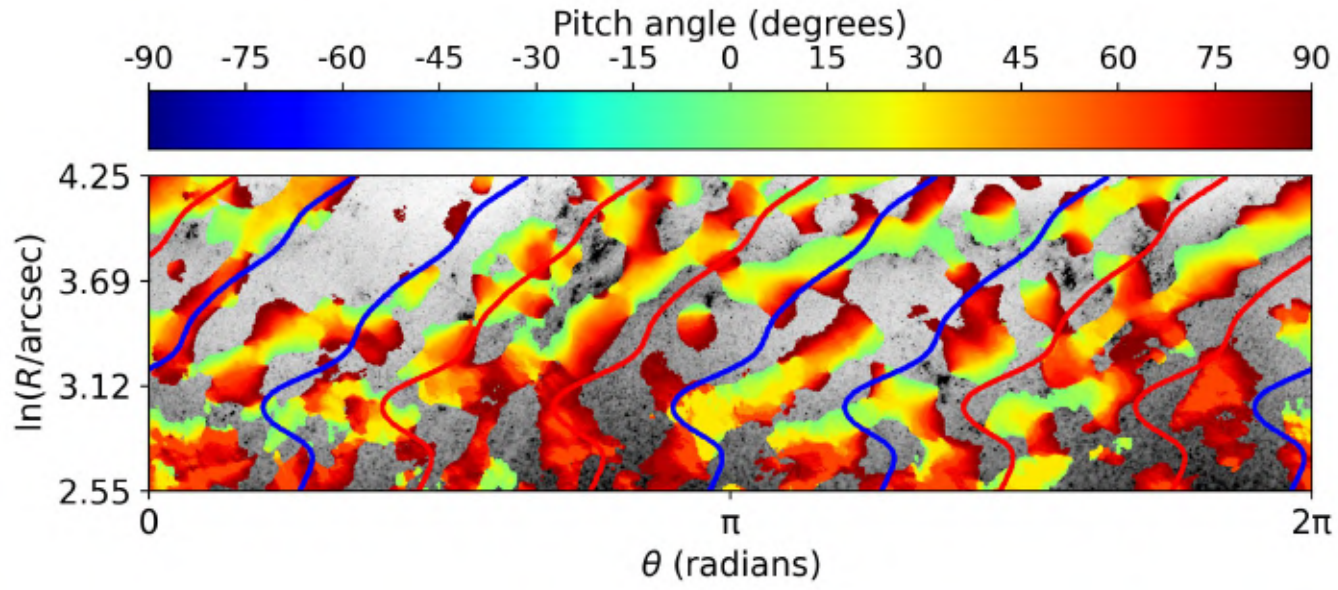

(c) Distribution of pitch angles obtained with the $D_w = 0.5$ filter.

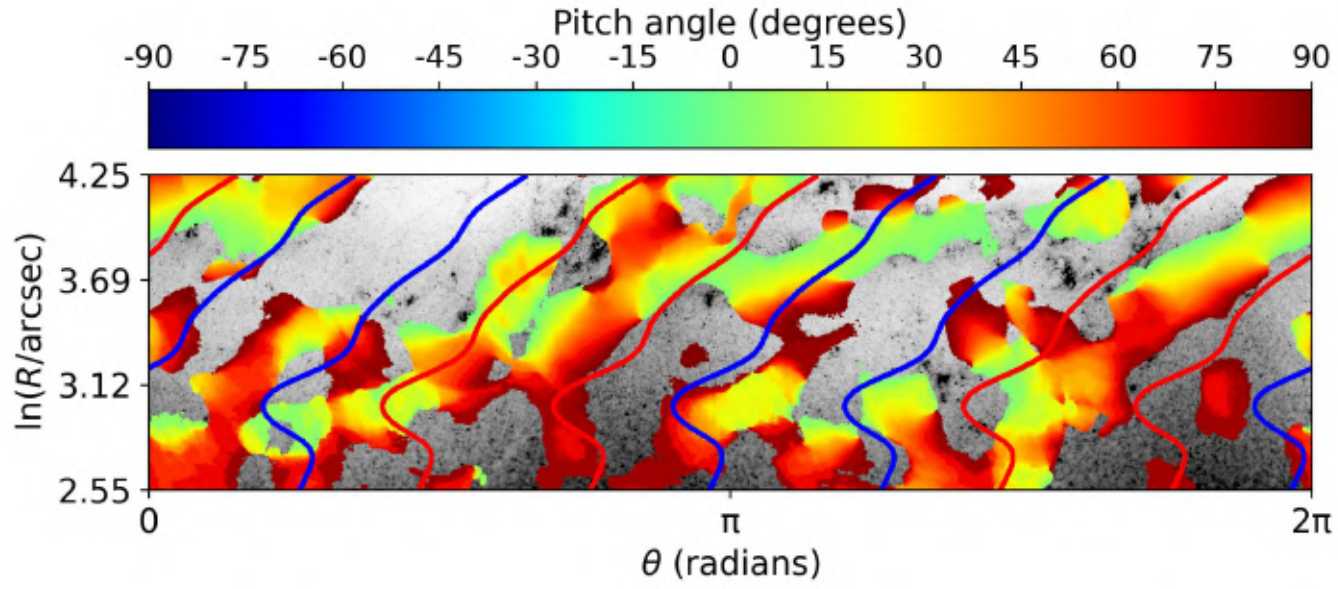

(d) Distribution of pitch angles obtained with the $D_w = 0.75$ filter.

**Figure A18.** Distribution of pitch angles obtained from the highest correlations at each position $(\ln(R/\mathrm{arcsec}), \theta)$ in the $F335M$ image of NGC 4303. The colorbar represents pitch angle values, ranging from $-90°$ (dark blue) to $90°$ (dark red). The colors of the points indicate the pitch angle associated with the maximum correlation, based on the idea that the strongest correlations occur when the structure has the same pitch angle as the filter at the corresponding position.

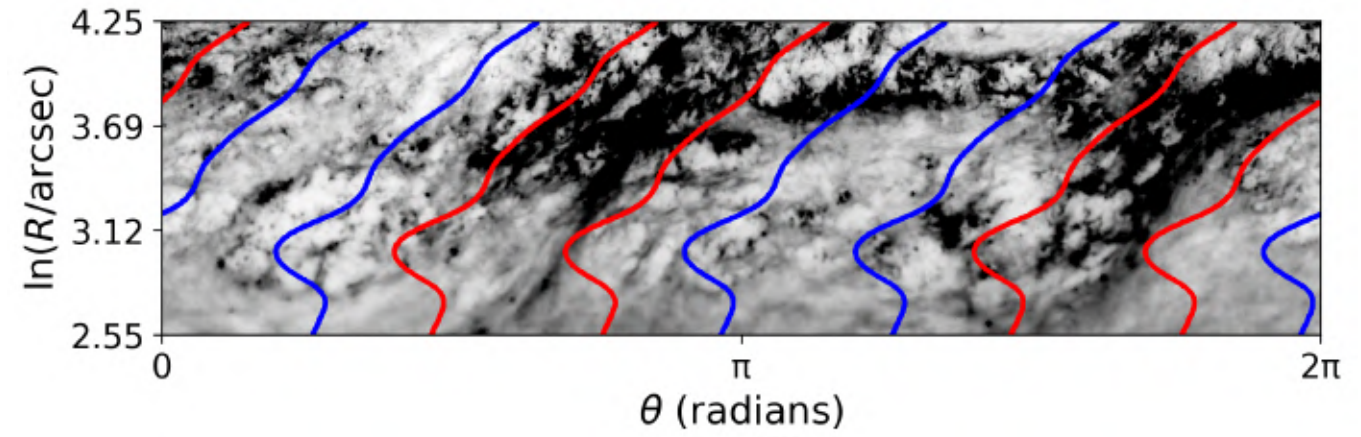

(a) Regions delimited in the $F1130W$ image of NGC 4303. The spiral arm regions are located between the red lines, while the interarm regions are between the blue lines.

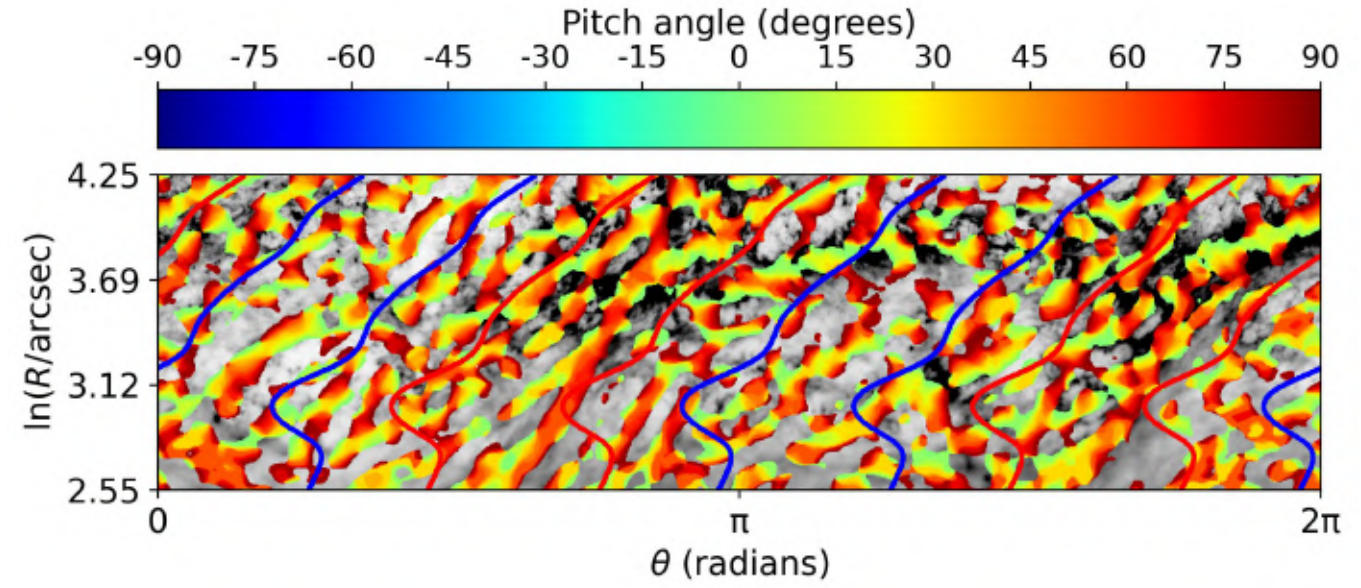

(b) Distribution of pitch angles obtained with the $D_w = 0.25$ filter.

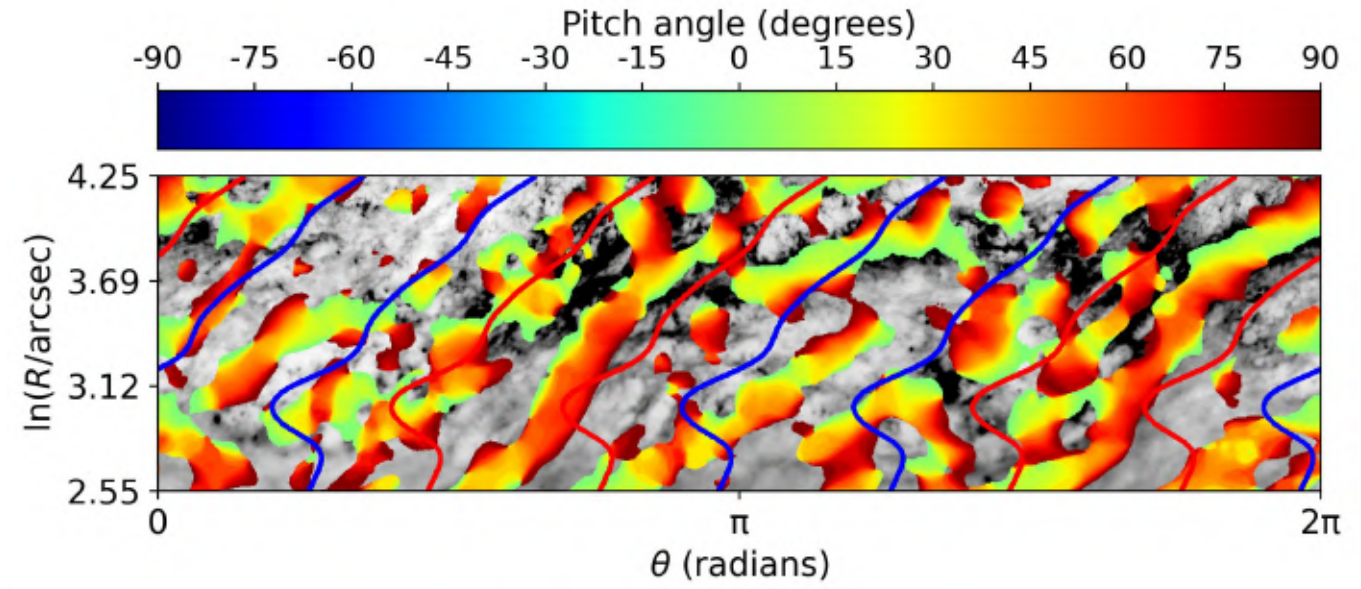

(c) Distribution of pitch angles obtained with the $D_w = 0.5$ filter.

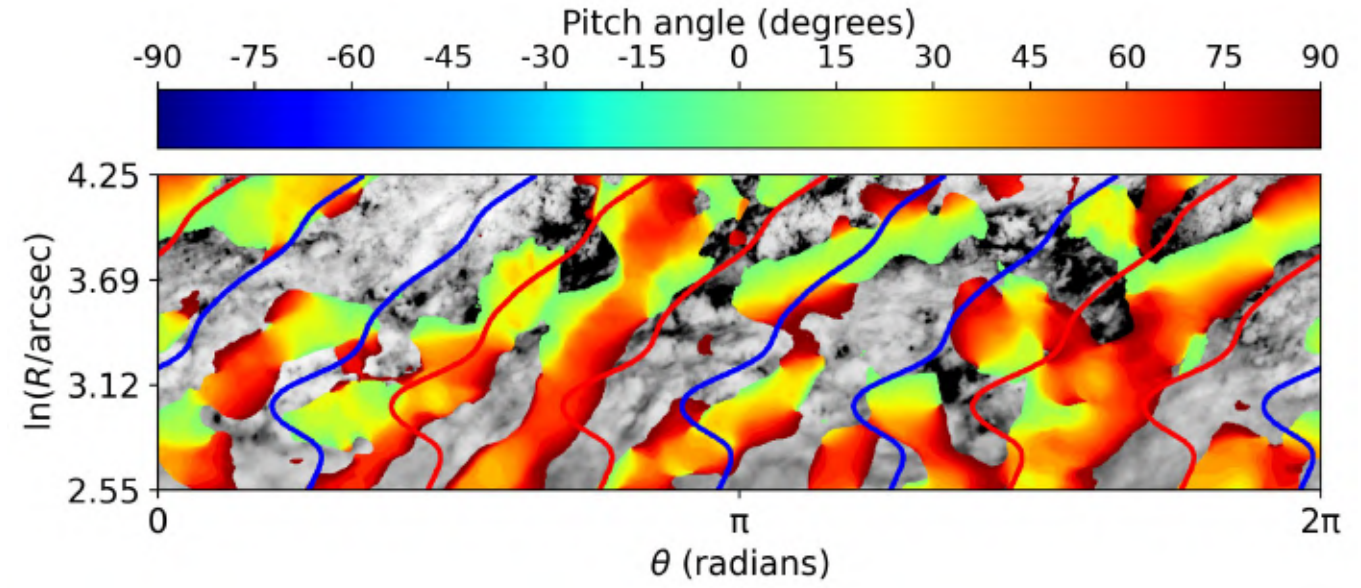

(d) Distribution of pitch angles obtained with the $D_w = 0.75$ filter.

**Figure A19.** Distribution of pitch angles obtained from the highest correlations at each position $(\ln(R/\mathrm{arcsec}), \theta)$ in the $F1130W$ image of NGC 4303. The colorbar represents pitch angle values, ranging from $-90°$ (dark blue) to $90°$ (dark red). The colors of the points indicate the pitch angle associated with the maximum correlation, based on the idea that the strongest correlations occur when the structure has the same pitch angle as the filter at the corresponding position.

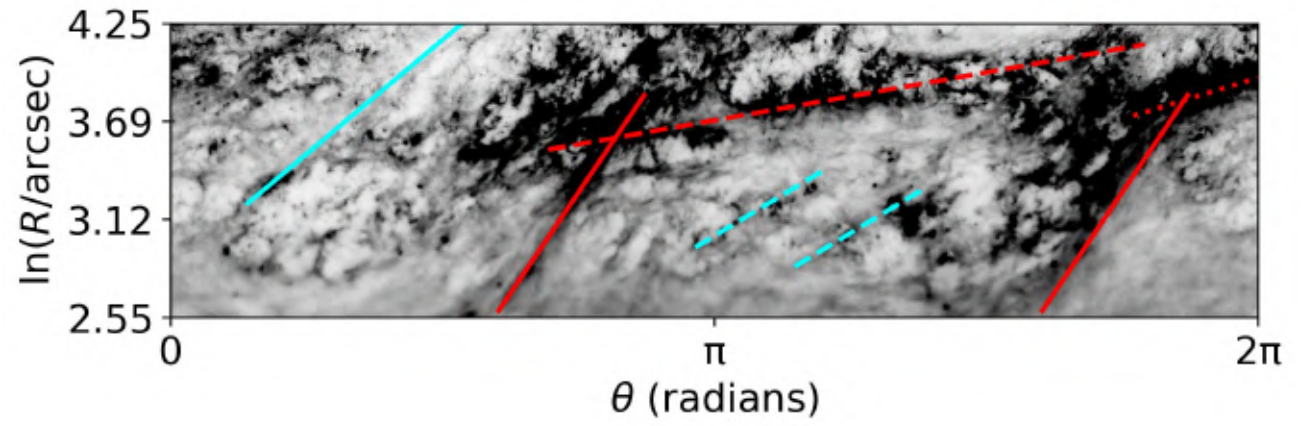

**Figure A20.** Lines overlaid on the $F1130W$ image of NGC 4303, showing inclinations that represent the pitch angles of the spiral structures. The solid red line corresponds to 56°, the dashed red line to 10°, the dotted red line to 17°, the solid blue line to 40°, and the dashed blue line to 31°.

### A5 NGC 4321

The galaxy NGC 4321, classified as SAB(s)bc, has S-type spiral arms and is described as having "two long symmetric arms dominating the optical disk" (Elmegreen & Elmegreen 1987). The analysis was carried out in the radial range $\ln(R_{min}/\mathrm{arcsec}) = 2.8$ to $\ln(R_{max}/\mathrm{arcsec}) = 4.3$ in the $(\ln(R/\mathrm{arcsec}), \theta)$ space (Figure A21), with the arm and interarm regions highlighted in Figures A23a and A24a.

Figure A22 shows the sum of the correlations as a function of pitch angle for the arm and interarm regions of NGC 4321, for the three different window diameters $D_w = 0.25, 0.5$, and $0.75$ (Figures A22a, A22b, and A22c, respectively).

This galaxy exhibits an intermediate bar, from which the spiral arms originate with a relatively high pitch angle compared to the outer regions, where the arms become progressively more tightly wound. The analyzed radial range mainly covers this transition between the bar and the arms, reflecting the more open geometry that characterizes the vicinity of the bar.

Accordingly, the intermediate window $D_w = 0.5$ (Figure A22b) indicates a pitch angle of $\alpha \approx -55°$ at the beginning of the main arm. Within the arm region, it was also possible to identify substructures such as branches, with pitch angles of $\alpha \approx -46°$ and $\alpha \approx -29°$. In the interarm region, we found extended substructures with $\alpha \approx -8°$. Regarding the feathers, a visual inspection suggests that most of them lie outside the analyzed radial range. Nevertheless, we were able to estimate the pitch angle of one feather emerging from the second main arm, with $\alpha \approx -76°$.

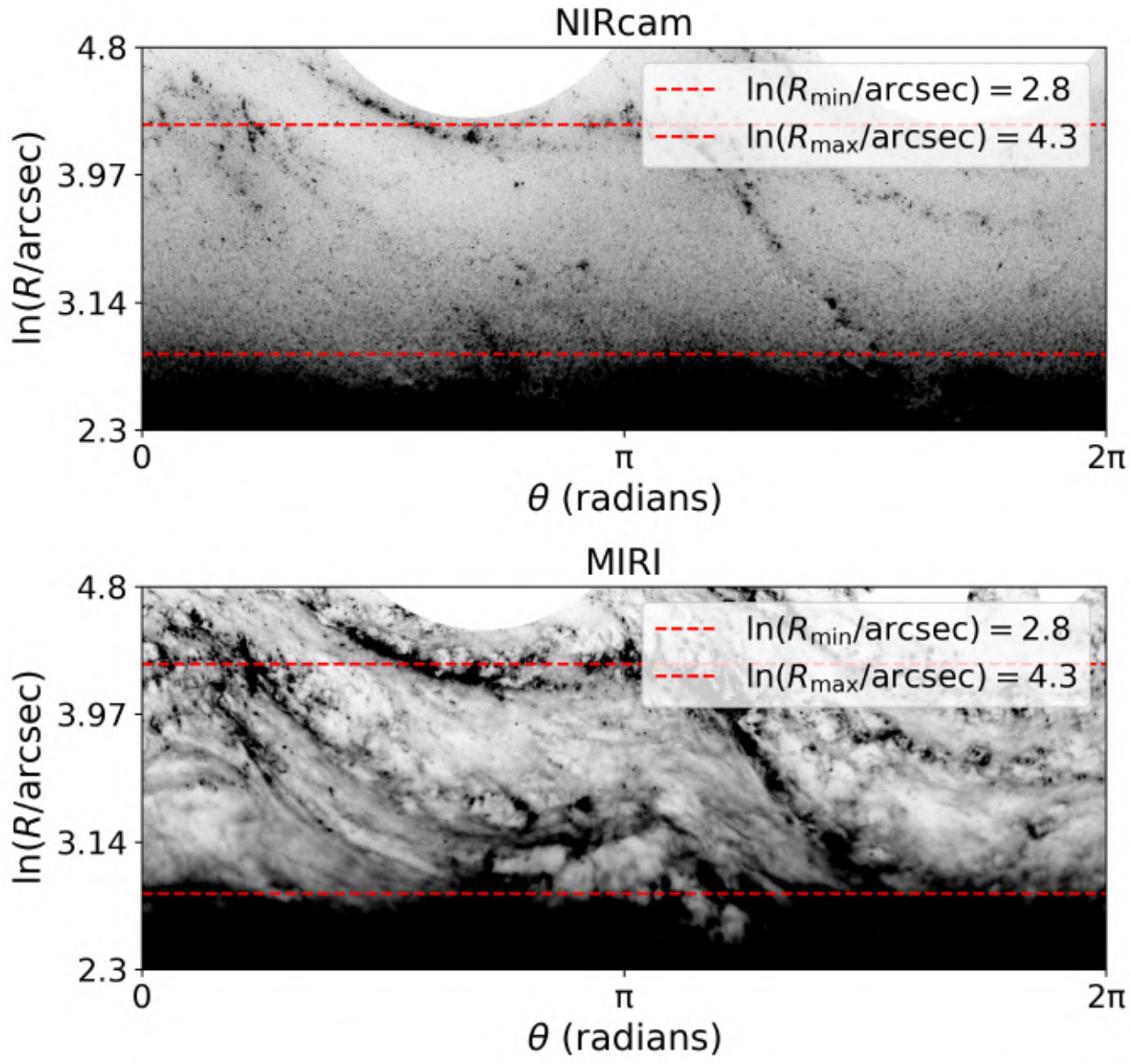

**Figure A21.** Diagrams $\ln(R/\mathrm{arcsec})$ vs. $\theta$ for the $F335M$ (NIRCam) and $F1130W$ (MIRI) images of NGC 4321. The dashed red lines indicate the values of $\ln(R_{min}/\mathrm{arcsec})$ and $\ln(R_{max}/\mathrm{arcsec})$ adopted in the analysis.

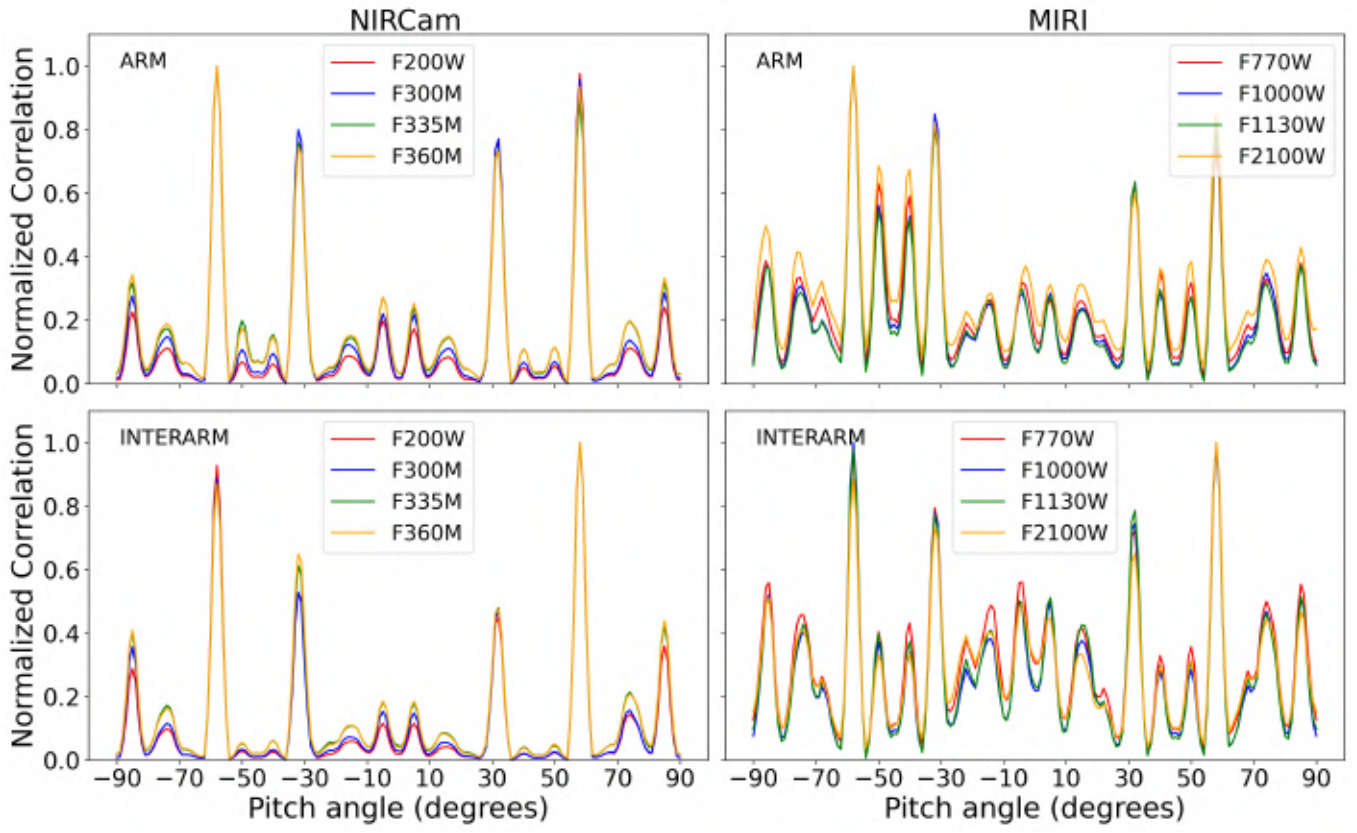

(a) Sum of the correlations calculated with $D_w = 0.25$ for the arm and interarm regions.

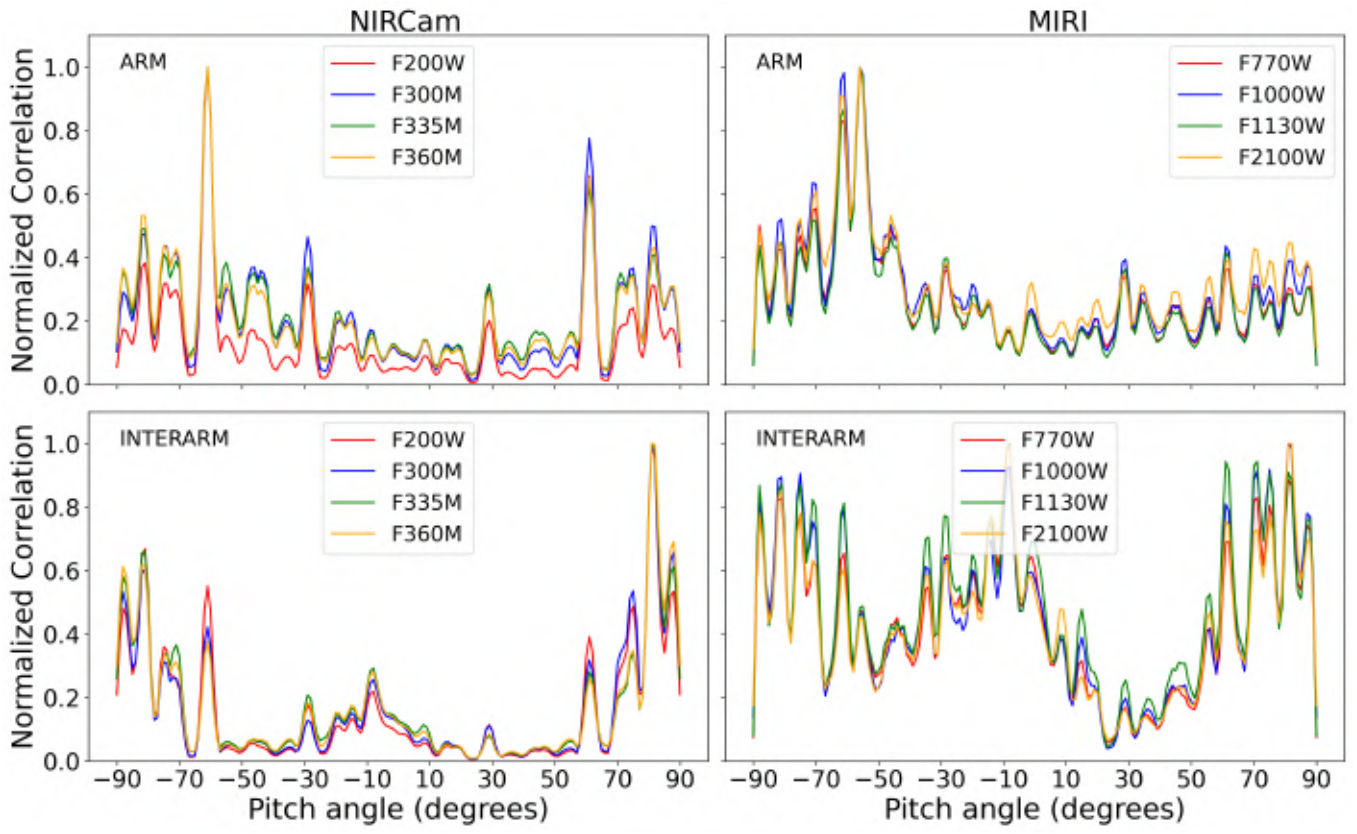

(b) Sum of the correlations calculated with $D_w = 0.5$ for the arm and interarm regions.

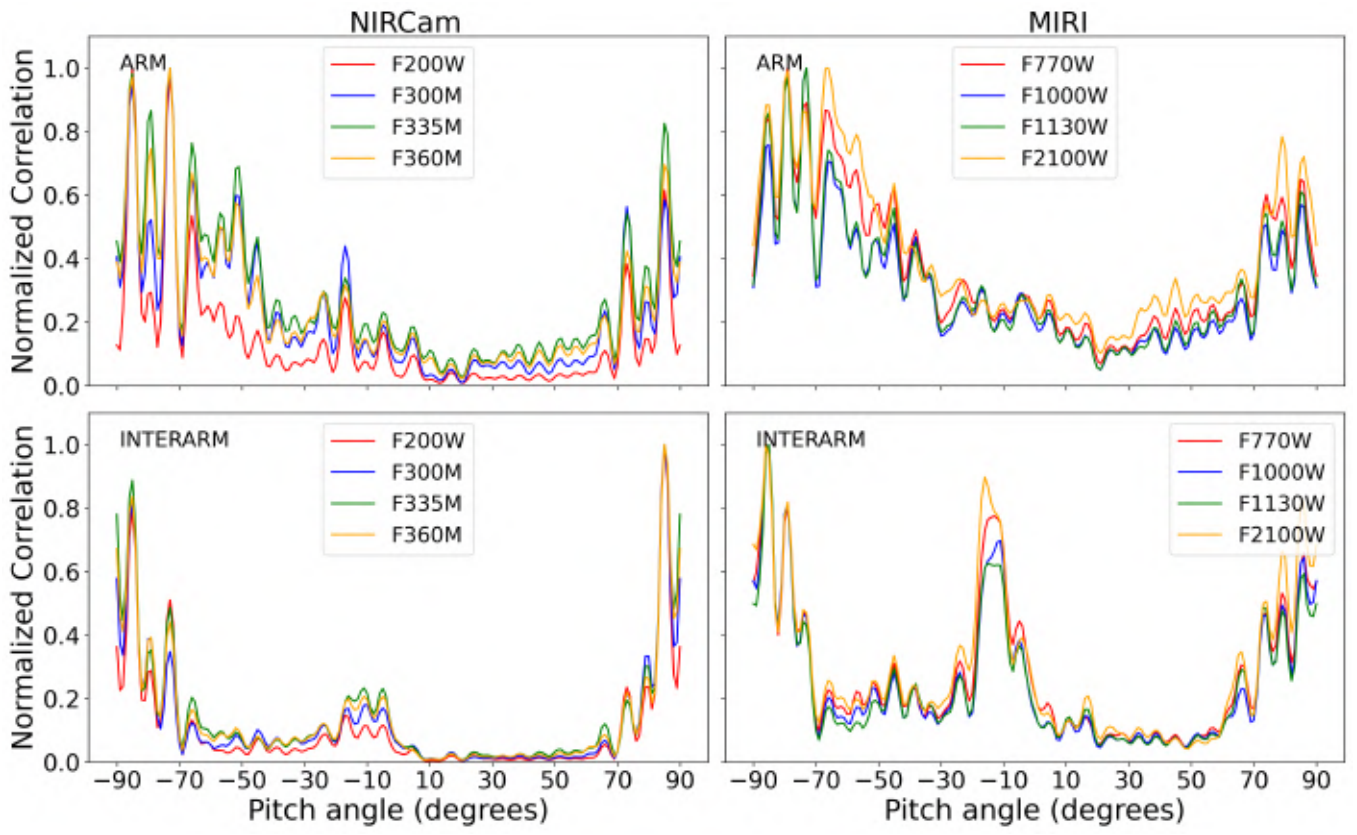

(c) Soma das correlações calculada com $D_w = 0.75$ para as regiões braço e interbraço.

**Figure A22.** Results for the sum of the correlations in the analyzed regions of NGC 4321. On the left, plots from the NIRCam images with filters $F200W$ (red), $F300M$ (blue), $F335M$ (green), and $F360M$ (yellow), where the upper panel corresponds to the arm region and the lower panel to the interarm region. On the right, plots from the MIRI images with filters $F770W$ (red), $F1000W$ (blue), $F1130W$ (green), and $F2100W$ (yellow), following the same division between arm (upper) and interarm (lower) regions.

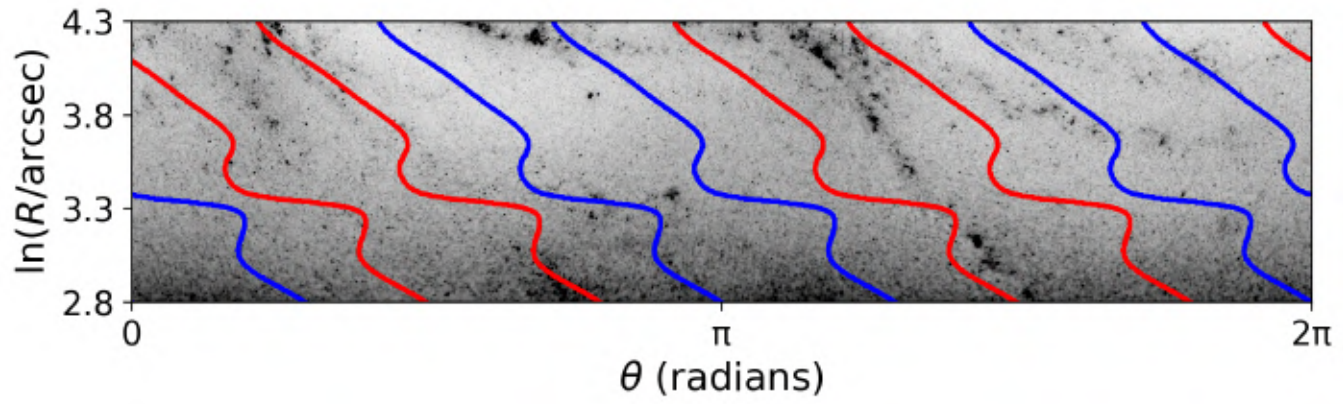

(a) Regions delimited in the $F335M$ image of the galaxy NGC 4321. The spiral arm regions are located between the red lines, while the interarm regions are between the blue lines.

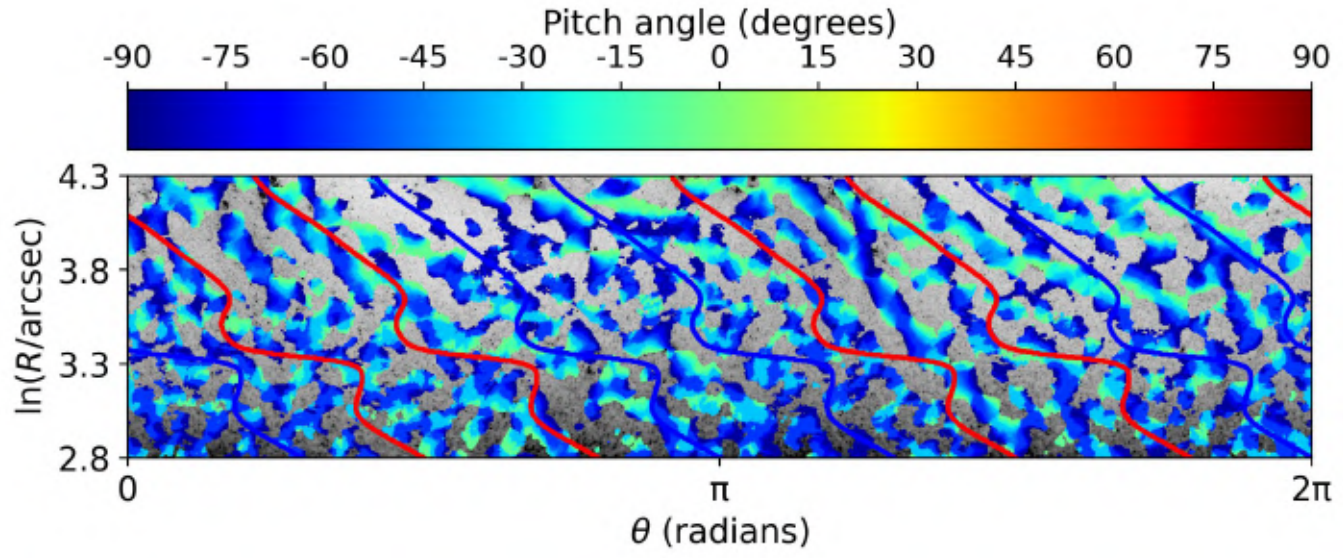

(b) Distribution of pitch angles obtained with the $D_w = 0.25$ filter.

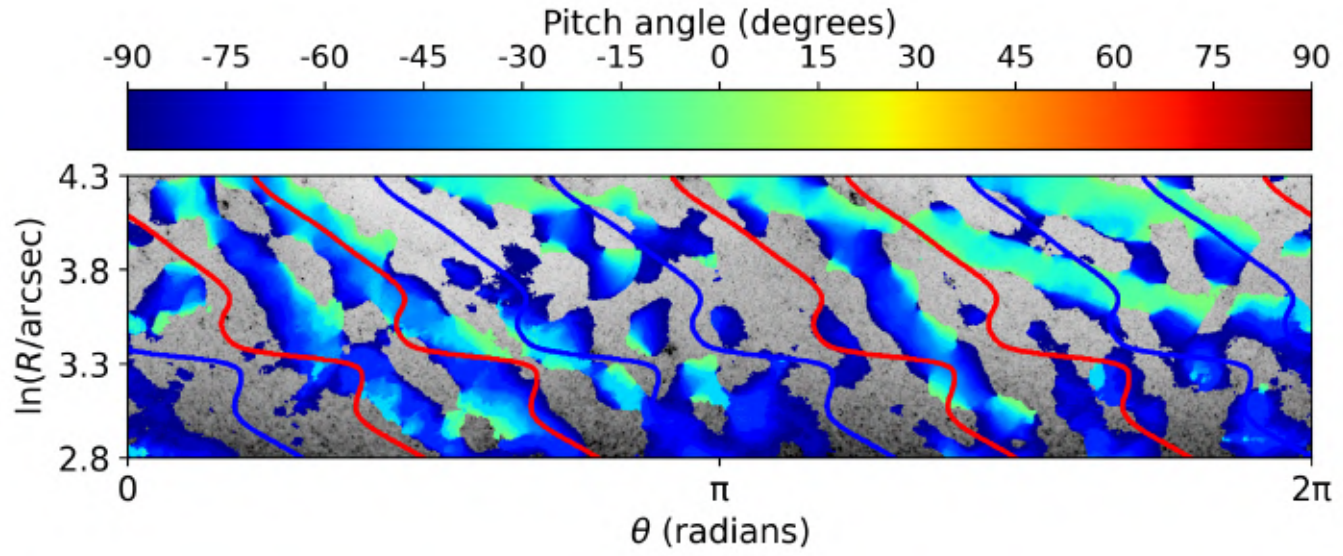

(c) Distribution of pitch angles obtained with the $D_w = 0.5$ filter.

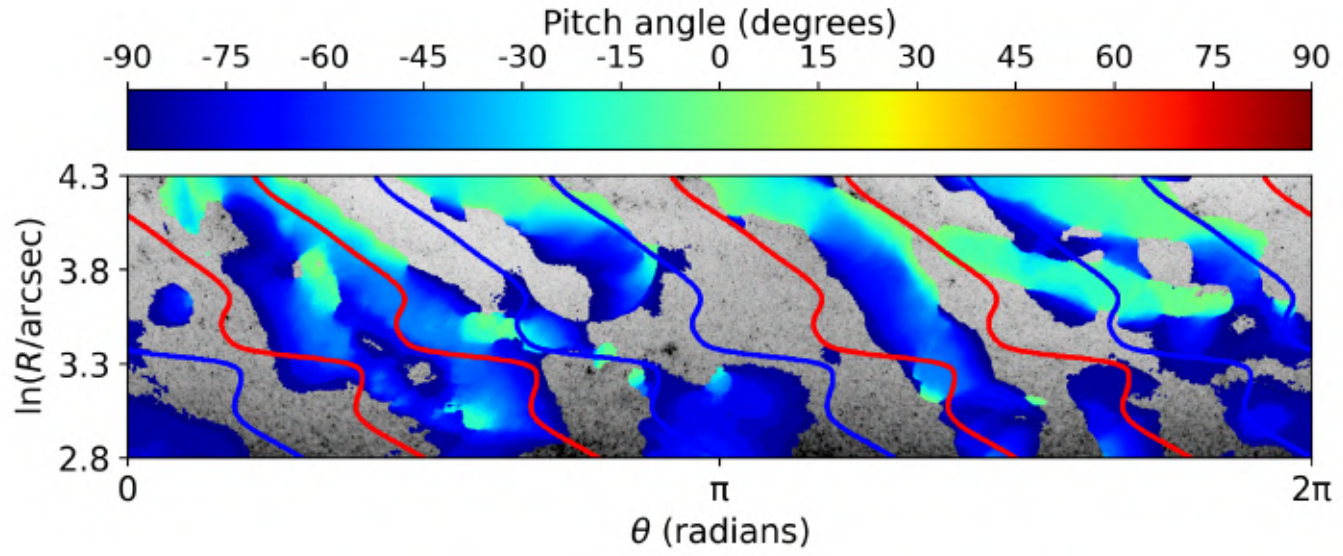

(d) Distribution of pitch angles obtained with the $D_w = 0.75$ filter.

**Figure A23.** Distribution of pitch angles obtained from the highest correlations at each position $(\ln(R/\text{arcsec}), \theta)$ in the $F335M$ image of NGC 4321. The colorbar represents pitch angle values, ranging from $-90°$ (dark blue) to $90°$ (dark red). The colors of the points indicate the pitch angle associated with the maximum correlation, based on the idea that the strongest correlations occur when the structure has the same pitch angle as the filter at the corresponding position.

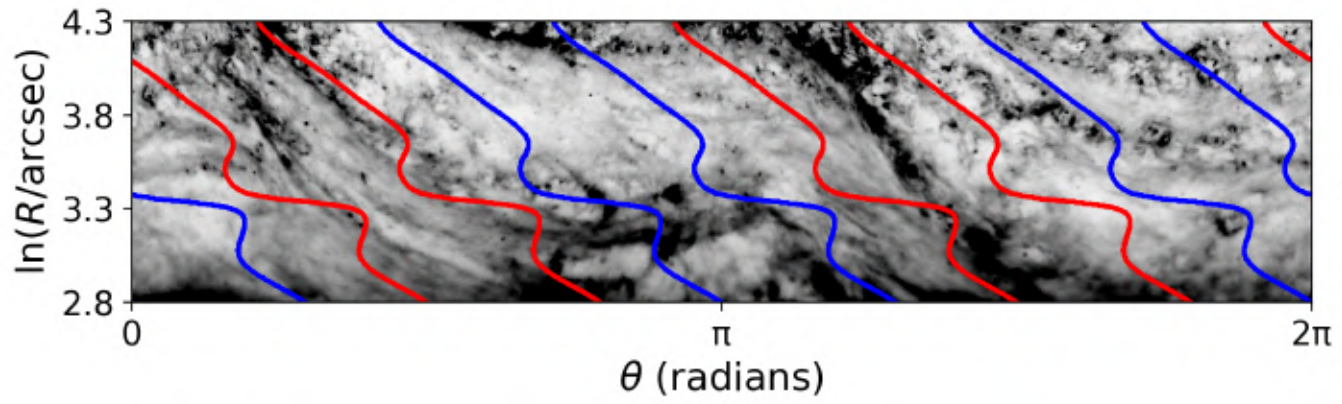

(a) Regions delimited in the $F1130W$ image of the galaxy NGC 4321. The spiral arm regions are located between the red lines, while the interarm regions are between the blue lines.

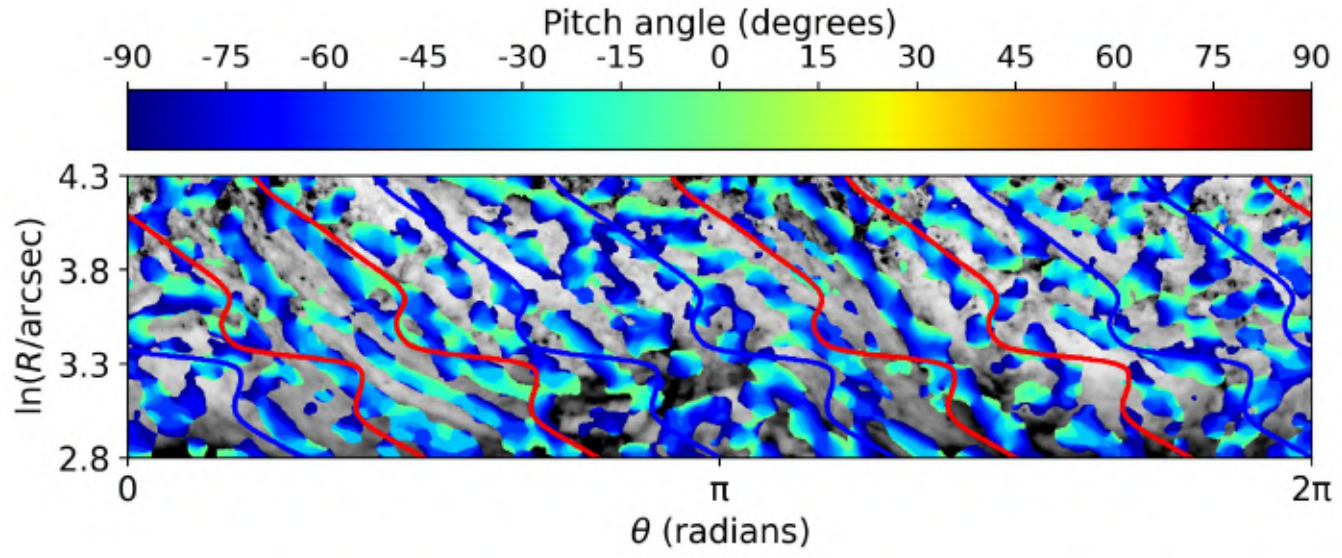

(b) Distribution of pitch angles obtained with the $D_w = 0.25$ filter.

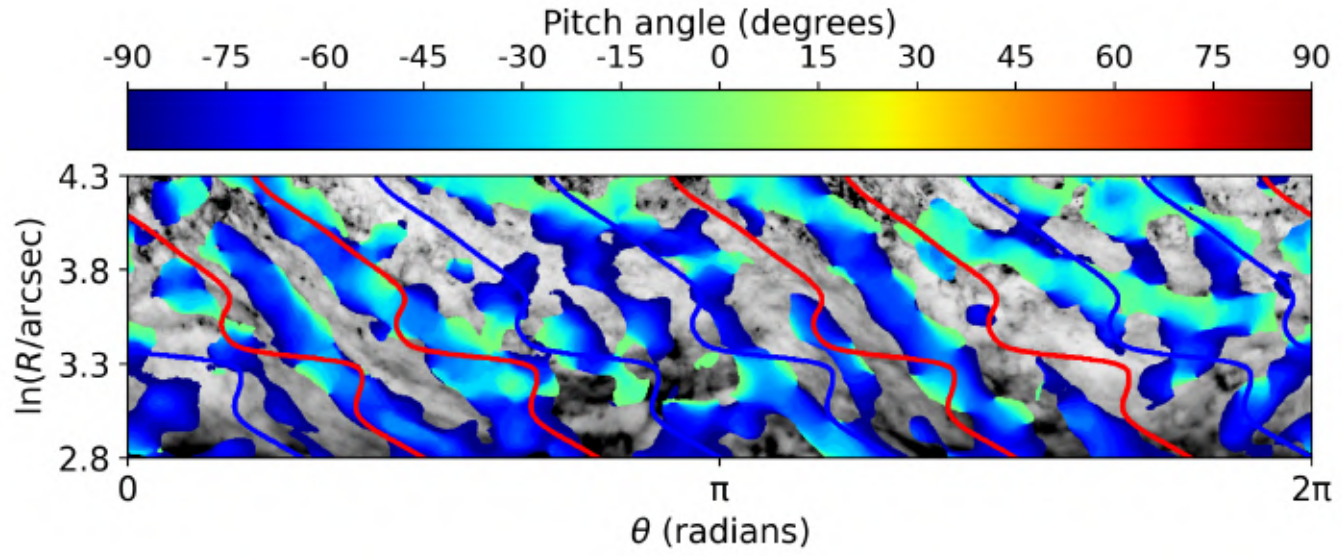

(c) Distribution of pitch angles obtained with the $D_w = 0.5$ filter.

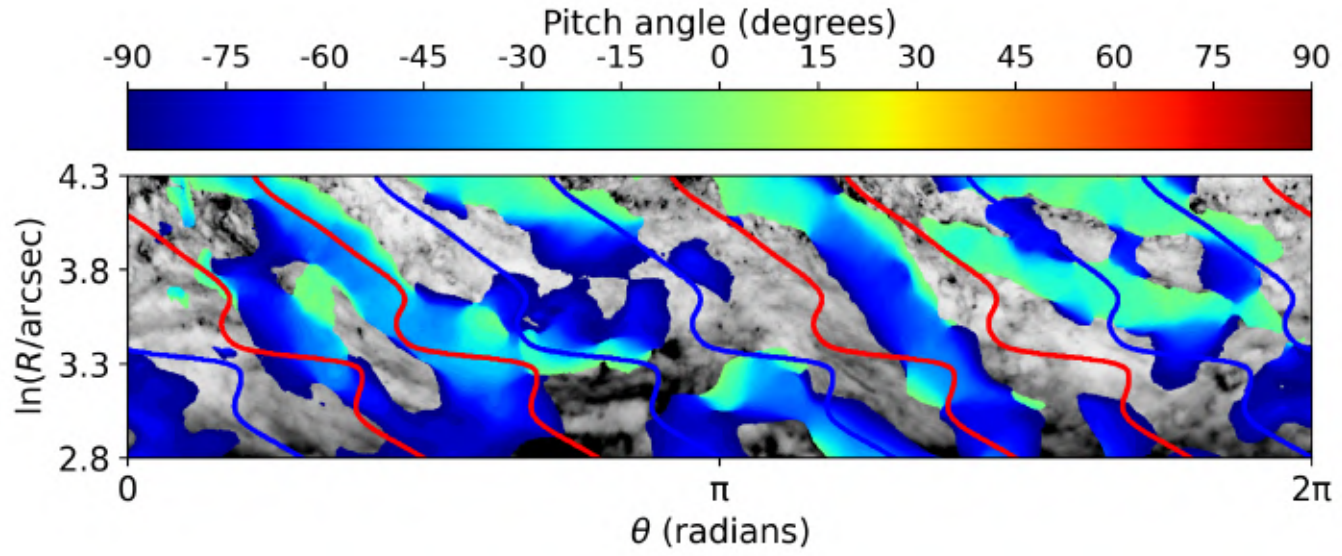

(d) Distribution of pitch angles obtained with the $D_w = 0.75$ filter.

**Figure A24.** Distribution of pitch angles obtained from the highest correlations at each position $(\ln(R/\text{arcsec}), \theta)$ in the $F1130W$ image of NGC 4321. The colorbar represents pitch angle values, ranging from $-90°$ (dark blue) to $90°$ (dark red). The colors of the points indicate the pitch angle associated with the maximum correlation, based on the idea that the strongest correlations occur when the structure has the same pitch angle as the filter at the corresponding position.

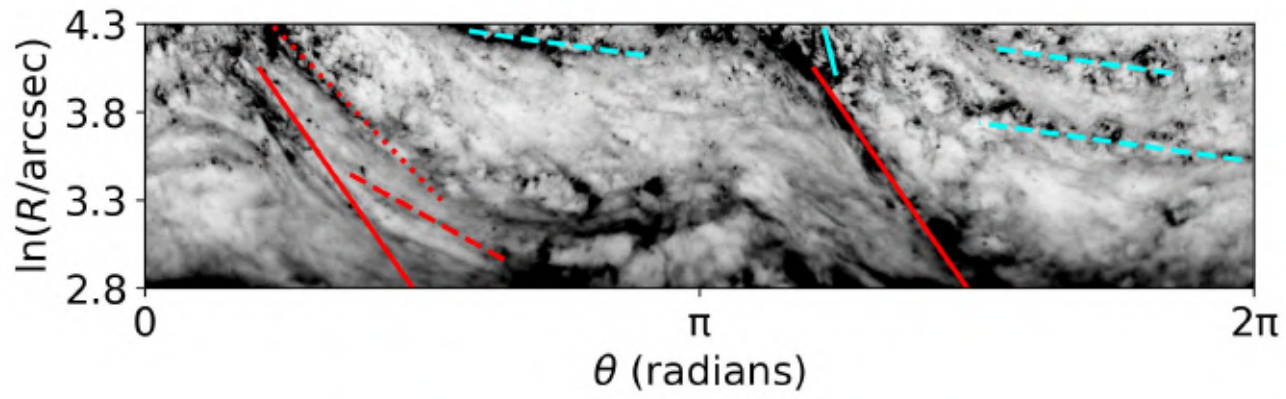

**Figure A25.** Lines overlaid on the $F1130W$ image of the galaxy NGC 4321, showing inclinations that represent the pitch angles of the spiral structures. The solid red line corresponds to $-55°$, the dashed red line to $-29°$, the dotted red line to $-46°$, the solid blue line to $-76°$, and the dashed blue line to $-8°$.

### A6 NGC 4535

The galaxy NGC 4535, classified as SAB(s)c, exhibits S-type spiral arms and, according to Elmegreen & Elmegreen (1987), shows "two symmetric inner arms; multiple long and continuous outer arms". The analysis was carried out in the interval $\ln(R_{min}/\mathrm{arcsec}) = 3.7$ to $\ln(R_{max}/\mathrm{arcsec}) = 4.05$ in the $(\ln(R/\mathrm{arcsec}), \theta)$ space (Figure A26), with the arm and interarm regions identified in Figures A28a and A29a.

Figure A27 shows the sum of the correlations as a function of pitch angle for the arm and interarm regions of NGC 4535 for the three different window diameters $D_w = 0.25, 0.5$, and $0.75$ (Figures A27a, A27b, and A27c, respectively).

Because the analysed radial range is relatively narrow, we emphasize the intermediate window, $D_w = 0.50$, for the arm geometry. At this scale, both the NIRCam and MIRI spectra (Figure A27b) show recurrent components near $\alpha \approx -14°$ and $\alpha \approx -2°$. Their spatial distributions associate these values with the first and second spiral arms, respectively, with the second arm being more tightly wound than the first.

In the interarm region, substructures with $\alpha \approx -59°$ and $\alpha \approx -76°$ are detected, which are more clearly visible with the $D_w = 0.25$ filter (Figures A28b and A29b).

These results highlight a clear separation between the pitch angles associated with large-scale spiral arms and those associated with interarm substructures. This distinction suggests that the pitch angle, when measured locally, encodes information about different physical regimes within the galactic disk, depending on the spatial scale and the nature of the structures analyzed.

This paper has been typeset from a TEX/LATEX file prepared by the author.

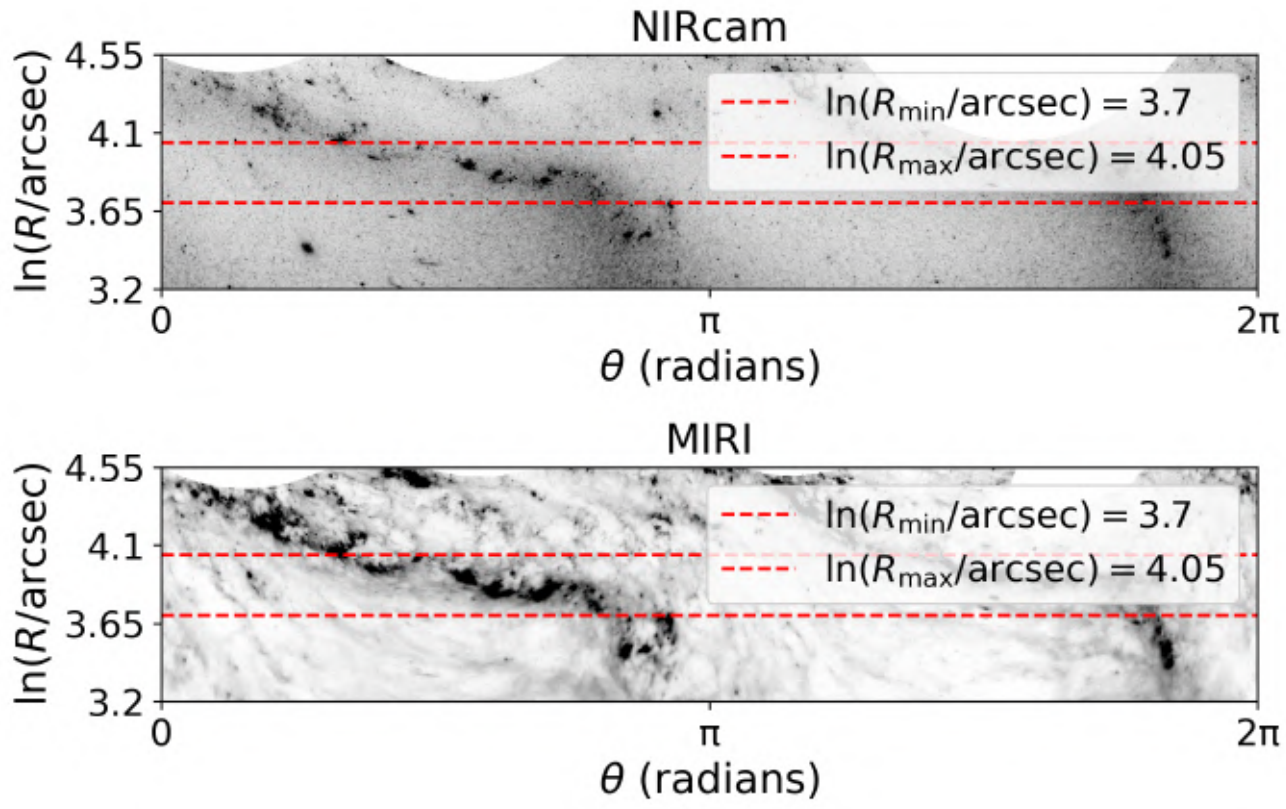

**Figure A26.** Diagrams $\ln(R/\mathrm{arcsec})$ vs. $\theta$ for the $F335M$ (NIRCam) and $F1130W$ (MIRI) images of NGC 4535. The dashed red lines indicate the values of $\ln(R_{min}/\mathrm{arcsec})$ and $\ln(R_{max}/\mathrm{arcsec})$ adopted in the analysis.

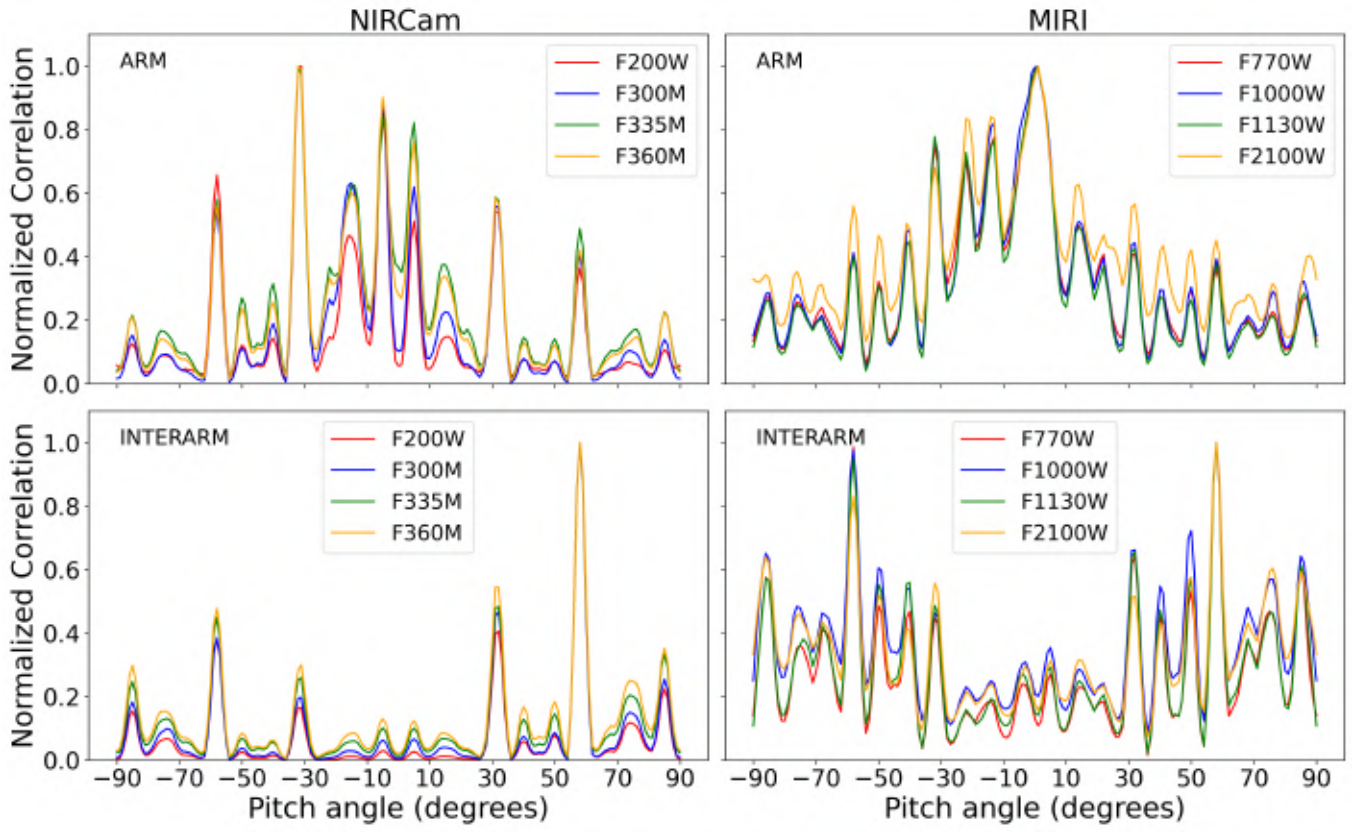

(a) Sum of the correlations calculated with $D_w = 0.25$ for the arm and interarm regions.

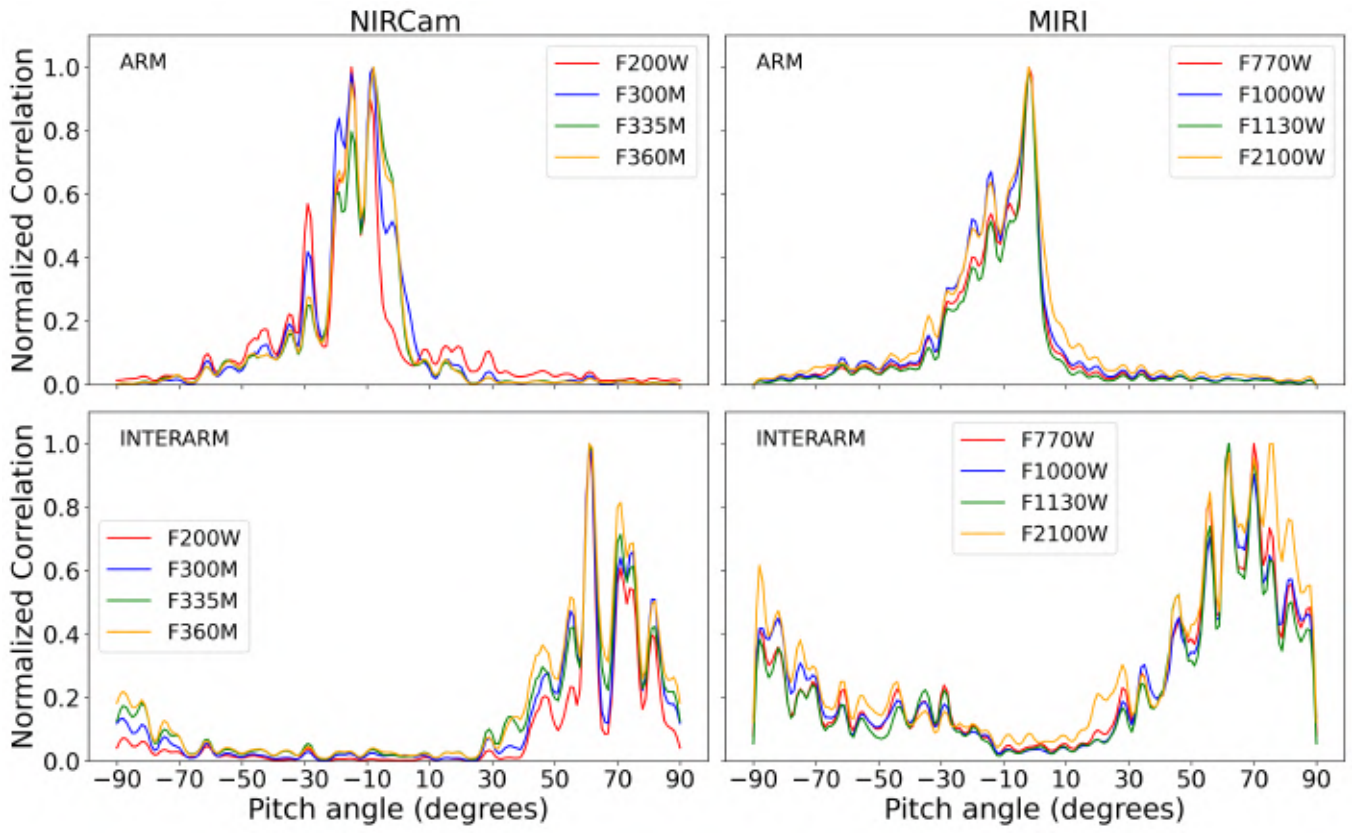

(b) Sum of the correlations calculated with $D_w = 0.5$ for the arm and interarm regions.

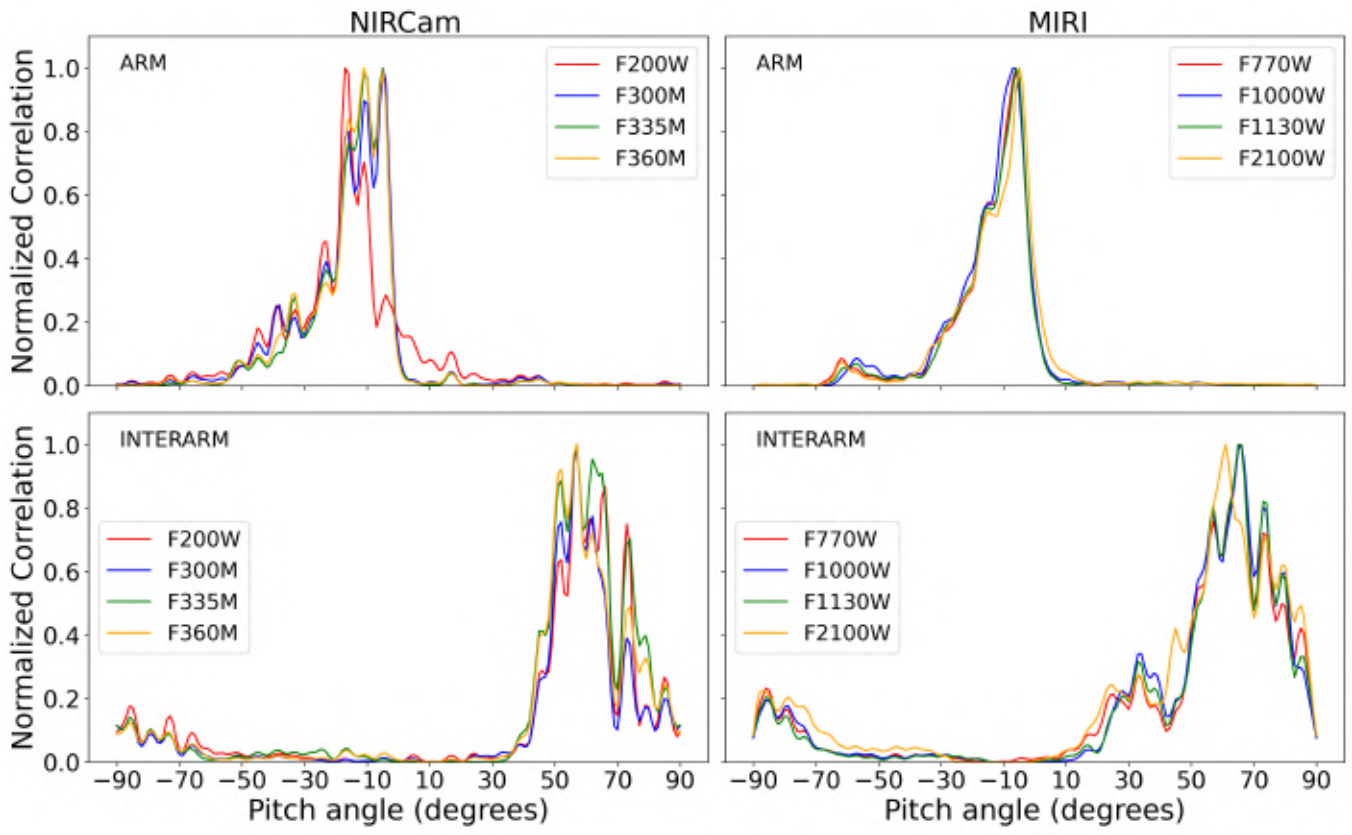

(c) Sum of the correlations calculated with $D_w = 0.75$ for the arm and interarm regions.

**Figure A27.** Results of the correlation sums in the analyzed areas of NGC 4535. On the left, plots from the NIRCam images with filters $F200W$ (red), $F300M$ (blue), $F335M$ (green), and $F360M$ (yellow), with the upper panel referring to the arm region and the lower panel to the interarm region. On the right, plots from the MIRI images with filters $F770W$ (red), $F1000W$ (blue), $F1130W$ (green), and $F2100W$ (yellow), following the same division between arm (upper) and interarm (lower) regions.

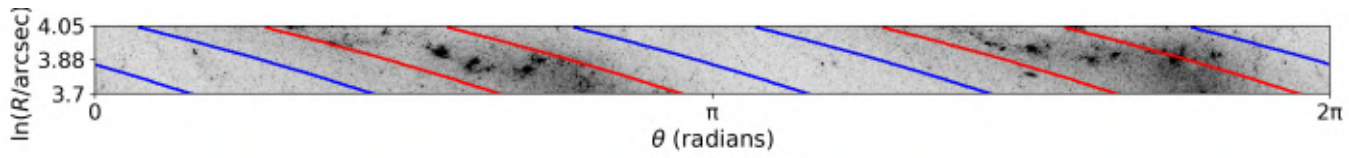

(a) Regions delimited in the $F335M$ image of the galaxy NGC 4535. The regions defined as the arm are located between the red lines, and the interarm regions are between the blue lines.

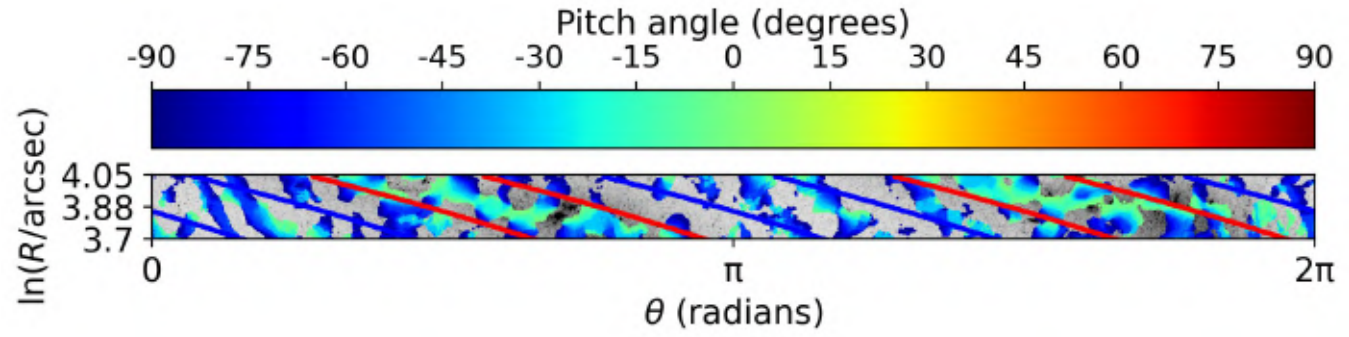

(b) Distribution of pitch angles obtained with the $D_w = 0.25$ filter.

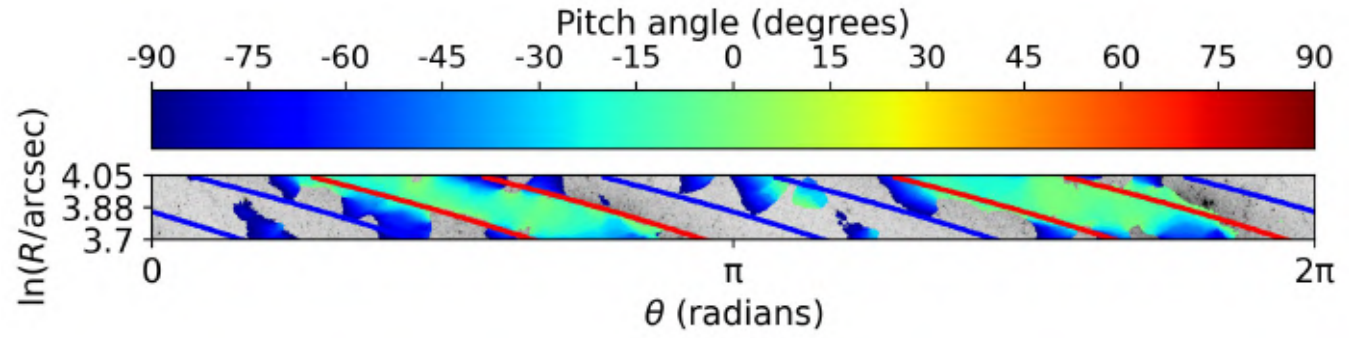

(c) Distribution of pitch angles obtained with the $D_w = 0.5$ filter.

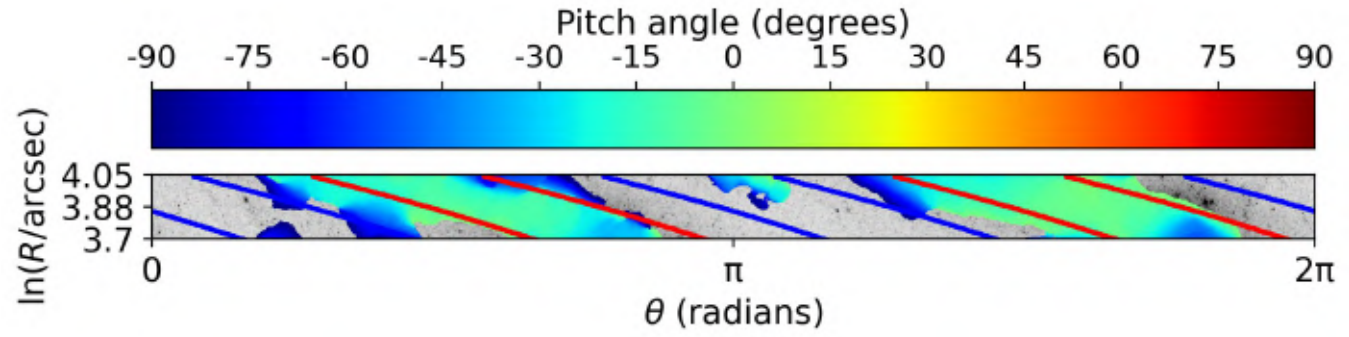

(d) Distribution of pitch angles obtained with the $D_w = 0.75$ filter.

**Figure A28.** Distribution of pitch angles obtained from the highest correlations at each position $(\ln(R/\text{arcsec}), \theta)$ in the $F335M$ image of NGC 4535. The colorbar represents pitch angle values, ranging from $-90°$ (dark blue) to $90°$ (dark red). The colors of the points indicate the pitch angle associated with the maximum correlation, based on the idea that the strongest correlations occur when the structure has the same pitch angle as the filter at the corresponding position.

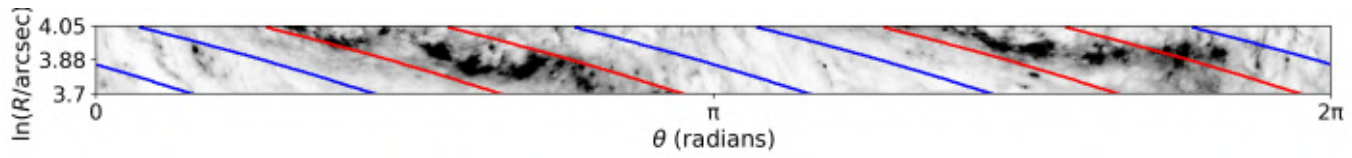

(a) Regions delimited in the $F1130W$ image of the galaxy NGC 4535. The regions defined as the arm are located between the red lines, and the interarm regions are between the blue lines.

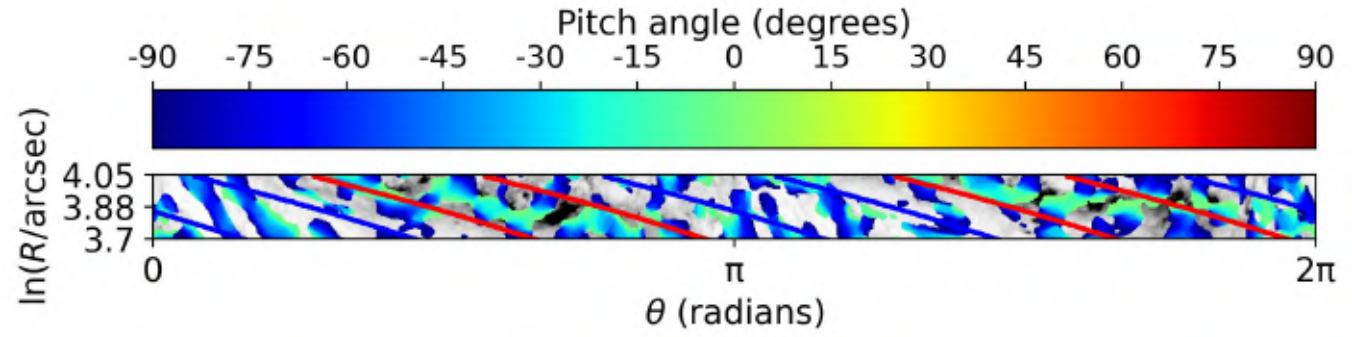

(b) Distribution of pitch angles obtained with the $D_w = 0.25$ filter.

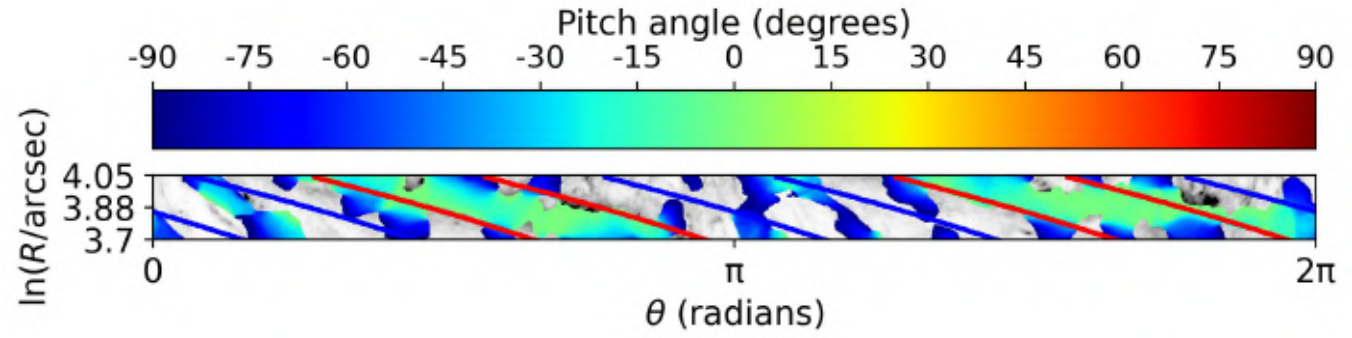

(c) Distribution of pitch angles obtained with the $D_w = 0.5$ filter.

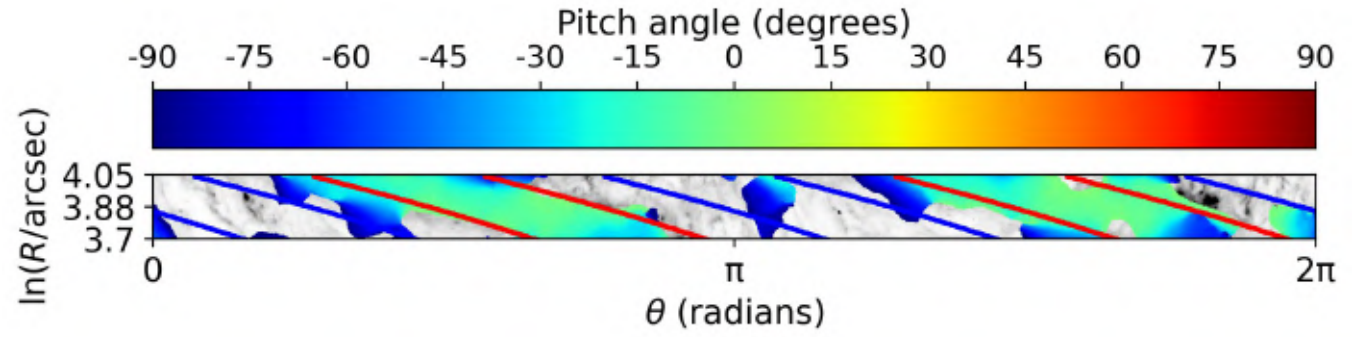

(d) Distribution of pitch angles obtained with the $D_w = 0.75$ filter.

**Figure A29.** Distribution of pitch angles obtained from the highest correlations at each position $(\ln(R/\text{arcsec}), \theta)$ in the $F1130W$ image of NGC 4535. The colorbar represents pitch angle values, ranging from $-90°$ (dark blue) to $90°$ (dark red). The colors of the points indicate the pitch angle associated with the maximum correlation, based on the idea that the strongest correlations occur when the structure has the same pitch angle as the filter at the corresponding position.

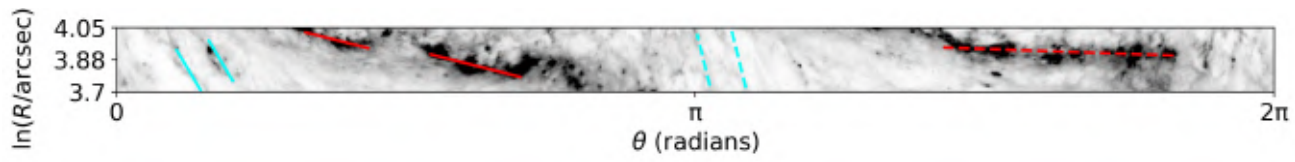

**Figure A30.** Lines overlaid on the $F1130W$ image of the galaxy NGC 4535, showing inclinations that represent the pitch angles of the spiral structures. The solid red line corresponds to $-14°$, the dashed red line to $-2°$, the solid blue line to $-59°$, and the dashed blue line to $-76°$.